\documentclass[journal,a4paper,final]{IEEEtran} 
\IEEEoverridecommandlockouts
\usepackage[T1]{fontenc}
\usepackage[italian,french,english]{babel}
\usepackage{amsmath,amsfonts,amssymb,bm,mathrsfs}
\usepackage{graphicx,color}
\usepackage{subcaption}
\usepackage[dvipsnames]{xcolor}
\usepackage[colorlinks=true,linkcolor=black,citecolor=black,urlcolor=black]{hyperref}
\usepackage{url}
\usepackage{cite}
\usepackage{fancyhdr}
\usepackage{subfiles} 

\graphicspath{{./Figures/}}
\newcommand{\Celsius}{\ensuremath{\mathrm{^\circ C}}} 
\newcommand{\doi}[1]{DOI:~\href{https://doi.org/#1}{#1}}
\newcommand{\unit}[1]{\ensuremath{\mathrm{#1}}}       

\newcommand{\HeaderL}{\parbox[t]{\textwidth}{E. Rubiola and F. Vernotte, \emph{The Companion of Enrico's Chart for Phase Noise and Two-Sample Variances}, IEEE Transact MTT vol~71 no.~7 pp.~2996--3025, July 2023\\This version includes corrections to the normalization of the Hadamard version, and to trivial mistakes.}}
\title{The Companion of Enrico's Chart\\for Phase Noise and Two-Sample Variances
  \author{Enrico Rubiola, \emph{Member, IEEE}, and François Vernotte}
  \thanks{Cite this article as \foreignlanguage{italian}{E. Rubiola} and \foreignlanguage{french}{F. Vernotte}, ``The Companion of Enrico's Chart for Phase Noise and Two-Sample Variances,'' \emph{IEEE Transact Microwave Theory Techniques} vol.~71 no.~7, pp.~2996--3025, July 2023.  \doi{10.1109/TMTT.2023.3238267}.}
  \thanks{This work is partially funded by the \foreignlanguage{french}{Agence Nationale de Recherche (ANR) Programme d'Inves\-tis\-se\-ment d'Avenir} under the following grants:
    ANR-11-EQPX-0033-OSC-IMP (Oscillator IMP project)
    ANR-10-LABX-48-01 (FIRST-TF network), and
    ANR-17-EURE-00002 (EIPHI); 
    and by grants from the \foreignlanguage{french}{Région Bourgogne Franche-Comté} intended to support the above.
  }
  \thanks{The authors are with 
    \foreignlanguage{french}{Université Marie and Louis Pasteur (UMLP)}, formerly \foreignlanguage{french}{Université de Franche Comté (uFC)}, 
    \foreignlanguage{french}{Centre National de la Recherche Scientifique (CNRS)},
    \foreignlanguage{french}{Institut FEMTO-ST},  
    and SUPMICROTECH-ENSMM\@.
    Mail: rubiola@femto-st.fr; francois.vernotte@femto-st.fr.
    ORCID: 0000-0002-5364-1835 (E.R.); 0000-0002-1645-5873 (F.V.).
  }
  \thanks{\foreignlanguage{italian}{Enrico Rubiola} is also with the Division of Quantum Metrology and Nanotechnology, \foreignlanguage{italian}{Istituto Nazionale di Ricerca Metrologica (INRiM)}, Torino, Italy.  Mail: e.rubiola@inrim.it;}
  \thanks{\emph{Corresponding author is \foreignlanguage{italian}{Enrico Rubiola}}, web page http://rubiola.org}
}

\begin{document}
\maketitle

\begin{abstract}
\boldmath
Phase noise and frequency (in)stability both describe the fluctuation of stable periodic signals, from somewhat different standpoints.  Frequency is unique compared to other domains of metrology, in that its fluctuations of interest span at least 14 orders of magnitude, from $10^{-4}$ in a mechanical watch to $10^{-18}$ in atomic clocks. The frequency span of interest is some 12--15 orders of magnitude, from $\mu$Hz to GHz Fourier frequency for phase noise, while the time span over which the fluctuations occur ranges from sub-\unit{\mu s} to years integration time for variances.  Because this domain is ubiquitous in science and technology, a common language and tools suitable to the variety mentioned are a challenge.

This article is at once (1) a tutorial, (2) a review covering the most important facts about phase noise, frequency noise and two-sample (Allan and Allan-like) variances, and (3) a user guide to ``Enrico's Chart of Phase Noise and Two-Sample Variances.''
In turn, the Chart is a reference card collecting the most useful concepts, formulas and plots in a single A4/A-size sheet, intended to be a staple on the desk of whoever works with these topics.  The Chart is available under Creative Commons 4.0 CC-BY-NC-ND license from Zenodo, \doi{10.5281/zenodo.4399218}.  A wealth of auxiliary material is available for free on the Enrico's home page \href{http://rubiola.org}{http://rubiola.org}.

\color{BrickRed}
This version includes the corrections for an unfortunate error in the normalization of the Hadamard version, and some corrections for trivial (albeit sometimes subtle) mistakes. 
\color{Black}

\unboldmath
\end{abstract}


\setcounter{tocdepth}{1}
\addtolength{\parskip}{-8pt}
\tableofcontents
\addtolength{\parskip}{+8pt}

\section{Introduction\label{sec:Introduction}}

\ifSubfilesClassLoaded{\setcounter{section}{0}\section{Introduction}}{}
\IEEEPARstart{P}{hase noise} and frequency instability are equivalent concepts, to the extent that frequency is the derivative of the instantaneous phase.  However, the choice of terminology, of mathematical tools and of experimental methods really depends on what aspect of the noise our system is sensitive to, or which aspect we focus on.
The Power Spectral Density (PSD) of the phase fluctuations of a periodic signal, as a function of the Fourier frequency, is the preferred tool to describe phase noise.  Similarly, the frequency instability is described by the PSD of frequency fluctuations, or more often by the two-sample (Allan or Allan-like) variance of the fractional frequency, as a function of the measurement time.

\begin{figure}[b]
  \begin{minipage}[t][24mm][t]{0.27\columnwidth}
    \vfill
    \includegraphics[scale=0.21]{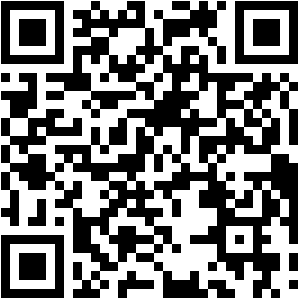}
  \end{minipage}%
  \begin{minipage}[t][11mm][b]{0.23\columnwidth}
    \href{https://doi.org/10.5281/zenodo.4399218}{\footnotesize\sffamily Download\\[-0.5ex]Enrico's Chart\\[-0.5ex]from Zenodo}
  \end{minipage}%
  \begin{minipage}[t][23mm][b]{0.23\columnwidth}
    \flushright
    \href{https://doi.org/10.1109/TMTT.2023.3238267}{\footnotesize\sffamily Download the\\[0ex]Companion article from\\[-0.5ex]IEEE Xplore}
  \end{minipage}%
  \begin{minipage}[t][24mm][b]{0.27\columnwidth}
    \flushright
    \includegraphics[scale=0.21]{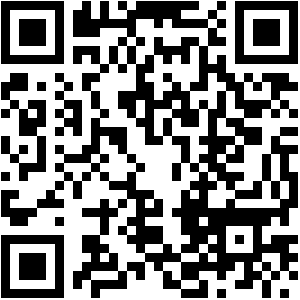}
  \end{minipage}
  \caption{QR codes to download Enrico's Chart and the Companion article.}
  \label{fig:QR-Code}
\end{figure}

\begin{figure*}[t]
  \centering
  \includegraphics[angle=0,width=\textwidth]{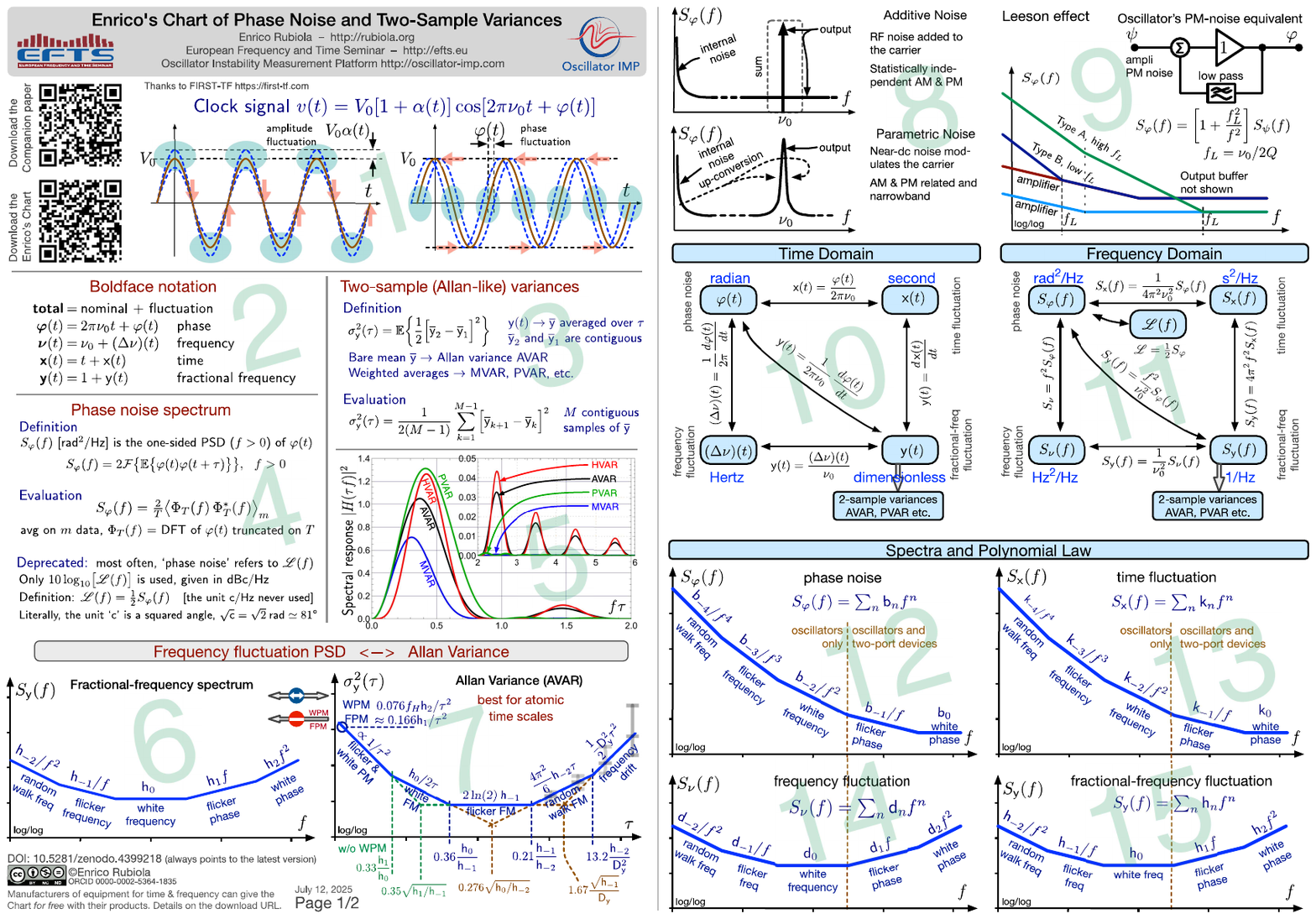}
  \caption{Enrico's Chart of Phase Noise and Two-Sample Variances, front side.}
  \label{fig:p1}
\end{figure*}

\begin{figure*}[t]
  \centering
  \includegraphics[angle=0,width=\textwidth]{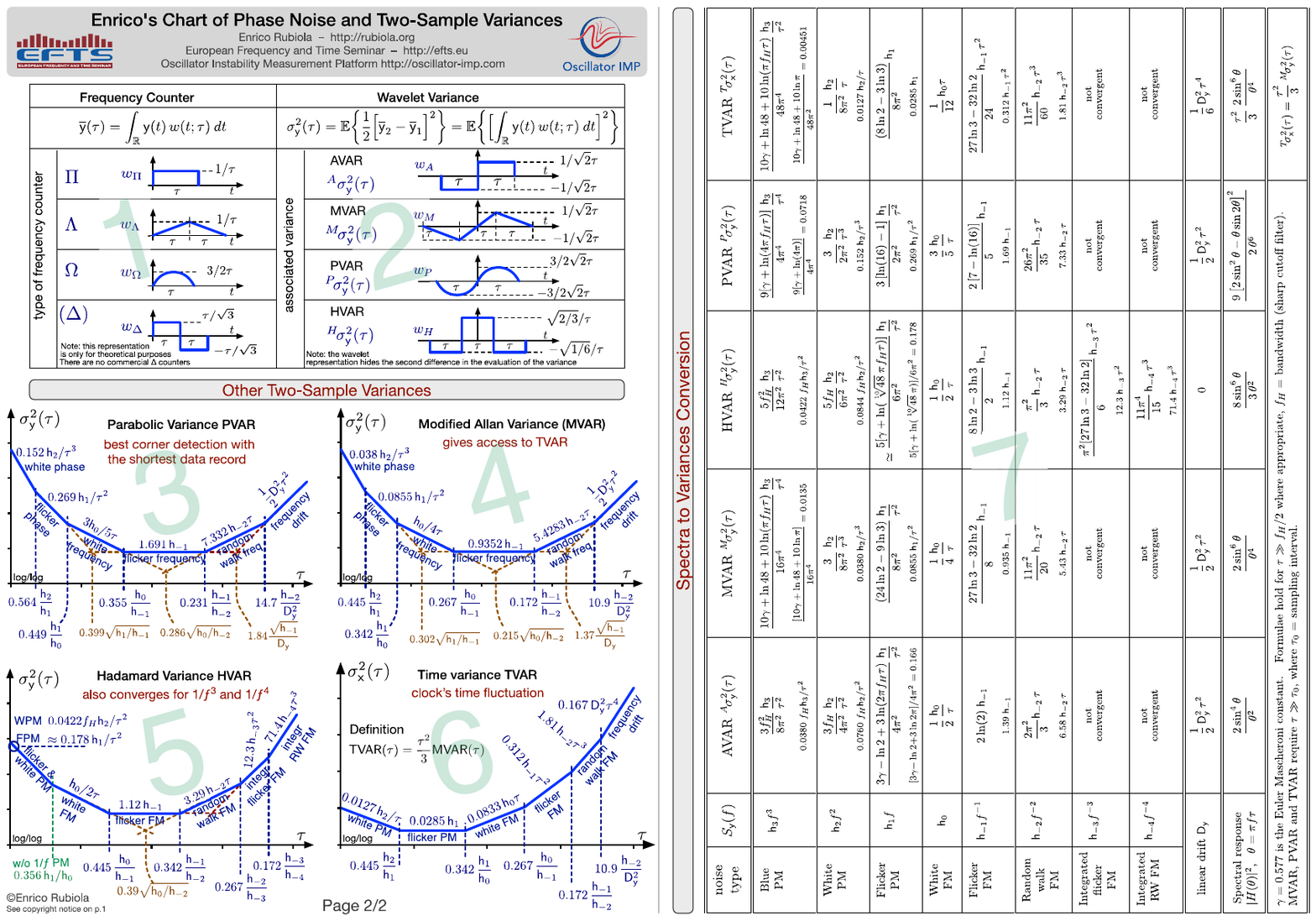}
  \caption{Enrico's Chart of Phase Noise and Two-Sample Variances, back side.}
  \label{fig:p2}
\end{figure*}

\subsection{A Taste of Phase Noise and Frequency Stability}\label{sec:Taste}
There are many reasons to be interested in phase noise and frequency instability, most of which are related with spectrum broadening, with timing uncertainty, and with reduced coherence time. 
The impact of phase noise and frequency instability is surprisingly ubiquitous, from everyday technology to fundamental science.  Let us go through some examples.

In radio systems, sensitivity or dynamic range may be limited by phase noise sidebands appearing (i) on the local oscillator of a superheterodyne receiver, causing neighboring strong signals to mask the desired IF signal (reciprocal mixing) \cite{Grebenkemper-1981}, \cite[Sec.~2.5]{Sosin-1971}, (ii) on strong adjacent transmitted signals themselves in channelized systems (the near-far problem, in which sideband noise from local users can overpower distant cellular base stations), or (iii) on strong reflected radar signals such as ground clutter, limiting the ranging accuracy and the detection of small objects in the clutter \cite[Chap.~6]{Skolnik-2008}, \cite{Leeson-1966b,Leeson-1964}.  In these systems, amplitude fluctuation (AM) is limited in oscillators and nonlinear power amplifiers, so phase noise (PM) is generally the dominant source of noise sidebands. Similar problems are found in LIDAR, RADAR's optical counterpart. 
Phase- and frequency-modulated systems are, of course, directly subject to noise in the frequency domain; examples include analog frequency-division FM multiplex systems and digital QAM modulation systems.

In digital systems and communications, and in microelectronics as well, the term `jitter' is often used for phase noise integrated over the appropriate bandwidth and converted into time fluctuations.  In turn, time fluctuations increase the bit error rate (see for example \cite{DaDalt-2018} and \cite{Li-2008}).  In such systems, the `quality' of the digital clock is often shown as the eye pattern (eye diagram), which is corrupted by the jitter (see \cite[Sec~1.2]{DaDalt-2018}, \cite[Sec.~4.1]{Li-2008} and `\href{https://en.wikipedia.org/wiki/Eye_pattern}{Eye pattern}' entry on Wikipedia).
Accurate clock synchronization is a major leap forward in 5G/6G wireless systems \cite{Lin-2018,Tataria-2021}.  

The community of power grids is looking at frequency stability for future power grids \cite{Conte-2021,Kruse-2021,ENTSOE-2016} because sustainability requires a spread of generator technologies with different inertia (latency between demand for increased/reduced power, and actual power delivery).  We believe that, at some point the ``generators' inertia'' will be identified with the integration time $\tau$ of the Allan variances.  Insufficient synchronization is blamed as a co-factor of the 2003 Northeast Blackout in the USA \cite{Amelot-2012}.
  
Unique requirements arose with the development of the Deep Space Network. The DSN required a blend of long-term stability for initial acquisition of weak signals and short-term stability for maintaining lock.  This new combined need led to the 1964 IEEE-NASA Symposium on Short-term Frequency Stability \cite{Chi-1964}, to a Special Issue of Frequency Stability published in 1966 in the Proc. IEEE \cite{Chi-1966}, and to the definitive 1971 Barnes et al.\ article \cite{Barnes-1971}. The latter is a precursor of the IEEE Standard 1139 \cite{IEEE-1139-2022}. That effort to update this standard continues today.  

The Allan variance tools also find applications in geodesy and astrometry \cite{Malkin-2016}.  The short-term fluctuations of oscillators used in Very-Large Baseline Interferometry (VLBI) cannot be compensated numerically, and limit the detection sensitivity.  
Time fluctuations are critical in gravitational wave experiments (LIGO) because of the tiny spacetime warp to be detected, and in the RF cavities of particle accelerators \cite[Chapters by F. Tecker, and H. Damereau]{Holzer-2018}, where they cause intensity loss or energy loss in the beam.

In quantum computing, the correlation time limits the lifetime of a qubit \cite{Ball-2016}.  

Optics supports a major trend in high-purity signals \cite{Diddams-2020}, and in atomic oscillators and clocks as well \cite{Udem-2002}, because the time interval associated with a phase angle is ${\sim}10^{-4}$ smaller than at microwave frequencies. The femtosecond laser, which enabled the first direct synthesis from RF to optics, played a major rôle in this trend.  The Nobel Prize in Physics was awarded to John L. Hall and Theodor W. Hänsch in 2005 for the femtosecond laser (together with Roy J. Glauber).  
See also \cite{Cundiff-2003,Kuse-2017,Xie-2017a}.
Phase noise challenges the networks intended for fundamental science and clock comparison between metrological labs using optical fibers shared Internet data traffic \cite{Calosso-2016b,Lisdat-2016,Delva-2017,Nicolodi-2014}.

In the new version of the International System of Units SI \cite{SI-2019}, virtually all measurement units rely on time.  Thus time fluctuations, and equivalently frequency fluctuations, impact all branches of metrology.  For example, the ampere, unit of electric current, relies on counting the electrons flowing in the unit of time.  And the Josephson voltage standard is based on the conversion from photon energy to voltage.  The Nobel Prize in Physics was awarded to Brian D. Josephson in 1973 for the theoretical prediction of this effect.

We now discuss Enrico's Chart, the subject of this paper, which brings the efforts to define and standardize phase noise and time-domain frequency instabilities to a concise two-page working summary.

\subsection{Enrico's Chart}
\emph{Enrico's Chart of Phase Noise and Two-Sample Variances} is a reference card collecting the most useful definitions, formulas and plots in the domain of phase noise and frequency stability.  Starting from the first draft in 2011, it has been improved from a loose single page on the back of the program of the European Frequency and Time Seminar\footnote{The EFTS \href{http://efts.eu}{http://efts.eu} is a full-week crash course with lectures and lab sessions funded in 2013, and held in Besançon every summer since.} to quite a dense front/back format.  At least 1000 plasticized prints have been distributed as learning material at the EFTS and at invited conferences, courses and seminars.  
The current form results from the scientific contribution of the two authors, and from feedback and amendments by the users.  The name on the Chart is mainly historical, but it reflects Enrico's tenacity in maintaining, updating, improving the graphic design, and in distributing the Chart.

\subsection{License and Distribution}
Enrico's Chart is a digital object (PDF file) available from Zenodo as \doi{10.5281/zenodo.4399218}, and released under Creative Commons 4.0 CC-BY-NC-ND license.  This DOI is called \emph{concept DOI} because it always resolves to the latest version.  Albeit Zenodo delivers a separate DOI for each version, version DOIs should only be used in special cases.  The QR codes of Fig.~\ref{fig:QR-Code} point to the Chart and to this article.  Redistribution is encouraged as the Internet link or as the QR code, not as a file.  

The `NC' and `ND' copyright attributions deserve a comment.  The `NC' restriction is mainly intended to prevent selling the Chart for profit.  In contrast, if Companies distribute the Chart \emph{for free} with their products, that is \emph{not considered commercial use}, and we \emph{encourage} this practice.  The equipment employed for measurements in time and frequency is a notable example.  The `ND' restriction preserves the academic independence of the Chart, preventing Companies from modifying it, or from building advertisements upon it.

For any other use, Enrico himself is the preferred contact.

\subsection{About this Article}
This article is intended (i) to accompany the Chart, (ii) as a short tutorial, and (iii) as learning material for lectures.  Besides, (iv) it can be cited as a summary of modern notation, provided the reader agrees with our choices. 

The reader having an already good understanding of time and frequency may go straight to Enrico's Chart and put this article aside for later reading, or refer to students and younger colleagues.  Less experienced people may appreciate this article as a tutorial or as a review.  To them, if they have been around for long enough, phase noise and frequency stability may relate to a well identified problem.

A reduced copy of the Chart is included (Fig.~\ref{fig:p1} and \ref{fig:p2}); the indication ``\emph{Region 1.n}'' in the text means that we refer to the region \emph{n} defined by the watermarks on page 1 (Fig.~\ref{fig:p1}) of the Chart, while ``\emph{Region 2.n}'' refers to the region \emph{n} on page 2 (Fig.~\ref{fig:p1}).  In any event, it is a good idea to have on hand a separate copy of the Chart.

\subsection[Suggested Readings]{Suggested Introductory Readings\label{bib:General}}
We advise starting from the latest version of the IEEE Standard 1139 \cite{IEEE-1139-2022}.  The appendices of this standard cover topics similar to ours, but presentation and standpoint are surprisingly different. This article is our personal view, whereas the IEEE Standard is the outcome of a rather large committee.
The Recommendations of the International Telecommunication Union (ITU) \cite{ITU-TF.538,ITU-G.8260} are a must too.
The European Telecommunications Standards Institute (ETSI) provides a series of standards related to the clock synchronization in networks \cite[Part 1-7]{ETSI-300-462-Series}.

Barnes et al.\ 1971 \cite{Barnes-1971} is one of the very first articles that all readers should study.  It defines the language and the notation still in use, it introduces the spectra, the variances and their relations, and it explains the early experimental methods.  The quantities $\varphi$, $\mathsf{x}$, $\Delta\nu$ and $\mathsf{y}$ were first defined there.
People interested in the birth of the scientific ideas that originated most concepts of phase noise and of frequency stability should review~\cite{Chi-1964}.  This is the Proceedings book of a one-time IEEE-NASA conference that took impetus from the DSN\@.
Rutman 1978~\cite{Rutman-1978} is a review article about the progress on the concepts of phase noise and frequency stability after the early ideas.  See also~\cite{Rutman-1991}.  

Turning our attention to books and booklets, the Riley handbook \cite{Riley-2008} is free available, sponsored by NIST\@.  Emphasis is on Allan and Allan-like variances, rather than noise spectra, making extensive use of the Stable32 software package\footnote{This and other software packages are discussed in Section \ref{sec:Software}.}. 
Owen 2004 \cite{Owen-2004} is a rather extensive practical guide about phase noise, albeit elderly.
Kroupa 1983 \cite{Kroupa-1983} is a collection of classical articles about phase noise and frequency stability, a few of which are cited elsewhere in this article.  The same author also published a monograph \cite{Kroupa-2012}.
Reference~\cite{Sullivan-1990} is another edited book collecting classical articles from NIST\@.
Rohde et al.~2021, \cite[Chapter 2 (136 pages)]{Rohde-2021} is a recent text entirely about phase noise and frequency stability. 

Additional material and large slideshows are available on the Enrico's home page http://rubiola.org.  Upon request, the original PPTX files may be released to qualified users.  We turn now to the Chart and Fig.~\ref{fig:p1}.

\ifSubfilesClassLoaded{\bibliography{Ref-short,References,Ref-local}}{}
\end{document}

\section{Phase Noise and Amplitude Noise\label{sec:PhaseNoise}}

\ifSubfilesClassLoaded{\setcounter{section}{1}\section{Phase Noise and Amplitude Noise}}{}

\subsection[The Clock Signal]{The Clock Signal (Region 1.1)\label{sec:Clock-signal}}
\begin{figure}[t]\centering
  \includegraphics[bb = 0cm 0cm 8.0cm 9.2cm,angle=-90,width=0.98\columnwidth]{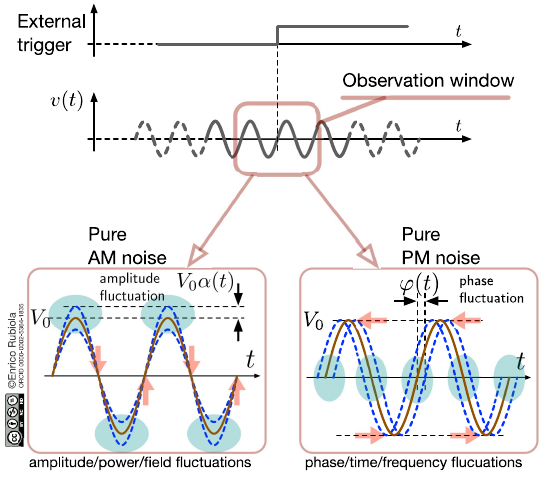}\par
  \vspace*{1ex}
  \caption{Clock signal \eqref{eqn:Clock-signal}, observed with an oscilloscope triggered by an external noise-free reference.   The shadowed (pink) arrows indicate where the signal is stationary, unaffected by noise.  The elliptical shadowed (cyan) regions emphasize where noise shows up most.}  
  \label{fig:Clock-signal}
\end{figure}

A pure sinusoidal signal affected by AM and PM noise\footnote{Amplitude Modulation noise and Phase Modulation noise.} (Fig.~\ref{fig:Clock-signal}) can be written as
\begin{equation}
  v(t) = V_0\bigl[1+\alpha(t)\bigr]\cos\bigl[2\pi\nu_0t+\varphi(t)\bigr] \,,
  \label{eqn:Clock-signal}
\end{equation}
where $V_0$ is the amplitude, $\nu_0$ the frequency, $\alpha(t)$ is the random fractional amplitude, and $\varphi(t)$ is the random phase.  This representation is general, not limited to electrical signals as the symbol $v(t)$ suggests.
In the IEEE Standard 1139, the amplitude is written as $V_0+\epsilon(t)$, and the fractional amplitude is defined as $\alpha(t)=\epsilon(t)/V_0$, which is equivalent to our notation.  However, \eqref{eqn:Clock-signal} is a simplification of \eqref{eqn:Clock-extended}.

The bandwidth $B$ of $\varphi(t)$ and $\alpha(t)$ deserves some attention.  
Because modulation needs sidebands with appropriate symmetry around $\nu_0$, the theoretical maximum $B$ is equal to $\nu_0$.  That said, in numerous cases of interest the quantity $B/\nu_0$ is rather small, likely of $10^{-4}\ldots10^{-2}$, and not necessarily the same for $\varphi(t)$ and $\alpha(t)$.  The autocorrelation function of noise spans over a time of the order of $1/B$, or equivalently $\nu_0/B$ oscillations. Thus, cycle-to-cycle noise is too small to be visible.  This is why we use the external trigger in Fig.~\ref{fig:Clock-signal}.  Otherwise, triggering on signal itself would require a dual time base with an ideally stable delay $>1/B$.

Naively, one may believe that $\overline{\alpha(t)}\approx0$ and $\overline{\varphi(t)}\approx0$ hold in general, where the `overline' means average, or to take these conditions as necessary.
In systems of practical interest we observe quite small amplitude and frequency fluctuations, thus we can assume that $|\alpha|{\,\ll\,}1$.
We have measured $|\alpha|$ of the order of $10^{-7}\ldots10^{-4}$ in quartz oscillators and in synthesizers \cite{Rubiola-2007-AM}.  The hypothesis that $|\varphi(t)|\ll1$~rad makes sense only for two-port systems, where $\varphi(t)/2\pi\nu_0$ is the fluctuation of the input-to-output delay.  Conversely, the phase of oscillators always contains divergent processes, so we switch our attention to $|\dot{\varphi}|/2\pi\nu_0$.  Practical value span from $10^{-4}$ for cheap watches to $10^{-16}$ (accuracy) and $10^{-18}$ (fluctuations) for the frequency-standard prototypes found in metrology labs.

\subsection[Phase Noise Spectrum]{Phase Noise Spectrum (Region 1.4)\label{sec:PN-spectrum}}
The \emph{variance}\footnote{The two sample variance we find later is a specialized instance of this general concept.} (generalized power) of a quantity $q$, denoted by $\sigma^2$, is the mean square of $q$.  Formally, $\sigma^2=\mathbb{E}\{|q-\mu|^2\}$, where $\mathbb{E}\{\,\}$ is the mathematical expectation, and $\mu=\mathbb{E}\{q\}$ is the average. 
The \emph{power spectral density}, denoted with $S(f)$, tells us how $\sigma^2$ is distributed in frequency.  
The variable $f$ is called `Fourier frequency' to differentiate it from the carrier frequency (constant).  The single-sided PSD ($f>0$) is generally preferred to the two-sided PSD with no need of saying so\@.
Referring to the quantity $q$, the subscript $q$ is optionally added, as in $\sigma^2_q$ and $S_q(f)$.

For our purposes, $S_\varphi(f)$ is the quantity that should be used to describe the phase noise.  In practical measurements $S_\varphi(f)$ is best \emph{evaluated} as 
\begin{equation}
\widehat{S_\varphi}(f) = \frac2T \Bigl< \Phi_T(f)\,\Phi_T^\ast(f) \Bigr>_m
\qquad[\unit{rad^2/Hz}]\,,
\label{eqn:Sphi}
\end{equation}
where the hat accent means `estimator' of $S_\varphi$ (the reader may ignore it at first reading), $\Phi_T(f)$ is the Fast Fourier Transform of $\varphi(t)$ sampled and truncated on an appropriate duration $T$, the superscript `$\ast$' means complex conjugate, the $\left<\;\right>_m$ operator is the average on $m$ acquisitions, and the factor of 2 is needed for energy conservation after deleting the negative frequencies\footnote{Because $\varphi(t)$ is a real function, its Fourier transform $\Phi(f)$ is Hermitian function, that is, $\Phi(f)=\Phi^\ast(-f)$.  Accordingly, $\Re\{\Phi(f)\}$ is even function of $f$, and $\Im\{\Phi(f)\}$ is odd function of $f$.  Thus, all the information is contained in the $f>0$ half-plane, and the $f<0$ half-plane is redundant and can be deleted.  In the case of the FFT, such redundant region is generally mapped to $f_s/2>f>f_s$, where $f_s$ is the sampling frequency.}.  
Equation \eqref{eqn:Sphi} is used in the Welch algorithm for the estimation of power spectra \cite{Welch-1967}.
The optional data overlapping used in the Welch algorithm, and the optional window function are not explicit in \eqref{eqn:Sphi}.  The most popular window functions are known under the names Bartlett, Blackman-Harris, flat-top, Hamming, Hann, Parzen and of course Welch.
The quantity $S(f)=\ldots$ defined by \eqref{eqn:Sphi}, with optional windowing but with no hat and no averaging ($m=1$), is called periodogram.

\subsection{Deeper Thoughts About the PSD\label{sec:PSD-thoughts}}
In statistics, the PSD $S(f)$ is formally defined as the Fourier transform of the autocovariance of a \emph{random process}.  In turn, the random process is a set of sample functions or distributions, called realizations, each of which is indexed by one outcome of a random experiment.  Time statistics and ensemble statistics are different concepts, and they are interchangeable only in the case of ergodic processes.

In contrast, some Authors take \eqref{eqn:Sphi} as the definition of $S_\varphi(f)$, with no `hat.'  The problem with this choice is that it avoids some key concepts of statistics, making the uncertainty difficult to understand.

Using the mathematical concept of process requires one to define the random experiment.  Since mathematics does not provide general rules for this, we need to make a choice.  For example, we may identify the random experiment with a large abstract class, or more pragmatically with the action of picking one oscillator from preproduction samples, from a batch, or from a larger part of a production.  In turn, a realization may be identified with the waveform $\varphi(t)$ obtained by comparing such oscillator with a noise-free reference.  This opens deeper questions, like the meaning and the scientific legitimacy of `typical' spectra (also stability, aging, drift, and other parameters) found in data sheets.

As a matter of fact, the right-hand side of \eqref{eqn:Sphi} can always be calculated from experimental data.  The question arises, whether or not the estimator converges to the PSD
\begin{align}
  \widehat{S_\varphi}(f)&\rightarrow S_\varphi(f)
  \qquad\text{for large $m$}\,.
  \label{eqn:Convergence}
\end{align} 
This is true for stationary processes (the statistical properties are independent of the origin of time).  That said, relevant processes often found in oscillators, like flicker and random walk of frequency, are not stationary in a strict sense.  Fortunately, evaluating \eqref{eqn:Sphi} such processes can be treated as stationary.

Finally, we notice some analogies between \emph{stationarity} and \emph{repeatability}, and also between \emph{ergodicity} and \emph{reproducibility}, with the caveat that stationarity and ergodicity are mathematical concepts, while repeatability and reproducibility are defined by the International Vocabulary of Metrology VIM \cite[Entry 2.21 and 2.25]{VIM-3} and have quite a technical meaning related to experimental outcomes.


\subsection{The Quantity $\mathscr{L}(f)$ and the Related Measurement Units}
Most often in the technical literature and data sheets, the term \emph{phase noise} refers to the quantity $\mathscr{L}(f)$, defined since the IEEE Standard 1139-1988 \cite{IEEE-1139-1988} as
\begin{equation}
  \text{definition:}\qquad
  \mathscr{L}(f)=\frac12 S_\varphi(f)\,.
  \label{eqn:L(f)}
\end{equation}
Plots and numerical values are always given as 
\begin{equation}
  10\log_{10}\mathscr{L}(f)
  \qquad[\unit{dBc/Hz}]\,.
  \label{eqn:L(f)dBc}
\end{equation}

After over 30 years in the field, we believe that $\mathscr{L}(f)$ is misleading and, if the community started from scratch, $S_\varphi(f)$ would be used instead.
The roots are found in a one-time symposium co-organized by IEEE and NASA in 1964 \cite{Chi-1964}.

Because $\varphi(t)$ is an angle, the dimension of $S_\varphi(f)$ is a square angle multiplied by time.  Accordingly, the appropriate unit is \unit{rad} for $\varphi(t)$ and \unit{rad^2s} for $S_\varphi(f)$, but the equivalent unit \unit{rad^2/Hz} is generally preferred.  In logarithmic units we use 
\begin{equation}
  10 \log_{10} S_\varphi(f) 
  \qquad[\unit{dBrad^2/Hz}]\,.
  \label{eqn:SphidB}
\end{equation}
It follows from \eqref{eqn:L(f)} that $\mathscr{L}(f)$ has the same dimension as $S_\varphi(f)$ but different units, like a mass in kg or in lb.  Accordingly, the unit associated to $\mathscr{L}(f)$ should be \unit{\mathfrak{A}^2/Hz} where \unit{\mathfrak{A}} in a never-used unit of angle that equals $\sqrt{2}\;\unit{rad}\simeq81^\circ$.
It worth mentioning that there is no reason to change symbol after switching unit, like in $M=1.5$ \unit{kg} and $M=3.3$ \unit{lb}.  So, why should we change the symbol from $S_\varphi$ to $\mathscr{L}$ because of the unit of angle?

The unit \unit{dBc/Hz} is even more confusing.
It is obvious that `\unit{c}' cannot be read `referred to the carrier,' as most people have in mind.  
Instead, taking \eqref{eqn:L(f)}-\eqref{eqn:L(f)dBc} literally, `$\unit{c}$' is a square unit of angle, $\unit{\mathfrak{A}^2}=2\;\unit{rad^2}$.  Notwithstanding this, `\unit{c}' and `\unit{c/Hz}' alone, with no `\unit{dB},' are never seen in the literature, and no unit of angle associated with $\mathscr{L}(f)$ either.

We advocate abandoning $\mathscr{L}(f)$ in favor of $S_\varphi(f)$ because $S_\varphi(f)$ is consistent with the International System of Units SI \cite{SI-2019}---with the minor caveat that the decibel is a non-SI unit accepted for use with the SI units; in contrast, $\mathscr{L}(f)$ is not consistent.  In order to prevent discontinuity in notation and units, $\mathscr{L}(f)$ may still be kept as an auxiliary scale.

\subsection[The Former, and Wrong Definition]{The Former Definition of $\mathscr{L}(f)$, and the Deprecated Terms `SSB Noise' and `Offset Frequency'\label{sec:L(f)}}

In the early time, $\mathscr{L}(f)$ was defined as  
\begin{equation}
  \mathscr{L}(f)=\frac{\text{noise power in 1 Hz bandwidth}}{\text{carrier power}}
  \qquad\text{(wrong!)}\,,
  \label{eqn:L(f)obsolete}
\end{equation}
and always given as $10\log_{10}\mathscr{L}(f)$ in \unit{dBc/Hz}.  
The historical meaning of the symbol `\unit{c}' is `referred to the carrier.'  For example, $-120$ \unit{dBc/Hz} means that `the noise sideband in 1 Hz bandwidth is 120 dB below the carrier power.'
Sadly, more than 30 years after the first version of the IEEE Standard 1139 \cite{IEEE-1139-1988}, the old definition is still in the mind of numerous engineers and physicists.  

The major problem is that phase modulation, and amplitude modulation as well, need sidebands with the appropriate symmetry with respect to the carrier.  The relationships between LSB and USB define the type of modulation, AM, PM or any combination of these.  Thus, the noise power in one sideband does not say what fraction goes to phase noise and to amplitude noise.
Consequently, terms `SSB noise' and `offset frequency' are unsuitable to describe the phase noise, and should be avoided.  
The variable $f$ in $S_\varphi(f)$ should be referred to as the `modulation frequency,' or as the `Fourier frequency' in formal mathematical language.

We point out that the use of \eqref{eqn:L(f)obsolete} for phase noise is
\begin{itemize}
\item\emph{conceptually incorrect}, because it does not divide PM noise from AM noise,
\item\emph{experimentally incorrect}, because $\mathscr{L}(f)$ is always measured with a phase detector using $\mathscr{L}(f)=\frac12 S_\varphi(f)$, instead of the noise-to-carrier ratio,
\item\emph{deprecated}, as a result of a major effort to adhere to the SI as the global system of units,
\item\emph{unsuitable} to describe phase noise exceeding a small fraction of a radian.
\end{itemize}
Large phase swings are common at low Fourier frequencies (long measurement time) and in optics, where the carrier frequency is high.
When $\int S_\varphi(f)\,df$ integrated over the full spectrum approaches or exceeds 1 \unit{rad^2/Hz}, \eqref{eqn:L(f)obsolete} breaks down because the noise sidebands in angular modulations come at the expense of the carrier power. 
Conversely, there is no reason to question the validity of \eqref{eqn:Sphi} and \eqref{eqn:L(f)}, even in the case of huge number of cycles.  In the presence of large angles, the correct measurement of  $S_\varphi(f)$ is only a matter of hardware design.

Finally, the power ratio as defined in the right-hand side of \eqref{eqn:L(f)obsolete} is definitely not evil---provided we don't call it `phase noise' and we don't use the same symbol $\mathscr{L}(f)$ defined by \eqref{eqn:L(f)}.  In some domains the noise per unit of bandwidth matters, regardless of the physical origin.  This is the case of the reciprocal mixing in the superheterodyne receiver \cite[Sec.~8.7.2]{Razavi-2012}, where a strong emission in neighbor channels leaks into the IF because of the local-oscillator noise sidebands.  Similarly, Doppler radar visibility of subclutter targets is typically determined by noise in the oscillator's sideband power density.  See  \cite{Leeson-1964,Leeson-1966b,Grebenkemper-1981}, and \cite{Barton-1998,Skolnik-2008} as general references on radars.  As noted previously, due to amplitude limiting, sideband noise in the cited examples is generally dominated by phase noise, leading to the persistence of the spectrum sideband usage in these applications.

\subsection{Suggested Readings}%
Digging in the old literature, the earliest use of the phase noise PSD is arguably Victor 1956 \cite{Victor-1956}.  In the NASA Symposium \cite{Chi-1964} both $S_\varphi(f)$ and the sideband-to-carrier ratio were used, the latter mainly for radar and receivers.  However, the symbol $\mathscr{L}(f)$ had still to come.  Its early occurrences in archival documents are Blair \cite{Shoaf-1974} and Shoaf, \cite[p.~167 and p.~179--180]{Blair-1974}, both in 1974.
Interestingly, the experimental techniques described in \cite{Chi-1964} rely on phase detectors, which points to $S_\varphi(f)$.  Horn \cite{Horn-1969} is the one and only remarkable counter-example we found, where the sideband-to-carrier ratio is measured with practically useful dynamic range.  
Note that the history of phase noise is beyond our scope and our skills; we recommend Leeson's review article \cite[Sec.~III--V]{Leeson-2016} and a blog page on Engineering and Technology History Wiki\footnote{D.B Leeson, \emph{First-Hand: Phase Noise}, \url{https://ethw.org/First-Hand: Phase_Noise}.} submitted by Leeson, which refers to additional material of historical interest.

Most of the references about phase noise relate to experimental methods, thus they are moved to Sec.~\ref{sec:Techniques}.  The early ideas found in \cite{Chi-1964} were better formalized in \cite{Barnes-1971} (see also \cite{Rutman-1978}).

There is a Special Issue of the Proceedings of the IEEE \cite{Chi-1966}, three useful books, Robins 1984 \cite{Robins-1984}, Da Dalt 2018 \cite{DaDalt-2018} and Li \cite{Li-2008}, and a free booklet from the NPL \cite{Owen-2004}.
There is very little about AM noise.  The reader should refer to \cite{Rubiola-2005-arXiv}.

\ifSubfilesClassLoaded{\bibliography{Ref-short,References,Ref-local}}{}
\end{document}

\section[Useful Quantities]{Useful Quantities (Region 1.10--1.11)}\label{sec:Quantities}

\ifSubfilesClassLoaded{\setcounter{section}{2}\section{Useful Quantities (Region 1.10--1.11)}}{}

\subsection{Phase Time Fluctuation (or Phase Time) $\mathsf{x}(t)$}
The \emph{phase time} is the phase fluctuation converted into time
\begin{equation}
  \text{definition:}\qquad
  \mathsf{x}(t)=\frac{\varphi(t)}{2\pi\nu_0} 
  \qquad[\unit{s}]\,.
  \label{eqn:x(t)}
\end{equation}
Its spectrum $S_\mathsf{x}(f)$ follows from the definition of $\mathsf{x}(t)$ using the property of the Fourier transform that the derivative operator $d/dt$ maps into a multiplication by $j2\pi f$, where $j^2=-1$, thus into a multiplication by $4\pi^2f^2$ in the PSD because of \eqref{eqn:Sphi}.  Accordingly, 
\begin{equation}
  S_\mathsf{x}(f)=\frac{1}{4\pi^2\nu_0^2}S_\varphi(f) 
  \qquad[\unit{s^2/Hz\equiv s^3}]\,.
  \label{eqn:Sx}
\end{equation}

\subsection{Frequency Fluctuation $(\Delta\nu)(t)$}
The instantaneous \emph{frequency fluctuation} is
\begin{equation}
  \text{definition:}\qquad
  (\Delta\nu)(t)=\frac{1}{2\pi}\frac{d\varphi(t)}{dt} 
  \qquad[\unit{Hz}]\,.
  \label{eqn:Delta-nu(t)}
\end{equation}
Enclosing $\Delta\nu$ in parentheses emphasizes the fact that $(\Delta\nu)$ is an unbreakable quantity.  The PSD is found using the property that $d/dt\rightarrow {\times}4\pi^2f^2$, thus
\begin{equation}
  S_{\nu}(f)=f^2S_\varphi(f) 
  \qquad[\unit{Hz^2/Hz\equiv Hz}]\,.
  \label{eqn:Snu}
\end{equation}
Albeit the notation $S_{\Delta\nu}$ is more common than $S_\nu$, the latter should be preferred because the PSD is already insensitive to the constant $\nu_0$, by definition, thus it shows only the fluctuation $\nu-\nu_0$.

\subsection{Fractional Frequency Fluctuation $\mathsf{y}(t)$}
The \emph{fractional frequency fluctuation} is 
\begin{equation}
  \text{definition:}\quad~
  \mathsf{y}(t)=\frac{(\Delta\nu)(t)}{\nu_0}
  \qquad[\unit{dimensionless}]\,.
  \label{eqn:y(t)}
\end{equation}
It follows from \eqref{eqn:x(t)}--\eqref{eqn:y(t)} that
\begin{equation}
  \mathsf{y}(t) =
  \frac{d\mathsf{x}(t)}{dt} =
  \frac{1}{2\pi\nu_0}\frac{d\varphi(t)}{dt}\,.
\end{equation}
Again, the PSD is found using $d/dt\rightarrow {\times}4\pi^2f^2$, thus
\begin{equation}
  S_\mathsf{y}(f) =
  \frac{f^2}{\nu_0^2}S_\varphi(f)
  \qquad[\unit{Hz^{-1}\equiv s}]\,.
\end{equation}

\subsection[The Polynomial Law, or Power Law ]{The Polynomial Law, or Power Law (Region 1.12--1.15)\label{sec:Poly-law}}
The polynomial fit, known as the \emph{power law} or \emph{polynomial law}, is widely used to model phase noise and related quantities.  It is often written as
\begin{align}
  \label{eqn:Poly-phi}
  S_\varphi(f)&=\sum_{n=-4}^{0} \mathsf{b}_n f^n\,,\\
  \label{eqn:Poly-x}  
  S_\mathsf{x}(f)&=\sum_{n=-4}^{0} \mathsf{k}_n f^n\,,
\end{align}
where the values of $n$ correspond to the noise types listed in Table~\ref{tab:NTypes}.  These noise types are found in oscillators, with additional negative-exponent terms sometimes needed, \mbox{$n<-4$}.  Limitations apply to two-port devices because the input-to-output delay is not allowed to diverge (CF Sec.~\ref{sec:Components} and \ref{sec:Oscillators}).

\begin{table}
  \caption{{\hspace*{-3em}}\\[1ex]\centering\textsc{Polynomial Law and Basic Types of Noise}}\label{tab:NTypes}
  \centering
\begin{tabular}{lcccc}
Noise type\rule[-.7em]{0pt}{2em} 
& $S_\varphi(f)$  
& $S_\mathsf{x}(f)$ 
& $S_{\nu}(f)$ 
& $S_\mathsf{y}(f)$\\\hline
Blue phase noise$^\star$\rule[-.7em]{0pt}{2em}
& $\mathsf{b}_1$ 
& $\mathsf{k_1}$ 
& $\mathsf{d}_{3}f^3$ 
& $\mathsf{h}_{3}f^3$ \\\hline
White phase noise\rule[-.7em]{0pt}{2em}
& $\mathsf{b}_0$ 
& $\mathsf{k_0}$ 
& $\mathsf{d}_{2}f^2$ 
& $\mathsf{h}_{2}f^2$ \\\hline
Flicker phase noise\rule[-1.3em]{0pt}{3em}
& $\dfrac{\mathsf{b}_{-1}}{f}$ 
& $\dfrac{\mathsf{k}_{-1}}{f}$ 
& $\mathsf{d}_{1}f$ 
& $\mathsf{h}_{1}f$\\\hline
White frequency noise\rule[-1.3em]{0pt}{3em} 
& $\dfrac{\mathsf{b}_{-2}}{f^2}$ 
& $\dfrac{\mathsf{k}_{-2}}{f^2}$ 
& $\mathsf{d}_{0}$ 
& $\mathsf{h}_{0}$\\\hline
Flicker frequency noise\rule[-1.3em]{0pt}{3em} 
& $\dfrac{\mathsf{b}_{-3}}{f^3}$ 
& $\dfrac{\mathsf{k}_{-3}}{f^3}$ 
& $\dfrac{\mathsf{d}_{-1}}{f}$
& $\dfrac{\mathsf{h}_{-1}}{f}$\\\hline
Frequency random walk\rule[-1.3em]{0pt}{3em} 
& $\dfrac{\mathsf{b}_{-4}}{f^4}$ 
& $\dfrac{\mathsf{k}_{-4}}{f^4}$ 
& $\dfrac{\mathsf{d}_{-2}}{f^2}$ 
& $\dfrac{\mathsf{h}_{-2}}{f^2}$\\\hline 
Integrated flicker FM noise$^\star$\rule[-1.3em]{0pt}{3em} 
& $\dfrac{\mathsf{b}_{-5}}{f^5}$ 
& $\dfrac{\mathsf{k}_{-5}}{f^5}$ 
& $\dfrac{\mathsf{d}_{-1}}{f^3}$
& $\dfrac{\mathsf{h}_{-1}}{f^3}$\\\hline
Integrated RW frequency$^\star$\rule[-1.3em]{0pt}{3em} 
& $\dfrac{\mathsf{b}_{-6}}{f^6}$ 
& $\dfrac{\mathsf{k}_{-6}}{f^6}$ 
& $\dfrac{\mathsf{d}_{-2}}{f^4}$ 
& $\dfrac{\mathsf{h}_{-2}}{f^4}$\\\hline 
\multicolumn{5}{l}{$\star$ Process not shown in the plots of $S(f)$ and $\sigma_\mathsf{y}^2(\tau)$}\rule[-1.3em]{0pt}{3em}
\end{tabular}
\end{table}

Transposing \eqref{eqn:Poly-phi}-\eqref{eqn:Poly-x} to frequency noise, the polynomial law is written as 
\begin{align}
  \label{eqn:Poly-nu}
  S_{\nu}(f)&=\sum_{n=-2}^{2} \mathsf{d}_n f^n\,,\\
  \label{eqn:Poly-y}
  S_\mathsf{y}(f)&=\sum_{n=-2}^{2} \mathsf{h}_n f^n\,.
\end{align}
Notice that, for a given process the exponents of $f$ differ by 2 from \eqref{eqn:Poly-phi}-\eqref{eqn:Poly-x} to \eqref{eqn:Poly-nu}-\eqref{eqn:Poly-y}, in agreement with the bounds of the sum.

In proper mathematical terms, \eqref{eqn:Poly-phi}-\eqref{eqn:Poly-y}
are Laurent polynomials, which is the extension of the regular polynomials to negative powers of the variable.

\subsection{The Quantities $\varphi$, $\mathsf{x}$, $\Delta\nu$ and $\mathsf{y}$ in Frequency Synthesis}
The ideal, noise-free synthesizer is the electrical analogue of a play-free gearbox.  It delivers an output frequency $\nu_o=(\mathcal{N}/\mathcal{D})\nu_r$, where $\mathcal{N}/\mathcal{D}$ is the rational number which defines the synthesis ratio, and $\nu_r$ is the reference frequency. 
Thus, the synthesizer transfers the quantities $\mathsf{x}(t)$ and $\mathsf{y}(t)$ from the reference input to the output, unchanged.
For example, shifting the reference by $+1.2$ ppm, the output frequency will be $+1.2$ ppm off the nominal value, thus 150 Hz higher if the output is set to 125 MHz.  Similarly, introducing a 100 ps delay with a line stretcher at the input results in the output shifted by 100 ps.  This value is the same at 5 MHz and 125 MHz output frequency.  

By contrast, the synthesis is ruled by $\varphi=(\mathcal{N}/\mathcal{D})\psi$ when we express the phase shift as an angle, $\psi$ at the input and $\varphi$ at the output.  For example, a $+1$ mrad shift of the 10 MHz reference results into a $+12.5$ mrad shift if the output is set to 125 MHz.  Accordingly, the phase noise spectrum is ruled by 
\begin{equation}
  S_\varphi(f)=\left(\frac{\mathcal{N}}{\mathcal{D}}\right)^{2} S_\psi(f)\,.
  \label{eqn:Synthesis}
\end{equation}

The above statements are simplistic, to the extent that we have not included the dynamic behavior, the noise bandwidth, and other phenomena.  The following limitations apply. 

\subsubsection{Digital dividers}
In digital signals, phase noise exists only on rising and falling edges, thus it is sampled at $2\nu_0$, twice its own frequency.   
Frequency division $\div\mathcal{D}$ results in lower sampling frequency, which originates aliasing.  Aliasing impacts only white noise because folding multiple aliases proportional to $1/f$ has negligible effect.   Thus, the white PM noise of the divided signal is ruled by $S_\varphi(f)=(1/\mathcal{D})S_\psi(f)$ instead of $(1/\mathcal{D}^2)S_\psi(f)$.

\subsubsection{Output stage}
For $\mathcal{N}/\mathcal{D}\ll1$, the scaled-down phase noise may hit the phase noise of the output stage.  When this happens, $S_\varphi(f)$ is limited by the output stage.  

\subsubsection{High-order multiplication}
Angular modulations are ruled by the property that the total power is constant, thus the sideband power comes at expense of the carrier power.  Carrier and $n$-th sideband amplitudes are described by the Bessel functions $J_n(m)$.  For $\mathcal{N}/\mathcal{D}\gg1$, the random phase in the full bandwidth may approach or exceed 2.4 rad, where $J_0(m)=0$.  When this happens, all the power goes in the sidebands and the carrier disappears.  This is known as the `carrier collapse' in multiplication.

\subsection{Phase-Type ($\varphi$-Type) and Time-Type ($\mathsf{x}$-Type) PM Noise}%
\label{sec:Types}
\begin{figure}[t]
  \begin{subfigure}[tl]{0.45\columnwidth}\flushleft
    \caption{\hspace*{0.0\columnwidth}\label{fig:Phi-type}}
    \vspace{1em}
    \includegraphics[bb = 0cm 20.3cm 7.0cm 29.7cm,angle=0,width=\textwidth]{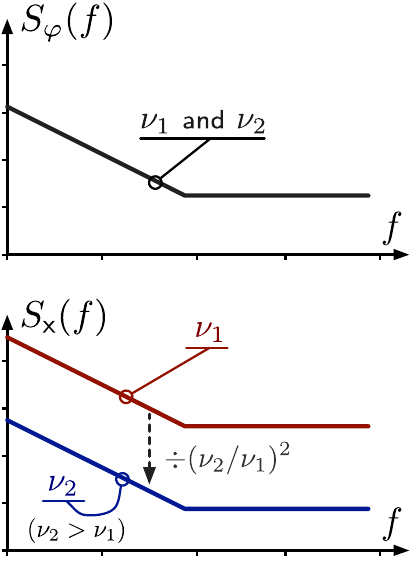}
  \end{subfigure}
  \hfill{\color{gray}\rule[-31mm]{0.2pt}{60mm}}\hfill
  \begin{subfigure}[tr]{0.45\columnwidth}\flushright
    \caption{\hspace*{0.0\columnwidth}\label{fig:X-type}}
    \vspace*{1em}
    \includegraphics[bb = 0cm 20.3cm 7.0cm 29.7cm,angle=0,width=\textwidth]{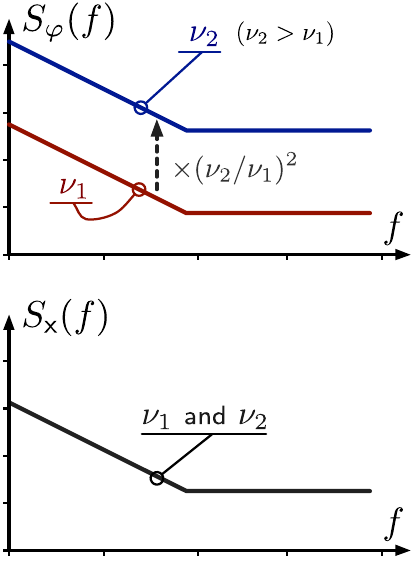}
  \end{subfigure}
  \vspace*{1ex}
  \caption{Basic phase-noise behavior of components and systems, (a) Phase-type, and (b) Time-type.  We show how the carrier frequency affects, or does not affect $S_\varphi(f)$ and $S_\mathsf{x}(f)$.  The effect of aliasing, not shown here, may change the white-noise scaling from $(\nu_2/\nu_1)^2$ into $\nu_2/\nu_1$.  In contrast, flicker and other colored noise types are immune from aliasing.}\label{fig:Phi-X-Type}
\end{figure}

A relevant aspect of the noise behavior of components and systems in the presence of large frequency changes is well described by the terms \emph{phase-type} and \emph{time-type} phase noise (Fig.~\ref{fig:Phi-X-Type}).  We first introduced this idea in \cite[Sec.~3]{Calosso-2017} to describe noise in the clock distribution of digital electronics.  The same terminology applies to 
\begin{itemize}
\item Amplifiers logic gates, etc., where the output frequency is the same as the input,
\item Frequency multiplication, division and synthesis, where the input frequency is scaled up/down by a rational number.
\item Frequency synthesis, where the output frequency is set to different values.
\end{itemize}

\subsubsection{Phase-Type ($\varphi$-Type) Phase Noise}Fig.~\ref{fig:Phi-type}.
The quantity $S_\varphi(f)$ is a parameter of the device or system, not affected by the carrier frequency.  The dominant noise processes impacts $S_\varphi(f)$ independently of $\nu_0$, and $S_\mathsf{x}(f)$ follows from \eqref{eqn:Sx}.
Additive noise in amplifiers (Sec.~\ref{sec:Components}), originating from the noise-to-carrier ratio \eqref{eqn:b0}, is like this.  Often flicker PM in amplifier is also this type, described by \eqref{eqn:bm1}.  Phase-type noise is also observed in frequency dividers, DDSs and DACs at low $\nu_o$ (output), where the scaled-down input noise is lower than the limit set by the output stage.

For example, the white PM noise of a RF amplifier in a reference condition is $S_\varphi(f)=\mathsf{b}_0=10^{-16}$ \unit{rad^2/Hz}, limited by noise figure and carrier power.  Sec.~\ref{sec:White-PM} explains how this can be calculated.  Such noise is of the phase type, shown in Fig.~\ref{fig:Phi-type}.  Using \eqref{eqn:Sx}, we get $S_\mathsf{x}=2.53{\times}10^{-32}$ \unit{s^2/Hz} when the carrier frequency takes the value $\nu_1=10$ MHz, and $S_\mathsf{x}(f)=1.62{\times}10^{-34}$ \unit{s^2/Hz} at $\nu_2=125$ MHz.  At $\nu_2$, the time fluctuation $\mathsf{x(t)}$ is a factor $\nu_1/\nu_2=0.08$ smaller than at $\nu_1$, and its PSD is a factor $(\nu_1/\nu_2)^2=0.0064$ smaller.

\subsubsection{Time-Type ($\mathsf{x}$-Type) Phase Noise}Fig.~\ref{fig:X-type}.
The quantity $S_\mathsf{x}(f)$ is a parameter of the device or system, not affected by the carrier frequency.
The dominant noise processes impacts on the time fluctuation $S_\mathsf{x}(f)$ independently of $\nu_0$, and $S_\varphi(f)$ follows from \eqref{eqn:Sx}.
This behavior is observed in ADCs, DACs and DDSs at high frequencies, where the PM noise is dominated by the jitter of the internal clock distribution.  Acoustic/seismic noise and other environment effects on coaxial cables and optical fibers are obviously like this.

For example, the clock distribution of a Cyclone III FPGA has a flicker $S_\mathsf{x}=\mathsf{k}_{-1}=4{\times}10^{-28}$ \unit{s/Hz}.  Otherway stated, $\sqrt{\mathsf{k}_{-1}}=20$ fs.  Such noise originates from the time fluctuation in the pipeline of logic gates, which is of the time type (Fig.~\ref{fig:X-type}).  Using \eqref{eqn:Sx}, we expect a flicker $\mathsf{b}_{-1}=1.6{\times}10^{-10}$ \unit{rad^2} ($-98$ \unit{dBrad^2}) at 100 MHz clock, $6.3{\times}10^{-10}$ \unit{rad^2} ($-92$ \unit{dBrad^2}) at 200 MHz, and $2.5{\times}10^{-9}$ \unit{rad^2} ($-86$ \unit{dBrad^2}) at 400 MHz.  The real experiment behind this example \cite[Fig.~6]{Calosso-2017} shows a rather small discrepancy from the $\times(\nu_2/\nu_1)^2$ law, $\approx0.5$ dB for $\mathsf{b}_{-1}$ at the three clock frequencies.

A PLL frequency multiplier is another example, where the output is generated by a VCO phase-locked to a 10 MHz reference after frequency division.  The division modulo sets the output frequency.  The PLL's own noise---not accounting for the external reference---is typically determined by the 10 MHz phase comparator.  Assuming that the comparator's flicker is $\mathsf{b}_{-1}=3.2{\times}10^{-10}$ \unit{rad^2} ($-95$ \unit{dBrad^2}), the output $\mathsf{b}_{-1}$ is of $-95$ \unit{dBrad^2} at 10 MHz output, $-89$ \unit{dBrad^2} at 20 MHz, $-83$ \unit{dBrad^2} at 40 MHz, etc., in 6 dB steps per factor-of-two.  This is exactly what we observe on the internal PLL of a Cyclone III FPGA, up to $-59$ \unit{dBrad^2} at 640 MHz, the maximum output frequency \cite[Fig.~6]{Calosso-2017}.

\subsubsection{Aliasing}
The two noise types described may be modified by aliasing, which changes the scaling rule from $(\nu_2/\nu_1)^2$ to $\nu_2/\nu_1$.  Aliasing only affects white noise, not flicker and other reddish noise types because the tails of the PSD are too low for the folded-down aliases to be significant.  Aliasing is typically seen in digital circuits, and in clock distribution of FPGAs and other complex devices as well, at low frequency \cite[Sec.~3]{Calosso-2017}.

\subsection[Notation]{Notation (Region 1.2)\label{sec:Notation}}
We use the sans serif font in $\mathsf{x}(t)$ and $\mathsf{y}(t)$, instead of the regular math font commonly found in the literature, to emphasize that $\mathsf{x}$ and $\mathsf{y}$ are special quantities defined by \eqref{eqn:x(t)} and \eqref{eqn:y(t)}.  This choice sets the regular $x$ and $y$ free for general use.  The same applies to the coefficients $\mathsf{b}_i$, $\mathsf{d}_i$, $\mathsf{h}_i$ and $\mathsf{k}_i$.

Working with digital systems, we are regularly faced with phase exceeding $\pm\pi$ because IQ detectors and digital dividers keep record of multiple cycles of the carrier.  For this purpose, we find it useful to describe the clock signal with the quantity written in boldface, which is the sum of the deterministic (or nominal) quantity plus the fluctuation
\begin{align}
&\text{phase:}&     \bm{\varphi}(t) &= 2\pi\nu_0t + \varphi(t)\\\
&\text{frequency:}& \bm{\nu}(t) &= \nu_0 + (\Delta\nu)(t)\\
&\text{time:}&      \bm{\mathsf{x}}(t) &= t + \mathsf{x}(t)\\
&\text{fractional frequency:}& \bm{\mathsf{y}}(t) &= 1 + \mathsf{y}(t)\,.
\end{align}
The quantity $\bm{\mathsf{x}}(t)$ is the most obvious.  To the layman, $\bm{\mathsf{x}}(t)$ is the readout of a clock, which is the sum of the `exact time' $t$ plus the `error' $\mathsf{x}(t)$.  A true layman would not consider relativity here, and would have no idea about the technical meaning of words like `error' and `uncertainty.'
The quantity $\bm{\nu}(t)$ is the instantaneous frequency, measured in a sufficiently short time, $\bm{\varphi}(t)$ is the total phase accumulated after IQ detection, and $\bm{\mathsf{y}}(t)$ differs from 1 by the small fractional fluctuation $\mathsf{y}(t)$.

\ifSubfilesClassLoaded{\bibliography{Ref-short,References,Ref-local}}{}
\end{document}

\section[Two-Port Components]{Two-Port Components (Region 1.8)}\label{sec:Components}

\ifSubfilesClassLoaded{\setcounter{section}{3}\section{. Two-Port Components (Region 1.8)}}{}

\subsection{Additive and Parametric Noise}\label{sec:Additive}
Most people find these concepts either quite simple, or rather confusing.
The point is that \eqref{eqn:Clock-signal} describes the clock signal as it is observed, hiding the physics of noise.  

To understand, we start with the example of a noise-free radio broadcasting $v_e(t) = V_e\bigl[1+\alpha_m(t)\bigr]\cos\bigl[2\pi\nu_0t+\varphi_m(t)\bigr]$, where $\alpha_m(t)$ is AM, and $\varphi_m(t)$ is PM\@.  AM and PM may be present simultaneously, as in the old analog television\footnote{More precisely, TV audio is FM for compatibility with audio broadcasting, which is equivalent to PM.} and in QAM modulations.  
The \emph{received} signal 
\begin{equation}
  v(t) = V_0\bigl[1+\alpha_m(t)\bigr]\cos\bigl[2\pi\nu_0t+\varphi_m(t)\bigr]+n(t)
  \label{eqn:Clock-extended}
\end{equation}
includes the noise $n(t)$, which is the receiver's own noise, atmospheric noise, and other forms of noise collected by the antenna.  However, the signal is \emph{detected} as \eqref{eqn:Clock-signal}.
In fact, $\varphi(t)$ is the sum of $\varphi_m(t)$ plus the image of $n(t)$ after phase detection.  Similarly, $\alpha(t)$ is the sum of $\alpha_m(t)$ plus the image of $n(t)$ after amplitude detection.
Random $n(t)$ is called additive noise, while random $\alpha_m(t)$ and $\varphi_m(t)$ are called parametric noise.
The relevant difference is that \emph{parametric noise} originates from the low-frequency signals $\varphi_m(t)$ and $\alpha_m(t)$ up converted to sidebands, while \emph{additive noise} originates around $\nu_0$.  If $\varphi_m(t)$ and $\alpha_m(t)$ are small\footnote{More precisely, small modulation angle $|\varphi_m(t)|\ll1$ is necessary for higher-order sidebands to negligible, and the PM to be accurately described by the first USB and LSB\@.  Conversely, only $|\alpha_m(t)|<1$ is formally necessary, but the $|\alpha_m(t)|\ll1$ condition makes sense in real systems.} and have bandwidth $B_m\ll\nu_0$ their image in the RF spectrum is a pedestal $\pm B_m$ wide centered at $\nu_0$.

As a relevant fact, white AM and PM noise results from both white $n(t)$ and white $\alpha_m(t)$ and $\varphi_m(t)$.  The former is additive, the latter is parametric.
Hence, the common belief that white AM/PM noise is only additive is incorrect.  
Conversely, `colored' noise ($1/f$, $1/f^2$, etc.) is generally of parametric origin.  A narrow-band $n(t)$ well centered at $\nu_0$ appearing as $1/f$ noise is totally unrealistic, albeit conceptually possible.

\subsection{Added Noise\label{sec:Added}}
The term `added noise' is seen in commercial phase noise analyzers to denote the phase noise \emph{added} by a two-port component under test, usually an amplifier, when the component is inserted in the signal path.  
This is an unfortunate choice because the term `added' is easily mistaken for `additive.'  
Of course, the `added' noise consists of white and flicker PM noise, thermal drift, aging, etc., while the `additive' noise is white or quasi white.

In proper terminology, added noise refers to a \emph{differential measurement} because the instrument measures the quantity $\varphi-\psi$, i.e., output minus input, and its statistical properties (spectra and variances, at will).  The oscillator noise $\psi$ is a common-mode signal, thus it is rejected.

\subsection{White Noise\label{sec:White-PM}}
Given a white noise $n(t)$ of PSD $N$ [W/Hz] added to a carrier of power $P$, the white PM noise's PSD is 
\begin{equation}
  \mathsf{b}_0=\frac{N}{P}
  \qquad\unit{[rad^2/Hz]}\,.
  \label{eqn:b0}
\end{equation}
Radio engineers often express the noise in terms of noise factor $F$.  Accordingly, $N$ is expanded as $N=FkT_0$, where $kT_0=4{\times}10^{-21}\:\unit{W/Hz}$ is the thermal energy at the standard temperature $T_0=290$ K (17 \Celsius).  The Noise Figure is defined by $\text{NF}=10\log_{10}(F)$.  For reference, the phase noise of a noise-free device ($F=1$, or $\text{NF}=0$ dB) in the presence of a 1-mW carrier (0 dBm) is $4{\times}10^{-18}\:\unit{rad^2/Hz}$, thus $-174\:\unit{dBrad^2/Hz}$ or $-177\:\unit{dBc/Hz}$.  Besides thermal energy, $n(t)$ may originate from shot noise.  In our experience, experimentalists in optics are led to confusion by the fact that because noise is represented as a temperature, while in optics the thermal energy is generally negligible.

Additional white PM noise, not accounted for in \eqref{eqn:b0} can result from parametric effects, as explained in Sec.~\ref{sec:Additive}

\subsection{Flicker Noise}
Flicker noise has PSD proportional to $1/f$.
It has been observed that the flicker PM of RF and microwave amplifiers 
is rather constant vs $P$ and $\nu_0$
\begin{equation}
  \mathsf{b}_{-1}=C\qquad\text{(constant vs $P$ and $\nu_0$)}\,.
  \label{eqn:bm1}
\end{equation}
It follows from \eqref{eqn:b0} and \eqref{eqn:bm1} that the corner frequency $f_c=\mathsf{b}_{-1}/\mathsf{b}_{0}$, where flicker equals white noise, depends on $P$.

Albeit the common integral $\int_a^b(1/f)\,df=\ln(b/a)$ diverges for $a\rightarrow0$ or $b\rightarrow\infty$, the practical result is surprisingly small.
To convince the reader, we evaluate $\ln(b/a)$ for the largest conceivable bounds, from the reciprocal of the age of the universe ($a=2.3{\times}10^{-18}$~Hz) to the reciprocal of the Planck time ($b=1.9{\times}10^{43}$~Hz).
The results is $\ln(b/a)=140.3$, i.e., 21.5 dB\@.  So, if the flicker coefficient is $\mathsf{k}_{-1}=10^{-24}$ \unit{s^2}, ($\sqrt{\mathsf{k}_{-1}}=1$ ps), the total 1-$\sigma$ fluctuation is $\sqrt{140.3\times(10^{-24}\:\unit{s^2})}=11.8$ ps.

The general literature suggests that the spectrum of flicker is $1/f^\eta$ with $\eta$ close to one.  It turns out that the input-to-output delay never grows too large even for $\eta>1$.  Try this yourself with $\eta=1.1$ and the integration bounds from $10^{-9}$ Hz (the reciprocal of 30 years, the supposed device's lifetime) to $10^9$ Hz noise bandwidth (above the highest Fourier frequency found in any commercial noise analyzer).

\subsection{Input-to-Output Delay}
The fluctuation (1-$\sigma$) of the input-to-output delay of a two-port device is given by
\begin{equation}
  \delta T = \sqrt{\int_a^bS_\mathsf{x}(f)\:df}\,,
  \label{eqn:deltaT}
\end{equation}
which follows from the time-fluctuation PSD integrated on the appropriate bandwidth $[a,b]$.  Common sense suggests that $\delta T$ does not diverge, nor does it grow disproportionately during the device life.  This limits the phase noise to finite-bandwidth white PM noise and to flicker PM noise.  

Environmental parameters, like humidity and thermal drift, have only localized effect in time, or are periodic.  Random walk, aging and other ever growing phenomena are possible, but their amount is generally small enough not to significantly affect the delay over the lifetime of the device.  Electrical engineers may be familiar with similar effects in voltage references or in the offset of analog components.

\subsection{Example of Noise in a Two-Port Component}
We consider an amplifier having $\mathrm{NF}=2$ dB and flicker coefficient $\mathsf{b}_{-1}=2{\times}10^{-11}$ \unit{rad^2} ($-107$ \unit{dBrad^2}), processing a 10 GHz carrier of power $P=62.5$ \unit{\mu W} ($-12$ dBm).
Let us calculate the noise spectrum and the rms delay fluctuation assuming a low cutoff $f_1=10^{-8}$ \unit{Hz} (reciprocal of 3 years, rounded), and a bandwidth $f_2=50$ MHz.


\subsubsection*{PM noise PSD} Using $\mathsf{b}_0=FkT/P$, in dB we get $10\log_{10}\left(\mathsf{b}_0\right)=+2-174+12=-160$ \unit{dBrad^2/Hz}, hence $\mathsf{b}_0=10^{-16}$ \unit{rad^2/Hz}.  The phase noise PSD is 
\begin{equation*}
  S_\varphi(f)
  =\mathsf{b}_0+\frac{\mathsf{b}_{-1}}{f}
  =10^{-16}+\frac{2{\times}10^{-11}}{f}
  \quad\unit{rad^2/Hz}.
\end{equation*}
Using $\mathsf{b}_0=\mathsf{b}_{-1}/f$, the corner frequency is $f_c=2{\times}10^5$ Hz.

\subsubsection*{Phase-time PSD}
Using \eqref{eqn:Sx} with $\nu_0=10$ GHz, we get 
\begin{equation*}
  S_\mathsf{x}(f)
  =\mathsf{k}_0+\frac{\mathsf{k}_{-1}}{f}
  =2.6{\times}10^{-38}+\frac{5.1{\times}10^{-33}}{f}
  \quad\unit{s^2/Hz}.
\end{equation*}

\subsubsection*{RMS delay fluctuation} This is evaluated using \eqref{eqn:deltaT}.\\
White noise $(\delta T)_0=\sqrt{\mathsf{k}_0(f_2-f_1)}=1.1{\times}10^{-15}$ s.\\  
Flicker noise $(\delta T)_{-1}=\sqrt{\mathsf{k}_{-1}\ln(f_2/f_1)}=4.3{\times}10^{-16}$ s.\\
Total $\delta T = \sqrt{(\delta T)_{0}^2+(\delta T)_{-1}^2}=1.2{\times}10^{-15}$ s.
\subsubsection*{Comment}  Common sense suggests that the delay of an amplifier can be of a fraction of a ns (several periods of the 10 GHz carrier) for a wideband device, with a thermal coefficient up to $10^{-3}/\unit{K}$ if no special design care is taken.  As a result, we expect that the thermal effects exceed the random noise, and that the random noise is visible  only beyond 10 Hz, where the temperature may be stabilized by the thermal capacitance (inertia).

\subsection[Suggested Readings]{Suggested Readings About Two-Port Components}
Reference \cite{Boudot-2012} is an extensive treatise of phase noise in amplifiers.
Arguably, \cite{Halford-1968} is the first article suggesting that the flicker noise (the parameter $\mathsf{b}_{-1}$) in RF and microwave amplifiers is independent of power and frequency over a rather broad range.

The double-balanced mixer is a tool of paramount importance in PM noise and frequency stability.  Reference \cite{Rubiola-2006-arXiv} is a useful tutorial, and \cite{Barnes-2011} provides experimental data about noise of commercial double-balanced mixers for phase noise measurements.

We recommend the book \cite{Kester-2004} to understand analog-digital conversion.  Reference \cite{Calosso-2017} is a tutorial on phase noise digital systems. 
The origin and the propagation of phase noise in frequency division are explained in \cite{Egan-1990,Kroupa-2001}.

Phase locking and frequency synthesis are a totally different kind of two-port systems.  
We suggest \cite{Egan-2008,Banerjee-2017,Shu-2005} for phase locking, \cite{Goldberg-1999,Kroupa-1999} for digital frequency synthesis, \cite{Calosso-2012-IFCS} for the phase noise in digital synthesis, and \cite{Rohde-2021} for a general treatise about modern synthesizes.

\ifSubfilesClassLoaded{\bibliography{Ref-short,References,Ref-local}}{}
\end{document}

\section[Oscillators and the Leeson Effect]{Oscillators and the Leeson Effect (Region 1.9)}\label{sec:Oscillators}

\ifSubfilesClassLoaded{\setcounter{section}{4}\section{Oscillators and the Leeson Effect (Region 1.9)}}{}

\begin{figure}[t]\centering
  \begin{subfigure}[t]{\columnwidth}\centering
    \caption{\hspace*{0.8\columnwidth}\label{fig:Leeson-A}}
    \includegraphics[bb = 0cm 0cm 3.15cm 5.95cm, scale=1, angle=-90]{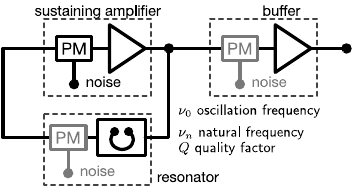}
  \end{subfigure}
  \begin{subfigure}[t]{\columnwidth}\centering
    \vspace*{0em}
    \caption{\hspace*{0.8\columnwidth}\label{fig:Leeson-B}}
    \includegraphics[bb = 0cm 0cm 3.0cm 6.1cm, scale=1, angle=-90]{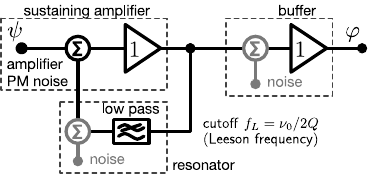}
  \end{subfigure}
  \caption{Phase noise mechanisms inside an oscillator.  
    (a) Conceptual schematic of the oscillator, where signal is the microwave/RF sinusoidal oscillation at the frequency $\nu_0$.
    (b) Phase-noise equivalent of the oscillator, where the signal is the random phase of the microwave/RF circuit shown above.  In this representation, phase noise is always additive, regardless of its physical origin.
    The grayed blocks in (a) and (b) represent the noise of the resonator and of the buffer, not included in equations.
  }  
  \label{fig:Leeson-effect}

\end{figure}

The oscillator (Fig.~\ref{fig:Leeson-A}) consists of a loop where the resonator sets the oscillation frequency $\nu_0$ and the sustaining amplifier compensates for the resonator loss.  Gain clipping (nonlinearity) is necessary to stabilize the amplitude, hence the typical dominance of PM over AM\@. The buffer isolates the loop from the load.  

Everyday experience suggests that, unlike the two-port components, the time fluctuation $\mathsf{x}(t)$ of an oscillator can be quite large.  The reason is that the oscillator accumulates the `error' of each cycle, however small it may be.
    
Resonator, sustaining amplifier and buffer, they all introduce phase noise (amplitude noise too, but this is a more specialized topic, not considered here).
The phase modulators in Fig.~\ref{fig:Leeson-A} are necessary to model phase noise of the various part of the oscillator in the general case, including white and colored processes.  However, these modulators make the mathematical treatise difficult.
The \emph{phase equivalent} model, introduced in \cite[Chapter~4]{Rubiola-2010-Cambridge} and shown in Fig.~\ref{fig:Leeson-B}, avoids such difficulty.  The signal circulating in Fig.~\ref{fig:Leeson-B} is the phase fluctuation of Fig.~\ref{fig:Leeson-A}\@.  In this representation:
\begin{itemize}
  \item Sustaining amplifier and buffer map into `phase amplifiers' of gain exactly equal one because they `copy' the input phase to the output, unchanged.  Otherwise stated, the amplifier's delay cannot be stretched/shrunk,
  \item Random phase modulation maps into noise \emph{added} in the signal path, regardless of the nature of the process,
  \item The phase model is linear because gain clipping has no effect on phase---at least, not first-order effects,
  \item The resonator (narrow-band 2nd-order filter) maps into a single-pole low-pass filter
\end{itemize}
This representation holds for large quality factor, say, $Q\gtrsim50$, which ensures that signals are sinusoidal.  Accepting this limitation, the approximation in Fig.~\ref{fig:Leeson-B} adequately models the more general oscillator in Fig.~\ref{fig:Leeson-A}.  Breaking the high-$Q$ hypothesis, the simple mathematical treatise provided below is no longer valid because distortion introduces coupling between amplitude and phase.

We give a short summary of the original derivation \cite[Sec.~4.4--4.5]{Rubiola-2010-Cambridge}.  The low-pass impulse response\footnote{The standard symbol of the impulse response is $h(t)$, and $H(s)$ its Laplace transform.  Here we use $b(t)\leftrightarrow B(s)$ to save $h(t)\leftrightarrow H(f)$ for the oscillator.} 
of the resonator is
  \begin{math}
    b(t)=(1/\tau)\,e^{-t/\tau}
  \end{math}, 
where $\tau=Q/\pi\nu_n$ is the relaxation time, $Q$ is the quality factor in actual load conditions, and $\nu_n$ is the natural frequency.  In practice, the oscillation frequency $\nu_0$ is so close to $\nu_n$ that they are interchangeable.  The Laplace transform of $b(t)$ is 
\begin{math}
  B(s)=(1/\tau)/(s+1/\tau)
\end{math}.
The cutoff frequency of the low-pass equivalent is called \emph{Leeson frequency}, equal to half the resonator's RF bandwidth $\nu_0/Q$
\begin{equation}
  f_L=\frac{\nu_0}{2Q}\,.
  \label{eqn:fL}
\end{equation}
Letting aside the buffer noise, we introduce the phase-noise transfer function of the oscillator
\begin{equation}
  \text{definition:}\qquad
  H(s)=\frac{\Phi(s)}{\Psi(s)}\,,
\end{equation} 
where $\Phi(s)$ and $\Psi(s)$ are the Laplace transforms of the output $\varphi(t)$ and of the input $\psi(t)$, respectively.  Inspecting the loop on the left of Fig.~\ref{fig:Leeson-B}, we find 
\begin{math}
  H(s)=1/[1-B(s)]
\end{math}.
The latter expression is derived with the classical rules of feedback, the same used in textbooks to analyze a simple control loop, of to derive the gain of the operational amplifier in non-inverting configuration. 
Expanding $H(s)$ and replacing $s\rightarrow j2\pi f$, we get  
\begin{math}
  \left|H(f)\right|^2=1+f^2_L/f^2
\end{math}.
The multiplication by $1/f^2$ which occurs at $f\,{<}\,f_L$ is the `soul' of the Leeson effect: the oscillator integrates the phase fluctuation $\psi(t)$ occurring in the loop.
The oscillator phase noise is 
\begin{math}
  S_\varphi(f)=|H(f)|^2S_\psi(f)
\end{math},
thus
\begin{align}
S_\varphi(f)&=\biggl[1+\frac{f^2_L}{f^2}\biggr]S_\psi(f)\,.
\label{eqn:Leeson}
\end{align}

\begin{figure}[t]\centering
  \begin{subfigure}[t]{\columnwidth}\centering
    \caption{\hspace*{0.8\columnwidth}\label{fig:Type-A}}
    \includegraphics[bb = 0cm -0cm 3.8cm 6.3cm, scale=1.1, angle=-90]{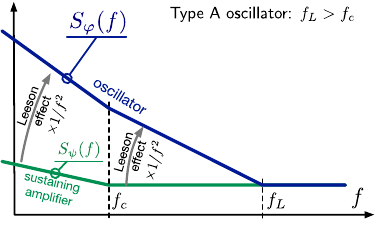}%
  \end{subfigure}
  \begin{subfigure}[t]{\columnwidth}\centering
    \vspace*{2em}
    \caption{\hspace*{0.8\columnwidth}\label{fig:Type-B}}
    \includegraphics[bb = 0.0cm 0cm 3.7cm 6.3cm, scale=1.1, angle=-90]{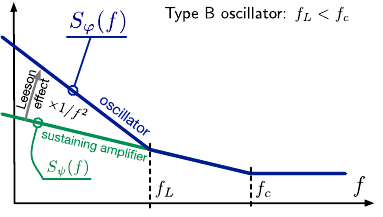}%
  \end{subfigure}
  \caption{Typical noise spectra occurring in oscillators. (a) Type A phase noise spectrum, often found in microwave oscillators.  (b) Type B phase noise spectrum, often found in RF and quartz oscillators.  Both (a) and (b) show the noise $S_\psi(f)$ of the sustaining amplifier, the flicker corner at frequency $f_c$, the Leeson corner at frequency $f_L$, and the oscillator noise $S_\varphi(f)$.  The fluctuation of the resonator's natural frequency and the noise of the output buffer are not included.}
  \label{fig:Leeson-spectra}
\end{figure}

Plugging the amplifier noise $S_\psi(f)=\mathsf{b}_{-1}/f+\mathsf{b}_0$ in \eqref{eqn:Leeson}, we find the two typical patterns of Fig.~\ref{fig:Leeson-spectra}.  Interestingly, the oscillator spectrum contains either $1/f^2$ or $1/f$ noise, not both.  The $1/f^2$ noise is common in microwave oscillators, characterized by high $\nu_0$ and low $Q$, thus $f_L\,{\gg}\,f_c$.  The $1/f$ noise is typical of RF and quartz oscillators, characterized by low $\nu_0$ and high $Q$.
  
The noise of the resonator and of the buffer, not included in \eqref{eqn:Leeson} and not shown in Fig.~\ref{fig:Leeson-spectra}, adds.  In most quartz oscillators, the $1/f^3$ noise due to the $1/f$ fluctuation of the resonator natural frequency exceeds the $1/f$ contribution of the electronics turned into $1/f^3$ via the Leeson effect \cite{Rubiola-2007-Xtal}.  In such cases, the Leeson effect is hidden below the resonator's fluctuations.

\subsection{Example of Noise in a Microwave Oscillator}
\begin{figure}[t]\centering
  \includegraphics[width=\columnwidth]{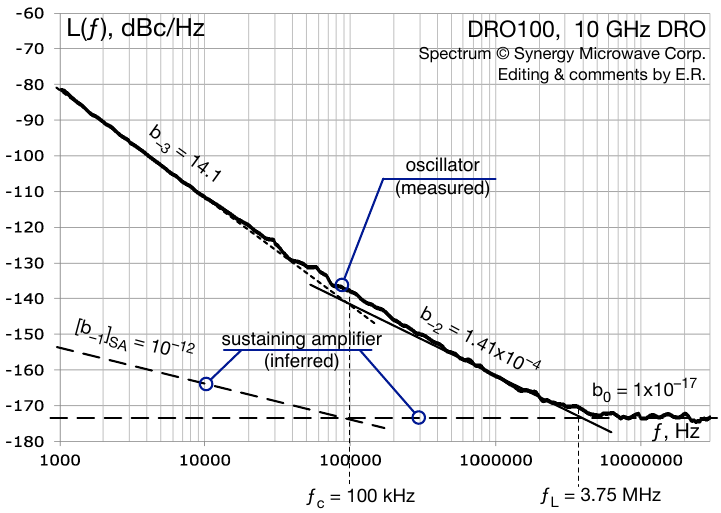}
  \caption{Example of phase noise of an oscillator.  The original plot is courtesy of Ulrich L. Rohde, \copyright Synergy Microwave Corp., used with permission.  
  We added the polynomial approximation (the units of $\mathsf{b}_i$ coefficients are omitted), the estimated amplifier noise, and other comments.}
  \label{fig:DRO-100}
\end{figure}

Fig.~\ref{fig:DRO-100} shows the phase noise of the DRO-100, a commercial 10 GHz oscillator based on a dielectric resonator. 
By comparison with Fig.~\ref{fig:Leeson-spectra}, this oscillator is clearly of type A, characterized by $f_c<f_L$.  The coefficients $\mathsf{b}_0$, $\mathsf{b}_{-2}$ and $\mathsf{b}_{-3}$ are obtained from graphical approximation of the measured spectrum with the theoretical straight line having slope $0$, $-2$ and $-3$.  The frequencies $f_c$ and $f_L$ are obtained graphically from the straight-line approximation.
We start our interpretation from the right-hand side ($f=10$ MHz) to the left.
  
The white PM noise $\mathsf{b}_0=10^{-17}$ \unit{rad^2/Hz} suggests that the power at the input of the sustaining amplifier (the location where the power is the lowest) is $P\approx$ 0.5 mW\@.  
The power in the resonator is substantially the same.  
This result is obtained using \eqref{eqn:b0}, guessing that the noise figure of such amplifier is of 1 dB, quite a plausible value for good amplifiers.
 
The white frequency noise $\mathsf{b}_{-2}/f$ crosses the white phase noise $\mathsf{b_0}$ at 3.75 \unit{MHz}.  This is clearly the Leeson frequency $f_L$.  Using \eqref{eqn:fL}, we estimate the resonator's quality factor $Q=1330$ in actual load conditions.

Finally, the flicker FM noise $\mathsf{b}_{-3}=14.1$ \unit{rad^2Hz^2} crosses the white FM noise at $f_c=100$ kHz.  Equation~\eqref{eqn:Leeson} with $f\ll f_c$ suggests that the flicker PM of the sustaining amplifier is of $10^{-12}$ \unit{rad^2} ($-120$ \unit{dBrad^2}).

\subsection[Suggested Readings]{Suggested Readings About Oscillators\label{bib:Oscillators}}
Our presentation is based on \cite{Rubiola-2010-Cambridge}, which includes the theoretical proof of the Leeson effect, the analysis of noise sources in the loop and at the output, and a chapter about reverse engineering of oscillators from phase noise.  Rather than focusing on the schematic, \cite{Rubiola-2010-Cambridge} analyzes the oscillator as a system.  A white paper \cite{Rubiola-2010-Leeson} extends the theory to amplitude noise.
The limitation of this approach is that it is suitable only to high-$Q$ resonator (say, $Q\gtrsim50$), where all relevant signals are sinusoidal. 

Numerous RF and quartz oscillator schemes make use of a negative-resistance amplifier in parallel to the resonator.  We propose two classical references about the noise of such amplifiers, Penfield 1960 \cite{Penfield-1960} and van der Ziel 1962 \cite{vanderZiel-1962}.

Other approaches deserve attention, chiefly:
(i) Models based on the limit cycle, derived from the Einstein's diffusion theory, i.e., Loh \& al\@. 2013 \cite{Loh-2013a,Loh-2013b}, and Demir \& al\@. 2000 \cite{Demir-2000}, or from the Langevin equations, Kärtner 1990 \cite{Kaertner-1990};
and (ii) the Impulse Sensitivity Function proposed by Lee \& Hajimiri \cite{Hajimiri-1998,Lee-2000}, which is good at describing low-$Q$ oscillators, like those commonly found in microelectronics.
Additionally, Pankratz 2014 \cite{Pankratz-2014} provides a large survey specific to oscillators in integrated circuits.

Digging in the origins, the widely-cited van der Pol oscillator \cite{vanderPol-1927} is about chaos in oscillations, rather than phase noise.
Edson 1960 \cite{Edson-1960} is arguably the first article that analyzes the phase noise in electronic oscillators, and Leeson 1966 \cite{Leeson-1966a} is the article that introduces the phase noise mechanism in feedback oscillators, later known as the Leeson effect.  A review article by D. B. Leeson is available \cite{Leeson-2016}.

\ifSubfilesClassLoaded{\bibliography{Ref-short,References,Ref-local}}{}
\end{document}

\section{The Allan Variance}\label{sec:AVAR}

\ifSubfilesClassLoaded{\setcounter{section}{5}\section*{The Allan Variance}}{}

The classical variance\footnote{This is a simplified notation.  More precisely, \eqref{eqn:classical-var} describes an estimator, thus it should be written as $\widehat{\sigma^2}=\ldots[x_i-\hat{\mu}]^2$, where $\hat{\mu}=\ldots$.} 
\begin{equation}
  \sigma^2=\frac{1}{n-1}\sum_{i=1}^{n}\, [x_i-\mu]^2\,,
  \label{eqn:classical-var}
\end{equation}
where $\mu=\frac{1}{n}\sum_{i=1}^{n}x_i$ is the average, fails at describing time divergent processes because (i) it depends on the averaging time used to take the samples $x_i$, and (ii) it depends on the number $n$ of samples.
Try this yourself, feeding $x_i=1.0001, 1.0002, 1.0003\ldots$, with $n=2, 4, 8\ldots$ in   \eqref{eqn:classical-var}. 
A solution consists of introducing the averaging time as a parameter, denoted with $\tau$, and to set $n=2$.  This is the minimum $n$ which gives a valid $\sigma^2$.  Welcome to the Allan variance.

\subsection[Definition and Evaluation]{Definition and Evaluation (Region 1.3)\label{sec:AVAR-eval}}
The two-sample (Allan) variance AVAR of the quantity $\mathsf{y}$ is defined as 
\begin{equation}
\text{2-sample variance:}\quad
\sigma^2_\mathsf{y}(\tau) = 
\mathbb{E}\biggl\{\frac12 \Bigl[\overline{\mathsf{y}}_2-\overline{\mathsf{y}}_1\Bigr]^2\biggr\}\,,
\label{eqn:AVAR}
\end{equation}
where $\mathbb{E}\{\:\}$ is the mathematical expectation, and the averages $\overline{\mathsf{y}}_1$ and $\overline{\mathsf{y}}_2$ are taken over contiguous time slots of duration $\tau$.  The quantity $\frac12\bigl[\overline{\mathsf{y}}_2-\overline{\mathsf{y}}_1\bigr]^2$ is the classical variance evaluated with two samples.  This is immediately seen by replacing $n=2$ in \eqref{eqn:classical-var}, then $\mu=(x_1+x_2)/2$; finally, we identify the generic $x_i$ with the average fractional frequency fluctuation $\overline{\mathsf{y}}_i$.
The quantity ADEV, a deviation, the square root of AVAR---and similarly MDEV, PDEV etc.\ defined later---can be seen as an estimator of the uncertainty of the quantity $\mathsf{y}$, accumulated in the time $\tau$ after reset or calibration.  

In experiments, $\mathbb{E}\{\:\}$ is replaced with the estimator
\begin{equation}
\widehat{\sigma^2_\mathsf{y}}(\tau) = 
\frac{1}{2(M-1)}
\sum_{k=1}^{M-1} 
\Bigl[\overline{\mathsf{y}}_{k+1}-\overline{\mathsf{y}}_{k}\Bigr]^2\,,
\label{eqn:AVAR-evaly}
\end{equation}
which is the average on $M-1$ realizations of $\overline{\mathsf{y}}_2-\overline{\mathsf{y}}_1$, thus it requires $M$ contiguous measures of $\overline{\mathsf{y}}$.

At first reading, one can take \eqref{eqn:AVAR-evaly} as the formula to evaluate AVAR, ignoring the `hat.'  Some Authors use \eqref{eqn:AVAR-evaly} as the definition of AVAR\@.  That said, keeping a clear difference between $\sigma^2_\mathsf{y}(\tau)$ and its estimate $\widehat{\sigma^2_\mathsf{y}}(\tau)$ is important in theoretical analysis and in the evaluation of the uncertainty.

It is often convenient to rely on time measurements $\mathsf{x}_k$, using $\mathsf{y}_k=(\mathsf{x}_{k+1}-\mathsf{x}_k)/\tau$.  Accordingly, \eqref{eqn:AVAR-evaly} rewrites as
\begin{equation}
  \widehat{\sigma^2_\mathsf{y}}(\tau) = 
  \frac{1}{2(M-1)}
  \sum_{k=1}^{M-1} 
  \Bigl[
\frac{ \mathsf{x}_{k+2} - 2\mathsf{x}_{k+1} + \mathsf{x}_{k}}{\tau}\Bigr]^2\,,
  \label{eqn:AVAR-evalx}
\end{equation}
which requires $M+1$ measures of $\mathsf{x}(t)$ spaced by $\tau$.

\subsection{Spectral Response (Region 1.5)\label{sec:Spectral-response}}
The Allan variance can be calculated from the spectrum using 
\begin{equation}
  \sigma^2_\mathsf{y}(\tau) = 
  \int_{0}^{\infty} \left| H_A(f;\tau) \right|^2 S_\mathsf{y}(f)\:df\,,
  \label{eqn:Spectral-response}
\end{equation}
where the transfer function
\begin{equation}
  |H_A(f;\tau)|^2=2\,\frac{\sin^4(\pi\tau f)}{(\pi\tau f)^2}
  \label{eqn:HA}
\end{equation}
is similar to an octave bandpass filter centered at $f\simeq0.45/\tau$.  Unfortunately, such filter suffers from significant side lobes (inset in the plot of Region 1.5).
Notice that $\sigma^2_\mathsf{y}(\tau)$ does not converge for white PM noise and flicker PM noise, unless a lowpass filter is introduced to limit the bandwidth, whose cutoff frequency is denoted with $f_H$.

\subsection{Overlapped Allan Variance}\label{sec:Overlapped}
\begin{figure}[t]\centering
  \includegraphics[bb = 0cm 0cm 8.7cm 13.3cm, angle=-90, width=0.98\columnwidth]{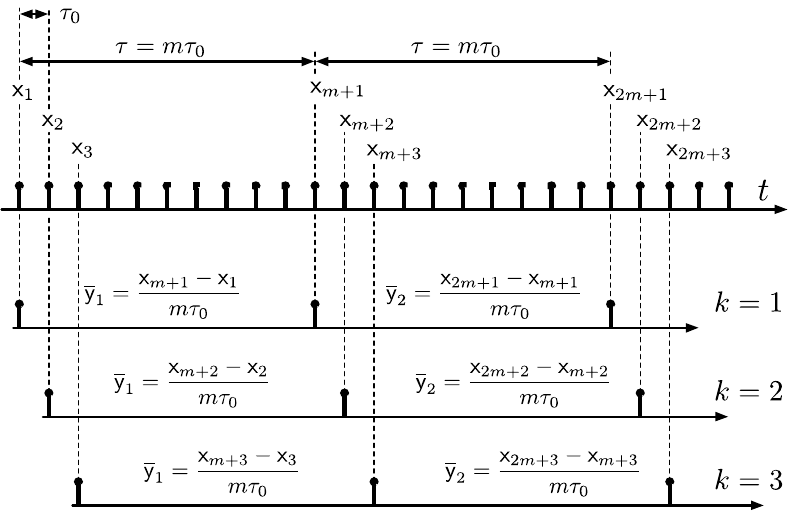}
  \caption{Evaluation of the overlapped Allan variance.}
  \label{fig:Ovl-avar}
\end{figure}

An efficient way to measure the Allan variance is to sample $\mathsf{x}(t)$ at the rate $1/\tau_0$, taking $\tau=m\tau_0$, integer $m$.  The $k$-th value of the fractional frequency is evaluated as 
\begin{equation}
\overline{\mathsf{y}}_k= \frac{\mathsf{x}_{k+m}-\mathsf{x}_{k}}{m\tau_0}\,.
\end{equation}
The overlapped Allan variance consists of using partially overlapped realizations of $\overline{\mathsf{y}}_2-\overline{\mathsf{y}}_1$ in \eqref{eqn:AVAR}, separated by the minimum amount $\tau_0$.  This concept is illustrated in Fig.~\ref{fig:Ovl-avar}.
Accordingly, \eqref{eqn:AVAR-evalx} becomes
\begin{align}
  &\tau=m\tau_0\nonumber\\
  \label{eqn:AVARovl}
  &\widehat{\sigma^2_\mathsf{y}}(m\tau_0) = 
  \frac{1}{2(M-1)}
  \sum_{k=1}^{M-1} 
  \Bigl[
  \frac{ \mathsf{x}_{2m+k} - 2\mathsf{x}_{m+k} + \mathsf{x}_{k}}{m\tau_0}\Bigr]^2\,,
\end{align}
which takes $2m+M-1$ samples, thus a measurement time $\mathcal{T}=(2m+M-2)\tau_0$.

A first advantage of overlapping is smaller uncertainty.  In fact, the confidence interval $\Delta \sigma_\mathsf{y}/\sigma_\mathsf{y}=\sqrt{2/\mathfrak{D}}$ is related to the number $\mathfrak{D}$ of degrees of freedom.
In turn, $\mathfrak{D}$ is equal to the number of samples of $\overline{\mathsf{y}}_{k+1}-\overline{\mathsf{y}}_{k}$ in the case of white PM noise (uncorrelated samples), it gets progressively smaller for slower noise phenomena, and it degenerates to 2 in the case of pure drift.
A simple example deserves attention.  Suppose we have a record of $3{\times}10^5$ samples spaced by $\tau_0=100$ ms, thus $\mathcal{T} = 3{\times}10^4~\unit{s} = 8.33~\unit{hours}$ acquisition time.  At $\tau=10^4$ s, we have only 2 realizations of $\overline{\mathsf{y}}_{k+1}-\overline{\mathsf{y}}_k$ using \eqref{eqn:AVAR-evalx}, while in the same conditions we have $10^5$ realizations if we opt for \eqref{eqn:AVARovl}.  At a deeper sight such realizations are highly correlated, but the larger number is still beneficial to the reduction of uncertainty.

A further advantage of overlapping is that it solves the erratic response of AVAR in the presence of cyclic disturbances of period $T\approx\tau/0.45$ ($2.2\tau$), like the diurnal temperature.  This happens because multiple realizations of AVAR are averaged, based on measures of $\overline{\mathsf{y}}_2-\overline{\mathsf{y}}_1$ progressively shifted by $m\tau_0$, as shown in Fig.~\ref{fig:Ovl-avar}.

The Allan variance with no overlap is seldom used, if ever. Generally, the term `Allan variance' \emph{refers to the overlapping algorithm} with no need of saying so.

\end{document}

\section{Other Options for the Two-Sample Variance\label{sec:Other-var}}

\ifSubfilesClassLoaded{\setcounter{section}{6}\section{Other Options for the Two-Sample Variance}}{}

\subsection[Frequency Counters and Averages]{Frequency Counters and Weighted Averages (Region 2.1)\label{sec:Counters}}
Introducing the classical Allan variance, we have defined $\overline{\mathsf{y}}$ as the uniform average of $\mathsf{y}(t)$ over the time $\tau$.  Other options make sense, based on the redefinition of the average $\overline{\mathsf{y}}$ as  
\begin{equation}
\overline{\mathsf{y}}(\tau) = \int_0^\infty \mathsf{y}(t)\,w(t;\tau)\:dt\,.
\label{eqn:w-avg}
\end{equation}
The weight function $w(t;\tau)$ takes different forms, among which the following deserve attention
\begin{itemize}
  \item $w_\Pi(t;\tau)$ gives the uniform average, the same used in Sec.~\ref{sec:AVAR} (there is still no change).  The classical reciprocal counter, also called $\Pi$ counter after \cite{Rubiola-2005-RSI}, uses $w_\Pi$.  The symbol $\Pi$ recalls the rectangular shape of $w_\Pi(t;\tau)$. 
  
  \item $w_\Lambda(t;\tau)$ gives the triangular average, which is calculated averaging on a sequence of highly overlapped rectangular averages.  The corresponding instrument is the \emph{$\Lambda$ counter} after \cite{Rubiola-2005-RSI}.  Of course, the Greek letter $\Lambda$ is chosen because of its triangular shape.  The benefit of such counter is a high rejection of the wideband white PM noise of the trigger at the counter input.
  Some commercial instruments implement the $\Lambda$ averaging, often without saying.  They can be identified from the `precision' (response to the trigger noise) proportional to $1/\tau\sqrt{\tau}$ instead of  $1/\tau$.
  
  \item $w_\Omega(t;\tau)$ relates to a frequency measurement implemented as a linear regression on phase-time data.
  The corresponding instrument is called $\Omega$ counter after \cite{Rubiola-2016-Omega}.
               The benefit of the $\Omega$ counter is the highest rejection of the white PM noise, by theorem.  
  The Greek letter $\Omega$ is the graphically closest to the parabolic shape of the frequency response.  Besides, the letter $\Omega$ indicates that this is the ultimate counter, to the extent that no other counter performs better rejection of white PM noise.
  
  The `precision' (response to the trigger noise) is proportional to $1/\tau\sqrt{\tau}$ as in the $\Lambda$ counter, just with 1.25 dB lower background noise (a factor of $3/4$).  Very few commercial instruments implement this algorithm.
  
  \item $w_\Delta(t;\tau)$ is equivalent to the difference between two contiguous measures taken with a $\Pi$ frequency counter.  This option is listed only for completeness, because to the best of our knowledge it is not implemented in commercial counters.
\end{itemize}
Like the overlapped Allan variance, $\mathsf{y}(t)$ is sampled at a suitable frequency $1/\tau_0$, thus it holds that $\tau=m\tau_0$.  As a consequence, the patterns of the $\Lambda$ and $\Omega$ counters shown in Region~2.1 hold for $m\gg1$, or equivalently $\tau\gg\tau_0$, so that the continuous approximation holds.

\subsection[Generalized Two-Sample Variances]{Generalized Two-Sample Variances (Region 2.2)\label{sec:Generalized-var}}
The definition \eqref{eqn:AVAR} is more general than the classical Allan variance.
In fact, feeding the weighted averages \eqref{eqn:w-avg} into \eqref{eqn:AVAR} results in different types of variance, which can all be described by the same formula
\begin{equation}
  \sigma^2_\mathsf{y}(\tau) = 
  \mathbb{E}\biggl\{\biggl[\int_0^\infty\mathsf{y}(t)\,w(t;\tau)\:dt\biggr]^2 \biggr\}\,.
  \label{eqn:All-variances}
\end{equation}
Notice that the difference $\overline{\mathsf{y}}_2-\overline{\mathsf{y}}_1$ is now included in the wavelet-like function $w(t;\tau)$.  The latter is similar to a wavelet, but for the normalization for finite-power signals (power-type signals) instead of finite-energy signals (energy-type signals).   

We have the following options
\begin{align*}
  w_A(t;\tau) &= w_\Pi(t-\tau;\tau) - w_\Pi(t;\tau)
    &\rightarrow &&&
    \text{AVAR} ~^{A\!}\sigma^2_\mathsf{y}(\tau)\\\
  w_M(t;\tau) &= w_\Lambda(t-\tau;\tau) - w_\Lambda(t;\tau)
    &\rightarrow &&&
    \text{MVAR} ~^{M\!\!}\sigma^2_\mathsf{y}(\tau)\\
  w_P(t;\tau) &= w_\Omega(t-\tau;\tau) - w_\Omega(t;\tau)
    &\rightarrow &&&
    \text{PVAR} ~^{P\!\!}\sigma^2_\mathsf{y}(\tau)\\
  w_H(t;\tau) &= w_\Delta(t-\tau;\tau) - w_\Delta(t;\tau)
    &\rightarrow &&&
    \text{HVAR} ~^{H\!\!}\sigma^2_\mathsf{y}(\tau)\,. 
\end{align*}

As with the frequency counters, the patterns of $w_M(t;\tau)$ and $w_P(t;\tau)$ shown in Region~2.2 are the limit for $\tau\gg\tau_0$, or equivalently for $m\gg1$, where the continuous approximations holds.  The formulae of Region~2.7 hold under this assumption.

Using commercial frequency counters to get a data stream $\overline{\mathsf{y}}_k$, we recommend attention to the contiguity of averages.  Such contiguity is implied in the waveforms of Region 2.2.  The case of MVAR is subtle because `contiguous' triangular averages overlap by one side, instead of touching one another by one edge.  Something like \makebox{${\wedge}\hspace*{-0.9ex}{\wedge}\hspace*{-0.9ex}{\wedge}\hspace*{-0.9ex}{\wedge}{\ldots}$} instead of \makebox{${\wedge}\hspace*{-0.32ex}{\wedge}\hspace*{-0.32ex}{\wedge}\hspace*{-0.32ex}{\wedge}{\ldots}$}  So, if the measurement of $\overline{\mathsf{y}}$ takes $2\tau$, the measurement of $\overline{\mathsf{y}}_2-\overline{\mathsf{y}}_1$ takes $3\tau$.  Reference \cite{Dawkins-2007} points out that some Keysight counters provide a data stream of the latter type, where continuous averages touch one another by one edge. 

Our notation 
$^{A\!}\sigma^2_\mathsf{y}(\tau)$,
$^{M\!\!}\sigma^2_\mathsf{y}(\tau)$,
$^{P\!\!}\sigma^2_\mathsf{y}(\tau)$ and 
$^{H\!\!}\sigma^2_\mathsf{y}(\tau)$, 
seems more elegant than AVAR, MVAR, PVAR and HVAR, but it is not used in the literature.

We draw the reader's attention to the fact that \eqref{eqn:All-variances} may be misleading in the case of HVAR because it hides the fact that HVAR is a second-difference variance.  HVAR is rather different from the other variances, chiefly in the fact that it converges for integrated flicker FM and random run FM ($1/f^3$ and $1/f^4$ FM), and it is blind to frequency drift.

\subsection{The Time Variance TVAR (Region 2.6)\label{sec:TVAR}}
The Time Variance TVAR is defined as
\begin{align}
  \sigma^2_\mathsf{x}(\tau) = \dfrac{1}{3}\: \tau^2 \; {}^{M\!\!}\sigma^2_\mathsf{y}(\tau)\,.
  \label{eqn:TVAR}
\end{align}
TDEV, the square root of TVAR, is an estimator of the uncertainty of the time elapsed after the duration $\tau$. TDEV is often used in telecom, e.g.\ to assess the Time Interval Error TIE, defined in \cite{ITU-G.8260}, and the holdover performance, i.e., the clock error accumulated between synchronizations \cite[Part 5.1: Timing characteristics of slave clocks\ldots]{ETSI-300-462-Series}.

\subsection{Additional Options} 
The following options are highly specialized instances of the Allan variance, beyond the scope of this article.  They  are mentioned for completeness, and left to experts.

\subsubsection{The Dynamic Allan Variance} 
A time series of $N$ data is sliced into $m$ sub-series of $n=N/m$ data.  Computing the AVAR for each sub-series, we end up with a 3D plot which shows the changes of AVAR vs time, most useful for diagnostic purposes.  This variance and its properties is found in a series of articles by the same team \cite{Galleani-2009,Galleani-2010,Galleani-2011,Galleani-2015,Galleani-2016}, but it does not seem to have been followed by other authors and research teams.

\subsubsection{The Total Variance} 
The time series is circularized by joining a copy with time reversed, as often done in the domain of spectral analysis.  Circularization cannot increase the number $\mathfrak{D}$ of degrees of freedom inherent in the experimental outcomes, but it makes their exploitation more efficient for the detection of certain phenomena.  This concept, described in \cite{Greenhall-1999,Howe-2000}, can be applied to all wavelet variances.

\subsubsection{Thêo1, ThêoH and ThêoBR}
These estimators are based on the idea that two measures of duration $\tau'\ll\tau$ whose centers are spaced by $\tau$ provide a precise estimation of the slow processes occurring at $\tau$, under the condition that $\tau'$ is long enough to average out the fast processes.  These ideas are found in \cite{Howe-2003,Howe-2006,Taylor-2010}.  The benefit is to extend the maximum $\tau$ beyond $\mathcal{T}/2$, where $\mathcal{T}$ is the duration of the data record.  For example, applying ThêoH to a long data record of an atomic time scale, it is possible to extend the plot up to $\tau=0.8\,\mathcal{T}$.  So, these variances are even more efficient than AVAR at estimating the fluctuations occurring at large $\tau$.  This benefit comes at the cost of higher computing burden and additional interpretation difficulty.

\subsection[Choosing the Most Appropriate Variance]{Choosing the Most Appropriate Variance\label{sec:Appropriate-var}}
For general use, the variances described are broadly similar to one another, and none is really ``the best'' or just ``bad.''  Each one has its own `personality,' which makes it more suitable in some specific case, and weaker in other cases.  Such personality follows from the wavelet-like patterns shown in Region 2.2.  A summary of the main options is given below.

\subsubsection{Normalization}
In signal processing, normalization for white noise is the most common option.  
A different choice is made here because the two-sample variances are issued from timekeeping.  The normalization is chosen for all the variances to have the same response $\sigma^2_\mathsf{y}(\tau)=\frac{1}{2}\mathsf{D}_\mathsf{y}^2\tau^2$ to the linear frequency drift $\mathsf{D}_\mathsf{y}$ (the ``linear drift'' row in the lower part of Region 2.7).
A consequence is that different variances have different responses to the other noise processes because the spectral response $|H(f\tau)|^2$ differs (Region~1.5).  Additionally, the corners separating the noise processes are not the same (Regions 1.7, and 2.3--2.5).  Comparing the plots requires a small effort of interpretation.  For example, the white FM noise $S_\varphi=\mathsf{b}_{-2}/f^2 \rightarrow S_\mathsf{y}=\mathsf{h}_0$ shows up as 
\begin{align*}
^{A\!}\sigma^2_\mathsf{y}(\tau)   
  &= \frac{1}{2}\:\frac{\mathsf{h}_0}{\tau} 
   = 0.50\:\frac{\mathsf{h}_0}{\tau} 
  &&\text{(Allan)}\\[1ex] 
^{M\!\!}\sigma^2_\mathsf{y}(\tau) 
  &= \frac{1}{4}\:\frac{\mathsf{h}_0}{\tau}
   = 0.25\:\frac{\mathsf{h}_0}{\tau} 
  &&\text{(Modified Allan)}\\[1ex]
^{H\!\!}\sigma^2_\mathsf{y}(\tau) 
  &= \frac{1}{2}\:\frac{\mathsf{h}_0}{\tau}
   = 0.50\:\frac{\mathsf{h}_0}{\tau} 
  &&\text{(Hadamard)}\\[1ex]
^{P\!\!}\sigma^2_\mathsf{y}(\tau) 
  &= \frac{3}{5}\:\frac{\mathsf{h}_0}{\tau}
   = 0.60\:\frac{\mathsf{h}_0}{\tau} 
  &&\text{(Parabolic)}\,,
\end{align*}
with a maximum difference of a factor of 2.4 (3.8 dB)\@.

\subsubsection{AVAR}This the best choice when we want to evaluate $\sigma^2_\mathsf{y}(\tau)$ up to the largest $\tau$ for a given data record of duration $\mathcal{T}$, that is, $\tau=\mathcal{T}/2$.  This is the typical case of atomic time scales, where the oscillators are continuously monitored, and we focus on the slow processes.  

Because $w_A(t;\tau)$ takes $2\tau$ for one realization of $\sigma^2_\mathsf{y}(\tau)$, averaging on $M\gg1$ realizations is made possible by overlapping the measures with $\tau_0\ll\tau$ (Sec.~\ref{sec:Overlapped}).

The uniform weight of $w_A(t;\tau)$ features the highest efficiency in picking up the energy of $\mathsf{y}$.  By contrast, AVAR is unsuitable to the measurement of white PM noise because $^{A\!}\sigma^2_\mathsf{y}(\tau)\simeq0.076\,f_{H}\mathsf{h}_{2}/\tau^2$ (Region 2.7), thus the result is highly dependent on the bandwidth $f_H$.  This is not a problem for the slow phenomena we mentioned.
Finally, it is worth mentioning that, in the presence of white FM noise only, AVAR gives the same result as the classical variance.

\subsubsection{MVAR} This is a good choice in the presence of wideband noise typical of fast processes.  MVAR originates from optics, where a precise and efficient measurement of white PM noise is a desired feature.

By contrast, MVAR is inferior to AVAR in the efficient use of $\mathcal{T}$ because the support of $w_M(t;\tau)$ is $3\tau$ wide instead of $2\tau$.  This may not be a problem when the physical phenomena we are interested in occur at small or moderate $\tau$, say hours.

\subsubsection{HVAR} This variance is useful for the measurement of strong slow phenomena occurring in some circumstances, for example in the absence of temperature stabilization.  In fact, unlike the other variances described here, it converges for integrated flicker FM noise and for integrated random walk FM noise, also called `random run.'  Such processes are the $\mathsf{h}_{-3}/f^{3}$ and $\mathsf{h}_{-4}/f^{4}$ terms of $S_\mathsf{y}(f)$, or equivalently the $\mathsf{b}_{-5}/f^{5}$ and $\mathsf{h}_{-6}/f^{6}$ terms of $S_\varphi(f)$.  By contrast, HVAR is blind to linear drift.  This makes HVAR a specialized tool, particularly useful when high drift makes it diffucult to estimate the other noise parameters. 

Finally HVAR, like AVAR, gives ambiguous response to white PM noise, as it depends on the instrument bandwidth $f_H$.  Like MVAR, $w_H(t;\tau)$ takes a time equals to $3\tau$.

\subsubsection{PVAR}
The computation of $\sigma^2_\mathsf{y}(\tau)$ at $\tau=m\tau_0$ requires a data record of duration $\mathcal{T}=k\tau$, where $k>2$ for AVAR and PVAR, and $k>3$ for MVAR and HVAR\@.   The minimum $\mathcal{T}$ depends on the noise process, on the confidence level required, and on the variance we choose.  Running the measurement, we start seeing white PM noise at short $\mathcal{T}$, then flicker PM, white FM etc.\ as $\mathcal{T}$ increases.
Now we take a different standpoint, asking which is the minimum $\mathcal{T}$ to detect a `new' noise phenomenon, out of the `previous,' faster one. For example, which is the shortest $\mathcal{T}$ to see that flicker FM is above the white FM, with 95\,\% probability?  
Among the four variances considered here, PVAR is the best at doing this from white PM to RW FM\@.  Yet, the corner between RW FM and frequency drift is still better detected by AVAR\@.

\subsection{Some Pieces of Advice}
Beginners should restrict their attention to ADEV and MDEV, the square root of AVAR and MVAR\@.  ADEV and MDEV are both available in commercial instruments, and both benefit from the size effect of a wide community.

For historical reasons, ADEV is definitely the manufacturers' preferred option.
Reading technical documentation, we recommend attention to a possible confusion between ADEV and MDEV under the term `Allan deviation,' with a possible `modified' omitted or implied.  The ITU-T Recommendation G.8260 \cite{ITU-G.8260} compares ADEV, MDEV and TDEV from the standpoint of telecommunications.  MDEV seems the favorite tool in telecommunications \cite{Bregni-2016}, and enables the direct calculation of TDEV using \eqref{eqn:TVAR}.  However, MDEV takes 50\% longer acquisition time $\mathcal{T}$.

Doing one's own measurements, MDEV is in most cases the best compromise.  In fact, MDEV improves on ADEV in the detection of fast noise processes (white and flicker PM), and is as suitable as ADEV to detect all the other noise processes.  
AVAR is still the best option for the measurement of the atomic clocks\footnote{In the international coordination of metrology under the guidance of BIPM, an atomic oscillator is a `clock' only after 6 months of uninterrupted contribution to UTC/TAI\@.  Some commercial oscillators are called clock by the manufacturer implying that they are suitable to contribute to UTC/TAI\@.  This is perfectly sound because hundreds of such Cesium clocks actually do this --- and Hydrogen masers as well.  However, the term clock is also used as a pretentious replacement for a precision oscillator even if reliability and long-term stability are insufficient for the oscillator to be even considered as a contributor to UTC/TAI.} intended for time scales, where increasing $\mathcal{T}$ is costly or impossible.
Finally, the reader should remember that normalization makes MVAR always `optimistic' compared to AVAR, as seen in the related columns in Region 2.7.

Looking at the future, PDEV may replace MDEV because it outperforms it in all parameters at no cost but computing power.  Likewise, Thêo may replace ADEV because of the more efficient use of $\mathcal{T}$ at representing longer values of $\tau$.

\subsection{Example of MDEV}
\begin{figure}[t]\centering
  \includegraphics[bb = 5mm 10mm 210mm 286mm,angle=-90,width=0.99\columnwidth]{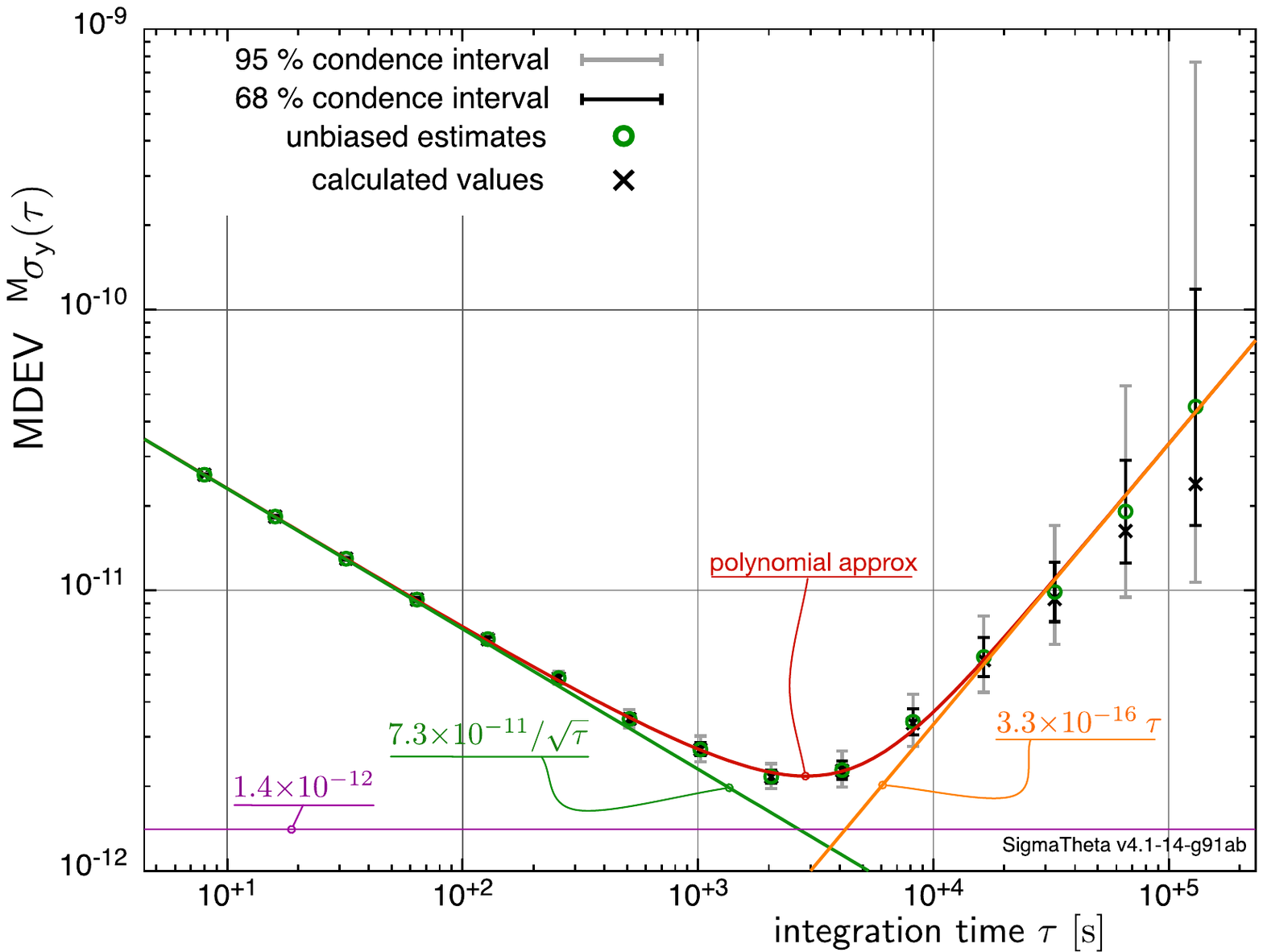}%
  \caption{Example of MDEV obtained with the SigmaTheta software tool.  Notice the difference between the calculated values of MDEV (black crosses) and the Bayesian estimates (green donuts), and the asymmetry of the uncertainty bars. Data are courtesy of Moustafa ``Mouss'' Abdel Hafiz, FEMTO-ST Institute.}
  \label{fig:MDEV-example}
\end{figure}

We measure the stability of a miniature Cs oscillator.  The oscillator is a laboratory prototype based on the Coherent Population Trapping (CPT) principle in a Cs microcell.  The actual experiment is described in \cite{Abdel-Hafiz-2022}, and the fabrication of the microcell in \cite{Hasegawa-2011}.
For technical reasons specific to the experiment, the frequency is sampled at 150 ms intervals and pre-processed to provide a stream of values of $\mathsf{y}_ks$ uniformly averaged on contiguous intervals of 1 s.  This sets $\tau_0=1$ s.  The total number of samples is 389998, for a total duration $\mathcal{T}\approx4.5$ days.  

Fig.~\ref{fig:MDEV-example} shows the MDEV of the experiment described, processed with the SigmaTheta software tool.  Starting from uniformly averaged data, a practical minimum of 8--10 samples is needed to approximate the triangular average (Regions 2.1 and 2.2.).  Thus, MDEV is plotted for $\tau=2^n\tau_0$ starting from $\tau=8\tau_0=8$ \unit{s}.  At the scale of this experiment, the H maser used as the reference can be considered ideally stable, and the noise of the instrument is negligible as well.

The SigmaTheta software package provides the following pieces of information (see Sections~\ref{sec:Confidence} and \ref{sec:Software})
\begin{itemize}
  \item The bare values of $^{M\!}\sigma_\mathsf{y}(\tau)$, shown as black crosses,
  \item The Bayesian estimates of $^{M\!}\sigma_\mathsf{y}(\tau)$, shown as green donuts, 
  \item The uncertainty bars,
  \item The identification of the most relevant noise processes, shown as the colored straight lines of slope $\smash{1/\sqrt{\tau}}$, constant vs $\tau$, and $\sqrt{\tau}$.
\end{itemize}
It is worth pointing out that the uncertainty bars systematically extend upwards more than downwards.  This is a kind of `signature' of the inverse problem, as opposed to the bare simulation approach.

\subsection[Suggested Readings]{Suggested Readings About Variances\label{bib:AVAR}}
\subsubsection{General references} The free booklet \cite{Riley-2008} is probably where most readers should start.  Sponsored and distributed by NIST, it provides an extensive coverage of most variances (AVAR, MVAR, HVAR etc.) and the evaluation of the confidence intervals, with numerous examples and plots made with Stable32.
Reference \cite{Stein-2010} is a review article about the Allan variance, and \cite{Enzer-2021} suggests that the Allan variance can be used as a diagnostic tool.
A wealth of information is available in a Special Issue of the IEEE Transact.\ UFFC celebrating the 50th anniversary of the Allan Variance \cite{Levine-2016}.  An historical review is available \cite{Allan-2016}, written by two of the most important contributors to the rise of this branch of knowledge.
Reference \cite{Allan-1966} is the original article that introduces the sample variances, later called Allan variance, and \cite{Allan-1981} introduces MVAR.

\subsubsection{$\Pi$ and $\Lambda$ counters, and the related statistics} Reference \cite{Rubiola-2005-RSI} is the first article which defines $\Pi$ and $\Lambda$ counters and the mathematical framework underneath, later extended to the case of non-overlapping triangular averages \cite{Dawkins-2007}.  However, the basic ideas were already in \cite{Kramer-2001,Kramer-2004}, yet without developping the statistical framework.  A wealth of practical knowledge about the architecture of high-resolution counters is available in a review article \cite{Kalisz-2004}.

\subsubsection{The $\Omega$ counter and the Parabolic Variance}
The linear regression of phase data is a rather obvious way to estimate a frequency.  It was used in the HP5371A HP5372A time interval analyzers\footnote{Information provided by Magnus Danielsson, NetInsight, Sweden.} in the late 1980s, and at Pendulum \cite{Johansson-2005}.
The name ``$\Omega$ counter'' comes from \cite{Rubiola-2016-Omega}, which introduces the related mathematical framework for frequency metrology.  
Feeding the linear-regression estimates into \eqref{eqn:AVAR}, we get PVAR\@.  This idea came independently from \cite{Benkler-2015} and \cite{Vernotte-2016-UFFC}, and the name PVAR was decided together by the two teams.  Reference \cite{Vernotte-2016-UFFC} digresses the advanced statistical properties of PVAR, including the Bayesian statistics and the minimum duration of the data record to detect a noise process.

\subsubsection{Aliasing\label{sec:AVAR-aliasing}} Reference \cite{Vernotte-1998b} provides theory and insight on  spectral aliasing, and \cite{Calosso-2016b} gives an interesting perspective about aliasing in the AVAR and MVAR, covering the effect of spectral bumps and `blue noise,' often found in long-range optical frequency-distribution systems.  Finally, \cite{Bernier-1987} provides useful insight in aliasing and cutoff frequency for MVAR\@.

\ifSubfilesClassLoaded{\bibliography{Ref-short,References,Ref-local}}{}
\end{document}

\section{Relations Between Phase Noise and Variance\label{sec:Conversion}}

\ifSubfilesClassLoaded{\setcounter{section}{7}\section{Relations Between Phase Noise and Variance}}{}

\subsection[Visual Inspection on Plots]{Visual Inspection of Plots (Regions 1.6--1.7)\label{sec:Visual-inspection}}
Regions 1.6--1.7 show a plot of $S_\mathsf{y}(f)$ beside AVAR $\sigma^2_\mathsf{y}(\tau)$ for the noise processes from white PM to random-walk FM\@.  The following facts deserve attention.
\begin{itemize}
  \item There is a kind of mirror symmetry between the plots of  $S_\mathsf{y}(f)$ and $\sigma^2_\mathsf{y}(\tau)$.  The fastest process, white PM, is on the right-hand side of $S_\mathsf{y}(f)$, and on the left-hand side of $\sigma^2_\mathsf{y}(\tau)$.  Vice versa, the FM random walk is on the left-hand side of $S_\mathsf{y}(f)$, and on the right-hand side of $\sigma^2_\mathsf{y}(\tau)$.
  \item The cutoff frequency $f_H$ has a dramatic effect on white PM noise, only a weak effect on the flicker PM noise, and virtually no effect on slower processes.
  \item The $1/\tau^2$ region of $\sigma^2_\mathsf{y}(\tau)$ is ambiguous, in that it represents both white PM and flicker PM noise.
  \item The conversion from $S_\varphi(f)$ to $\sigma^2_\mathsf{y}(\tau)$ is always possible, while the opposite suffers from limitations.  This is emphasized by the road signs between Regions 1.6 and 1.7. 
  \item The corner $\tau$ where the processes cross one another may occur rather far from the values one expects intuitively.
\end{itemize}
If we have both phase noise and variance measures, the conversion from $S_\varphi(f)$ to $\sigma^2_\mathsf{y}(\tau)$ is a great way to check on consistency.  However, this is possible only if $S_\varphi(f)$ extends to sufficiently low $f$ to reveal the slow processes shown by $\sigma^2_\mathsf{y}(\tau)$.  Regardless, the conversion is an exercise of interpretation we recommend.

\subsection[PM Noise to Variance Conversion]{Conversion from PM Noise to Allan Variance (Region 2.7)\label{sec:PM-to-AVAR}}
\begin{figure}[t]\centering
  \includegraphics[bb = 0cm 10.5cm 7.2cm 29.7cm, width=0.8\columnwidth]{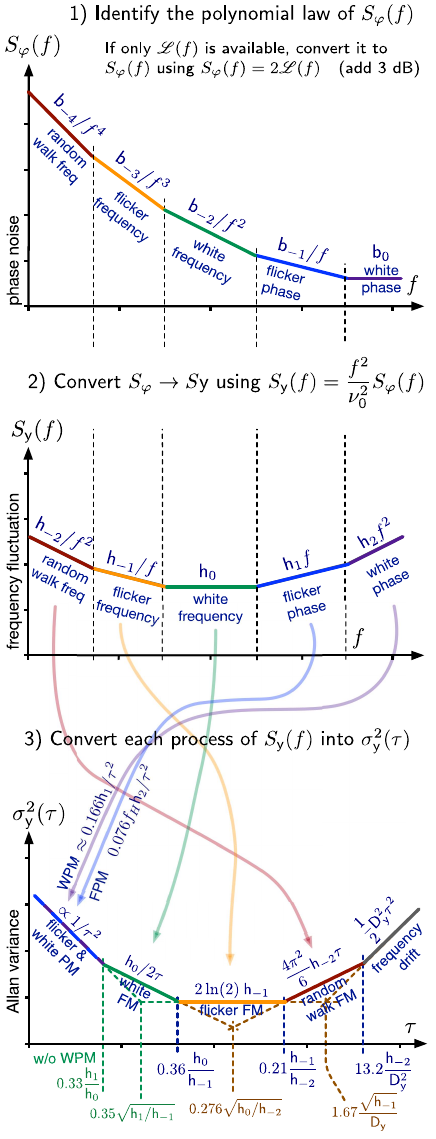}
  \caption{Conversion from phase noise to Allan variance.  
  The colors of the straight line approximation recall the frequency, from reddish (low) to bluish (high).
  The conversion from phase noise to the other two-sample variances is an obvious extension.}
  \label{fig:S-to-sigma2}
\end{figure}

The reader should first refer to Region 1.11 for the conversion $S_\varphi(f)\rightarrow S_\mathsf{y}(f)$.  Then, to Region 2.7 for $S_\mathsf{y}(f)\rightarrow\sigma^2_\mathsf{y}(\tau)$, based on  \eqref{eqn:Spectral-response}-\eqref{eqn:HA}.
A simple procedure is shown in Fig.~\ref{fig:S-to-sigma2}, and detailed below.
\begin{enumerate}\addtocounter{enumi}{-1}
\item Start with a log-log plot of phase noise.  Since in most cases the phase noise is given as $10\log_{10}[\mathscr{L}(f)]$, you have to convert it into $S_\varphi(f)$ using 
$S_\varphi(f)=2\mathscr{L}(f)$, i.e., add 3 dB (Region 1.11)

\item Approximate the true spectrum with the straight lines which represent the polynomial law.  Proceed from the right-hand side (white PM noise) to the left, not vice versa.  This best done with a drawing app inserting a straight line of exact slope (0, $-1$, $-2$ etc.) and shifting it to fit the plot.  
The coefficients $\mathsf{b}_0$, $\mathsf{b}_{-1}$,  $\mathsf{b}_{-2}$, etc.\ are the value of the corresponding straight line found at $f=1$ Hz.
Even with little training, visual inspection is efficient at approximating the noise process and ignoring the artifacts.
Sliding old-fashioned set drafting squares on a printed spectrum is a good alternative to a drawing app.
Worked-out examples are available in \cite[Chapter~6]{Rubiola-2010-Cambridge}.

\item Convert $S_\varphi(f)$ into $S_\mathsf{y}(f)$ using
  \begin{align}\nonumber
    \mathsf{h}_n = \frac{1}{\nu_0^2} \, \mathsf{b}_{n-2}\,,
  \end{align}
which is equivalent to $S_\mathsf{y}(f) = \frac{f^2}{\nu_0^2} S_\varphi(f)$
(Region 1.11).  Newcomers may be surprised by the small value of the $\mathsf{h}_n$ coefficients, due to $\nu_0^2$ in the denominator.

\item Sketch the Allan variance using the pattern of Region~1.7 (right) and the formulas found in the `AVAR' column of the Table in Region~2.7.  Each process (white PM, flicker PM, etc.) requires its own formula, found in Region 2.7.
\end{enumerate}
It goes without saying that the method described also applies to the other two-sample variances discussed in Sec.~\ref{sec:Other-var}, just picking up the appropriate column in Region~2.7.  The extension to non-integer slopes is found in \cite{Walter-1994,Vernotte-2021-IM}.

\subsection[Variance to Spectrum Conversion]{Variance to Spectrum Conversion (Regions 1.6, 1.7, 2.7)\label{sec:Var-to-S}}
The variance to spectrum conversion is not possible in the general case, but we can get useful information using the formulas given in Region 2.7 assuming that the spectrum is smooth and follows the polynomial law (see \cite{Greenhall-1998}).

A first limitation is that in the case of AVAR and HVAR, both white and flicker PM show up as $\sigma^2_\mathsf{y}\propto1/\tau^2$, thus they cannot be divided.  Additionally, AVAR and HVAR are almost unusable in the white PM region because $\sigma^2_\mathsf{y}\propto f_H$.  The cutoff $f_H$ does not appear explicitly in MVAR, PVAR and TVAR\@.  
However, the effect of $f_H$ is hidden in the sampling process itself and then in the data. Since data are sampled at $\tau_0$ interval, the Nyquist frequency is $\frac{1}{2\tau_0}$.  Thus, if $f_H>\frac{1}{2\tau_0}$, there will be spectral aliasing and the white noise level will be overestimated.

A second limitation is related to resolution.  In fact, the inherent resolution of $\sigma^2_\mathsf{y}(\tau)$ is one octave in $\tau$, as it follows from the bandwidth of the main lobe in Region~1.5.  Thus, $\sigma_\mathsf{y}(\tau)$ is usually plotted for $\tau$ in geometric series like $1, 2, 4, 8\ldots$.  Conversely, the frequency resolution of $S_\varphi(f)$ depends on the acquisition time, and plots usually represent $S_\varphi(f)$ with a resolution of 50--100 points/decade, that is, 15--30 points per octave.

\subsection[The Cutoff Frequency $f_H$]{The Cutoff Frequency $f_H$ and the Sampling Interval $\tau_0$}
The cutoff frequency $f_H$, extensively used in the spectrum-to-variance conversions is often a source of confusion.  Notice that an anti-aliasing filter is necessary, otherwise the variance does not converge in the presence of white and flicker PM noise \cite{Bernier-1987}.

The phase time $\mathsf{x}(t)$ has finite bandwidth which results from the measurement process and from the architecture of the instrument.  This is generally described as the noise equivalent bandwidth, denoted with $f_H$.  

The variance is evaluated after sampling $\mathsf{x}(t)$ at an appropriate frequency $1/\tau_0$, which requires that $f_H<1/(2\tau_0)$.  If this condition is not met, aliasing takes place.  The white PM noise is folded $2f_H\tau_0$ times to the first Nyquist zone, and the observed white noise level is $2f_H\tau_0\mathsf{k}_0$ instead of $\mathsf{k}_0$.  Flicker noise too is subject to aliasing, even though the impact on the results may be smaller.  References \cite{Bernier-1987,Vernotte-1998a,Vernotte-1998b} detail the specific problems related to our domain.

\color{black}

\subsection{Example ($f_H$)}
\begin{figure}[t]\centering
  \begin{subfigure}[t]{\columnwidth}\centering
    \caption{\hspace*{0.8\columnwidth}}
    \includegraphics[bb = 0cm 0cm 4.6cm 8.8cm,angle=-90,scale=0.71]{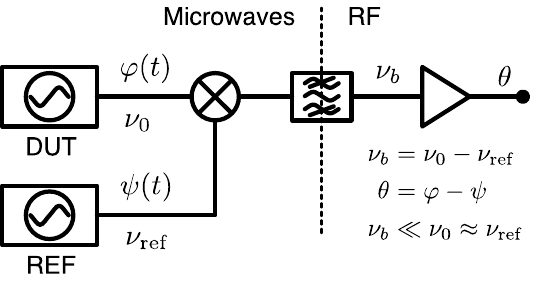}
  \end{subfigure}
  \begin{subfigure}[t]{\columnwidth}\centering
    \vspace{1em}
    \caption{\hspace*{0.8\columnwidth}}
    \includegraphics[bb = 0cm 0cm 4.6cm 8.8cm,angle=-90,scale=0.71]{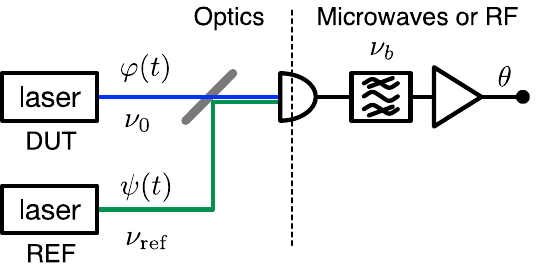}
  \end{subfigure}
  \caption{Beat method, applied (a) to electrical signals, and (b) to optical signals.  See Section~\ref{sec:Beat} for the general description.}
  \label{fig:Beat-method}
\end{figure}

Let us take an example from optics, where we beat two 1550-nm lasers in a photodiode with the scheme of Fig~\ref{fig:Beat-method}, getting a RF tone at $\nu_b=60$ MHz.  For technical reasons we choose to filter such signal with $\pm0.5$ MHz bandpass centered at $\nu_b$.  The filter halfwidth is $2.6{\times}10^{-9}$ of the optical carrier, thus the laser instability must be ${<}\,5{\times}10^{-10}$ for $\nu_b$ to be decently centered in the filter pass band.  The noise bandwidth is equal to the filter half-width, i.e., $f_H=500$ kHz.

\subsubsection{First option} we measure the beat note with a counter sampling at $\tau_0=1$ ms interval (for example, the old good K\&K counters, or the more recent version made by Lange Electronic).  In the conditions described, aliasing increases the white PM noise level by a factor $2f_H\tau_0=1000$.

\subsubsection{Second option} we measure the same beat note with an instrument based on direct digitization (Sec.~\ref{sec:Analyzer-digital} and Fig.~\ref{fig:Analyzer-Digital}) or a Tracking DDS (Sec.~\ref{sec:TDDS} and Fig.~\ref{fig:TDDS}).  Inside the instrument, the ADCs are preceded by antialiasing filters, and a lowpass reduces the bandwidth of $\mathsf{x}(t)$ after detection.  Thus the condition $f_H<1/(2\tau_0)$ is met, and there is no aliasing.  In some instruments $f_H$ can be set by the user.

\ifSubfilesClassLoaded{\bibliography{Ref-short,References,Ref-local}}{}
\end{document}

\section{Confidence Intervals}\label{sec:Confidence}

\ifSubfilesClassLoaded{\setcounter{section}{8}\section{Confidence Intervals}}{}
Assessing the confidence intervals is all about understanding the interplay between the `true value' $\sigma^2_\mathsf{y}(\tau)$ and its estimate $\widehat{\sigma_\mathsf{y}^2}(\tau)$.  The latter is evaluated with \eqref{eqn:AVAR-evaly}, or equivalently \eqref{eqn:AVAR-evalx} for AVAR, and with similar formulae for other variances.  Since everything in this Section applies separately to each value of $\tau$, we let $\tau$ implied in $\widehat{\sigma_\mathsf{y}^2}$ and $\sigma_\mathsf{y}^2$.

\subsection{Direct Problem}\label{sec:Direct}
The \emph{direct problem} is similar to computer simulations, where we add noise to a deterministic phenomenon.
Without noise, the right-hand side of \eqref{eqn:AVAR-evaly} gives $\sigma^2_\mathsf{y}$.  In this case, the term `true value' for $\sigma^2_\mathsf{y}$ is fully legitimate because uncertainty is zero.  Introducing the noise, the same equation gives the estimate $\widehat{\sigma_\mathsf{y}^2}$, together with the associated Probability Density Function, or PDF, 
\begin{math}
  p(\widehat{\sigma_\mathsf{y}^2}|\sigma_\mathsf{y}^2)
\end{math}.  
The notation $p(\:|\:)$ stands for conditional probability, and the vertical bar `|' reads `given' or `knowing,' as in $p(\widehat{\sigma_\mathsf{y}^2})$ \emph{given} $p(\sigma_\mathsf{y}^2)$.

\subsection{Inverse Problem}\label{sec:Inverse}
The simulation approach does not help to grab the hidden reality $\sigma^2_\mathsf{y}$ from the outcome $\widehat{\sigma_\mathsf{y}^2}$.  The answer is the \emph{inverse problem}, which consists of inferring the true  $\sigma^2_\mathsf{y}$ and the associated PDF
\begin{math}
  p(\sigma_\mathsf{y}^2|\widehat{\sigma_\mathsf{y}^2})
\end{math}.  

Otherwise stated, the direct problem targets the
\begin{math}
  \text{`cause}\rightarrow\text{effect?'}
\end{math}
relationship, where `?' emphasizes the unknown. 
In contrast, the inverse problem targets 
\begin{math}
  \text{`cause?}\leftarrow\text{effect'}
\end{math}.
Interestingly, the `true value' $\sigma^2_\mathsf{y}$ in the direct problem and the `inferred true value' $\sigma^2_\mathsf{y}$ in the inverse problem are different things, but they are generally denoted with the same symbol in the literature.

The evaluation of the confidence intervals in the inverse problem may be too complex for some readers.  These readers may keep only the two results summarizedbelow, skip Section~\ref{sec:Bayes}, and go straight to the choice of the appropriate software package.
\begin{itemize}
  \item When a small number $M-1$ of realizations $\overline{\mathsf{y}}_{k+1}-\overline{\mathsf{y}}_{k}$ is available, the `error bars' are large and asymmetric.  The half bar directed upwards is wider  than the downwards half bar.  This is seen in the example Fig.~\ref{fig:MDEV-example} for $\tau\ge4000$~s, and qualitatively illustrated in Region~1.7.
  
  \item With large $M$, there is no practical difference between direct problem and inverse problem.  The confidence intervals are small and symmetric, approaching $1/\sqrt{M-1}$, as seen in Fig.~\ref{fig:MDEV-example} for $\tau\le2000$~s.
\end{itemize}

\subsection{Application of the Bayesian Statistics to Allan Variances}\label{sec:Bayes}
Everything starts with the Bayes theorem, which states that 
\begin{equation}
  p(\theta|\xi) = \frac{\pi(\theta) \, p(\xi|\theta)}{\pi(\xi)}\,,
  \qquad \pi(\xi\neq0)\,,
\end{equation}
where $p(\:)$ \ denotes the \emph{posterior} PDF, and  $\pi(\:)$ denotes the \emph{prior} PDF\footnote{More specifically, $\pi(\theta)$ is the a-priori knowledge before any measurement, and $\pi(\xi)$ is an unknown function we don't care about because it is independent of $\theta$.  The prior $\pi(\xi)$ can be determined using the property that $\int p(\theta|\xi)\:d\theta=1$.}.
It is worth mentioning that some authors use the same symbol for both prior probability and posterior probability. 
In the measurement of frequency stability, the experimental value  $\widehat{\sigma_\mathsf{y}^2}$ is identified with  $\xi $, and the unknown `true'  $\sigma_\mathsf{y}^2$ with  $\theta$.  Thus, we have to infer a confidence interval on  $\sigma_\mathsf{y}^2$.

The central limit theorem suggests that the $\overline{\mathsf{y}}_k$ are Gaussian, as they result from a lot of data.  Thus, we assume that their differences are Gaussian centered.  Thus, \eqref{eqn:AVAR-evaly} indicates that $\widehat{\sigma_\mathsf{y}^2}$ is described by a $\chi^2$ distribution with $\mathfrak{D}$ degrees of freedom\footnote{In the literature about statistics the degrees of freedom are more often denoted with $\nu$, but in our notation $\nu$ is used for the carrier frequency.}.  Such distribution is denoted with $\chi^2_\mathfrak{D}$.  Of course, with  $M$ values of $\overline{\mathsf{y}}_k$ it holds that $\mathfrak{D}\leq M-1$, where the equality indicates that all the terms $(\overline{\mathsf{y}}_{k+1}-\overline{\mathsf{y}}_k)$ of the sum are statistically independent.  Greenhall and Riley provides a very useful method to evaluate $\mathfrak{D}$ \cite{Greenhall-2003}. 

Since the random variable $\xi $ is $\chi^2_\mathfrak{D}$ distributed, the cumulative density function (CDF) of $\xi$ knowing $\theta$, denoted with $F(\xi|\theta)$, is also known in analytic form.  The inverse CDF, available in the major mathematical libraries, enables the user to compute the confidence interval.  

The above is for the direct problem.  The inverse problem can be solved thanks to the relevant property that a  $\chi^2_\mathfrak{D}$ distribution is defined by one and only parameter, $\mathfrak{D}$.  It has been proved that such distributions are ``fiducial'' distributions \cite{Fisher-1935}, which means that the equality
\begin{equation}
F(\theta|\xi) = 1-F(\xi|\theta) \qquad\text{(fiducial)}
\label{eqn:fiducial}
\end{equation}
holds in both frequentist inference and Bayesian inference, provided that a $1/\theta$ prior (prior of total ignorance) is chosen \cite{Lindley-1958,Vernotte-2012}.  This implies that the confidence intervals given by frequentist or Bayesian methods are the same and are easy to compute.

Note that, because the $\chi^2_\mathfrak{D}(x)$ distribution is strongly asymmetric with steep rise at small $x$ and slow decay at high $x$, it results from \eqref{eqn:fiducial} that $p(\theta|\xi)$ must have a steep side at some high  $x$ and a slow decay towards $x=0$.  The consequence is that the error bars on a log-log plot of Allan variance are extend upwards more than downwards.  This behavior is more remarkable at small $\mathfrak{D}$.

\ifSubfilesClassLoaded{\bibliography{Ref-short,References,Ref-local}}{}
\end{document}

\section{Measurement Techniques\label{sec:Techniques}}

\ifSubfilesClassLoaded{\setcounter{section}{9}\section{Measurement Techniques}}{}
\subsection{Saturated-Mixer Phase Noise Analyzer}
\begin{figure}[t]\centering
  \includegraphics[bb = 0cm 0cm 9.2cm 13.1cm,angle=-90,width=0.98\columnwidth]{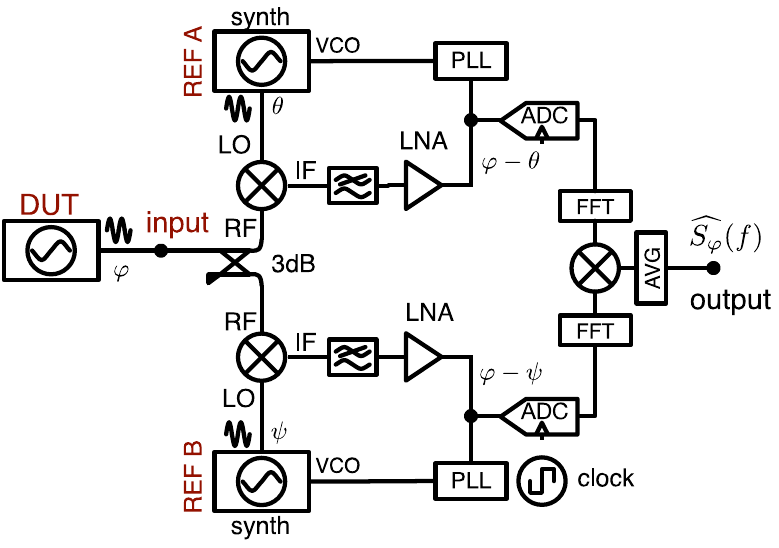}
  \caption{Traditional phase noise analyzer.  The mixers, saturated at both input, work as phase-to-voltage converters.  Two equal channels are used to reject the background noise after averaging on multiple cross spectra.}
  \label{fig:Analyzer-1}
\end{figure}

Fig.~\ref{fig:Analyzer-1} shows the block diagram of a traditional phase-noise analyzer.  The instrument consists of two equal channels where a double balanced mixer compares the phase $\varphi$ of the oscillator under test (DUT) to the phase $\theta$ and $\psi$ of the references.  The error signal sent to the LNA is equal to $k_d(\varphi-\theta)$, or to $k_d(\varphi-\psi)$, where $k_d$ is the mixer's phase-to-voltage gain.  
The typical value of $k_d$ is between 0.1 V/rad and 1 V/rad, depending on the signal level, technology, and frequency.

This operating mode requires that $\varphi-\theta=\pi/2$ within $\pm0.1$ rad, and likewise $\varphi-\psi$, which is ensured by the PLL\@.  The error signal, corrected for the equation of the PLL, is proportional to the DUT phase fluctuation vs the references.  A problem with mixers is that the power range rather narrow, from a minimum of $\approx5$ dBm to a maximum of 17--20 dBm, depending on technology.  The power range can be extended adding an amplifier at each RF input.  

Assuming that the two channels are independent (separate mixers and separate synthesizers, and no crosstalk), the average cross spectrum is proportional to the DUT phase noise because the background noise of the individual channel is rejected by the averaging process.  It is worth mentioning that averaging is essential to reject the noise of the synthesizers.  The synthesizers are necessary for the instrument to be suitable to a wide range of frequency, but they are generally noisier than oscillators.  

Instruments of this type are commercially available from Anapico, Berkeley Nucleonics Corp., Holzworth, Keysight Technologies, NoiseXT/Arcale, and Quantic Wenzel (a brand of Wenzel Associates).  The frequency synthesizers may be inside the instrument, or external modules.  The input power splitter may differ from the 3 dB directional coupler shown in Fig.~\ref{fig:Analyzer-1}.  We have seen Y resistive networks (Keysight), and V resistive network (Holzworth).

\subsection{Mathematical Tools for Spectral Analysis\label{sec:Math-tools}}%
The estimation of power spectra based on the Fast Fourier Transform (FFT) is common to all phase noise analyzers, analog or digital.  The FFT is a fast algorithm for the evaluation of the Discrete Fourier Transform (DFT), and gives the same result.  The algorithm used in most software packages and libraries is the `Fastest Fourier Transform of the West' FFTW3 \cite{Frigo-2005}.  With a bit of humor, we note that the FFTW comes from the Eastern USA\@.
The book \cite{Brigham-1988} is Enrico's favorite reference about the FFT, while \cite{Babitch-1974} is about the early use of the FFT in phase noise measurements.

The most commonly used algorithm for the estimation of the PSD, introduced in \cite{Welch-1967} and known as the Welch algorithm, stands on the FFT\@.  In several software packages and libraries, the PSD function takes a name which recalls Welch.  This is the case of Matlab, Octave and Python (SciPy).  The classical article \cite{Harris-1978} discusses extensively the windows (taper) functions used in spectral analysis.

Reference \cite{Bennett-1948} introduces the spectral analysis of quantized signal, and \cite{Widrow-1996,Widrow-2008} provide an extensive treatise of quantization noise.
\color{black}

\subsection[Background-Noise Rejection]{How the Background-Noise Rejection Works\label{sec:Rejection}}
The noise rejection is based on the \emph{cross-spectrum method}, described below.
Referring to Fig.~\ref{fig:Analyzer-1}, we define
\begin{align}
  x=\varphi-\theta\quad\leftrightarrow\quad X&=\Phi-\Theta \label{eqn:xspx}\\
  y=\varphi-\psi  \quad\leftrightarrow\quad Y&=\Phi-\Psi\,, \label{eqn:xspy}
\end{align}
where `$\leftrightarrow$' stands for Fourier transform inverse-transform pair, time and frequency are implied, and $x$ and $y$ should not be mistaken for $\mathsf{x}$ and $\mathsf{y}$. 
For us, $\varphi$ is the \emph{signal}, $\theta$ and  $\psi$ are the \emph{noise} to be rejected.  The quantities $\theta$ and $\psi$ account for all the noise sources, i.e., references, synthesizers, mixers, low-noise amplifiers, etc.

The cross spectrum averaged on $m$ acquisitions is
\begin{align}
  S_{yx} 
  &= \frac{2}{T} \Bigl< YX^* \Bigr>_m\label{eqn:xsp1}\\[0.8ex]
  &= \frac{2}{T} \Bigl< (\Phi-\Psi)\,(\Phi-\Theta)^* \Bigr>_m\label{eqn:xsp2}\\[0.8ex]
  &= \frac{2}{T} \Bigl< \Phi\Phi^* - \Phi\Theta^* - \Psi\Phi^* + \Psi\Theta^* \Bigr>_m\,.
  \label{eqn:xsp3}
\end{align}
It is seen on Fig.~\ref{fig:Analyzer-1} that $\varphi$, $\theta$ and $\psi$ are statistically independent (separate and independent hardware), and likewise $\Phi$, $\Theta$ and $\Psi$.  Thus, the mathematical expectation of $\Phi\Theta^*$, $\Psi\Phi^*$ and $\Psi\Theta^*$ is zero, and
\begin{align}
  S_{yx} 
  &\rightarrow \frac{2}{T} \Bigl< \Phi\Phi^* \Bigr>_m \equiv S_\varphi\qquad\text{for large $m$},
  \label{eqn:xsp4}
\end{align}
which is the same as \eqref{eqn:Sphi}. 
Averaging on $m$ acquisition, the background noise of the instrument is rejected by a factor of $\sim1/\sqrt{m}$, i.e., 5 dB per factor-of-ten in $m$. The actual rejection depends on the estimator.

Most phase noise analyzers use the estimator
\begin{align}
  \widehat{S_{yx}}&=\frac{2}{T}\Bigl|\bigl<YX^*\bigr>_m\Bigr|\,.
  \label{eqn:abs}
\end{align}
This estimator is biased and suboptimal \cite{Rubiola-2010-xSp}.
The residual noise terms $\left<\Phi\Theta^*\right>_m$, $\left<\Psi\Phi^*\right>_m$ and 
$\left<\Psi\Theta^*\right>_m$ are proportional to $1/\sqrt{m}$. 

Reference \cite{Rubiola-2010-xSp} shows that the real-part estimator
\begin{align}
  \widehat{S_{yx}}= \frac{2}{T} \Re\Bigl\{ \bigl< YX^* \bigr>_m \Bigr\}
  \label{eqn:real}
\end{align}
is the best option.  
First, it follows from \eqref{eqn:xsp3} that $\Phi\Phi^*\in\mathbb{R}$, the other terms are complex, and the background noise is equally split between $\Re\{YX^*\}$ and $\Im\{YX^*\}$.  Thus, \eqref{eqn:real} is obviously advantageous vs.~\eqref{eqn:abs} 
because it keeps the entire signal and discards the unnecessary noise in $\Im\{YX^*\}$.
The residual single-channel is proportional to $1/\sqrt{2m}$, improving by a factor of 2.  Otherwise stated, the measurement time for a given noise rejection is shorter by a factor of 4.  A further advantage of \eqref{eqn:real} is that it is unbiased.

Sadly, \eqref{eqn:abs} is often the one and only option, and when \eqref{eqn:real} is available, it is not the default.  

The cross-spectrum method derives from radioastronomy \cite{Hanbury-Brown-1952} and from the early attempts to assess the frequency fluctuations of Hydrogen masers \cite{Vessot-1964}.
Reference \cite[Section IV]{Walls-1976} is arguably the first application to the measurement of phase noise.
It goes without saying that the cross-spectrum method is of broader usefulness than just phase noise.  See \cite{Rubiola-2010-xSp} for a tutorial, \cite{Baudiquez-2020} for the use of Bayesian statistics, needed to assess the statistical uncertainty when $m$ is small, and \cite{Baudiquez-2022} for the extension to multiple instruments simultaneously measuring the same physical quantity.

\subsection{Limitations of the Cross-Spectrum Method}\label{sec:Limitations}%
The noise limit $\sim1/\sqrt{m}$ is often shown on the analyzers' screen under fancy names like `correlation factor,'  `number of correlations,' or `Xcorr,' together with $\mathscr{L}(f)$.  However, such noise rejection is useful only if the correlated disturbances are sufficiently small.
In fact, a disturbing term $\delta\leftrightarrow\Delta$ introduced in \eqref{eqn:xspx}-\eqref{eqn:xspy} changes the instrument readout from $S_\varphi$ to $S_\varphi+\varsigma S_\delta$, where $\varsigma=\pm1$ is a sign which depends on whether $\delta$ has same sign or opposite sign in the two channels.  This is a systematic error of unknown sign $\varsigma$.  Consequently, the common belief that the background noise of the instrument always results in the \emph{over-estimation} of the DUT noise is \emph{untrue}.

The obvious benefit of large $m$ is high sensitivity\footnote{Strictly speaking, this is an improper use of the term sensitivity.}, i.e., high capability to measure small $S_\varphi$.  The inevitable drawback is that a proportionally small $S_\delta$ can spoil the measurement.  A gross error occurs when $S_\delta\approx S_\varphi$, but $S_\varphi+\varsigma S_\delta>0$ still holds.  Worse, if $S_\varphi+\varsigma S_\delta<0$, the result is a blatant nonsense.

Among the causes for the disturbances $\delta$, we mention the thermal energy in the input power splitter, the effect of AM on the mixers, and the RF crosstalk inside the instrument.

Most power splitters are either 4-port directional couplers terminated at one port, or Y resistive network.  Such splitters, inherently, add an amount of anticorrelated noise originated from the thermal energy associated to the internal dissipation.  Thus, $S_\varphi<kT/P$ (thermal energy divided by the carrier power) at the splitter output is a physically legitimate outcome, but it goes with a measurement error.
The experimental evidence and the full theoretical proof is found in \cite{Rubiola-2000-RSI}.
The splitter's thermal noise may result in the collapse of the cross spectrum in the measurement of oscillators \cite{Nelson-2014}.  Different options for the power splitter are discussed in \cite{Hati-2016}.  Another perspective on the power splitters is proposed in \cite{Gruson-2017}, with original results.
In \cite{Gruson-2020} we propose a method for the measurement of the bias error due to both power splitter and internal crosstalk.  

The correction for the splitter's thermal energy is trivial if the instrument measures the carrier power.  NoiseXT/Arcale informed us that this correction now has been implemented in the DNA\footnote{Informal discussion with L. Adrien at the IEEE EFTF IFCS Joint Meeting, Paris, April 2022.}\@.

\subsection{Metrologist's Perspective of Phase Noise  Uncertainty}\label{sec:Metrologist}
The essential concepts we need to understand uncertainty are defined by the International Vocabulary of Metrology \cite{VIM-3}, generally known as VIM\@.
The VIM is published by the BIPM, the international organization in charge to ensure worldwide unification of measurements\footnote{The international coordination of metrology is a complex topic even for specialists because it has both scientific and political implications under the \emph{Metre Convention}.  The reader interested can find all information on the BIPM site \url{https://bipm.org}.}.

The statistical noise limit $\sim1/\sqrt{m}$ falls into \emph{Type A evaluation of measurement uncertainty} \cite[entry 2.28]{VIM-3}.  


Other components of uncertainty (crosstalk, AM noise, thermal energy, etc.) cannot be calculated from the measurement outcomes.  They fall into \emph{Type B evaluation of measurement uncertainty} \cite[entry 2.29]{VIM-3}, which relies on the analysis of the system.


Literature of the past 10 years \cite{Nelson-2014,Hati-2016,Gruson-2017,Gruson-2020} and three workshops \cite{L(f)-2014,L(f)-2015,L(f)-2017} point to substantial errors and discrepancies in the measurement of commercial oscillators exhibiting very-low phase noise.  The concept of \emph{null measurement uncertainty} applies \cite[entry 4.29]{VIM-3}, which is the uncertainty in the special case of signal approaching zero.  Broadly speaking, this is the extrapolation of uncertainty to the measurement of a noise-free oscillator, and tells us the minimum detectable amount of phase noise.

We encourage the reader to get awareness of the \emph{definitional uncertainty} \cite[entry 2.27]{VIM-3}, the component of uncertainty resulting from the lack of knowledge of details in the definition of the DUT; and of the \emph{influence quantity}, a quantity affecting the relation between the indication and the measurement result.  However, the direct application of these concepts to phase noise is not straightforward.

Reading the instruments' specs, we often find the \emph{uncertainty} (or improperly, \emph{error}), and the \emph{sensitivity} parameters.  For example, we may read that the uncertainty is $\pm3$ dB for $f<1$ kHz, and $\pm2$ dB for $f\ge1$ kHz.  Similar information may be more detailed, in the form of a table.  Forgiving the `$\pm$' symbol, we interpret this as the type B uncertainty.  The sensitivity takes the form of a table or a spectrum, indicating $\mathscr{L}(f)$ for relevant values of $\nu_0$ and $f$; for example, $-160$ dBc/Hz at 1 GHz carrier and 100 kHz `offset' (actually, modulation frequency).  We are inclined to interpret this parameter as the `null measurement uncertainty.'  Note that the term `sensitivity' found in this context is quite different from the VIM definition \cite[entry 4.12]{VIM-3}.  
That said, we observed that users tend to trust a measured noise level below the `sensitivity' of the instrument, provided it is still higher than the `$\sim1/\sqrt{m}$' statistical limit seen on the screen.
This practice is encouraged by ads showing a `typical' background noise, clearly lower than specs.


\subsection{The Beat Method}\label{sec:Beat}
The method, shown in Fig.~\ref{fig:Beat-method}, makes use of the leverage effect which results from beating $\nu_0$ down to $\nu_b=\nu_0-\nu_\text{ref}$.
Mathematically,
\begin{multline}
  2\cos\bigl[2\pi\nu_0t+\varphi(t)\bigr]\cos\bigl[2\pi\nu_\text{ref}t+\psi(t)]\\[0.8ex]
  = \cos\bigl[2\pi\nu_bt+\varphi(t)-\psi(t)\bigr]\qquad
\end{multline}
after deleting the $\nu_b+\nu_\text{ref}$ term with the lowpass filter.
The beat note preserves the frequency fluctuations $\nu_0-\nu_\text{ref}$ and the phase fluctuation $\varphi-\psi$, and stretches the time fluctuation by a factor $\kappa\simeq\nu_0/\nu_b$.

This method is often used to beat microwave signals down to the RF region, where we can use digital instruments.  In optics, this is the preferred option to measure the fluctuations of a stabilized laser with electrical instruments.

\subsection{The Discriminator Method}
\begin{figure}[t]
  \begin{subfigure}[t]{\columnwidth}\centering
    \caption{\hspace*{0.8\columnwidth}}
    \centering
    \includegraphics[bb = 0cm 0cm 5.9cm 12.4cm,angle=-90,scale=0.64]{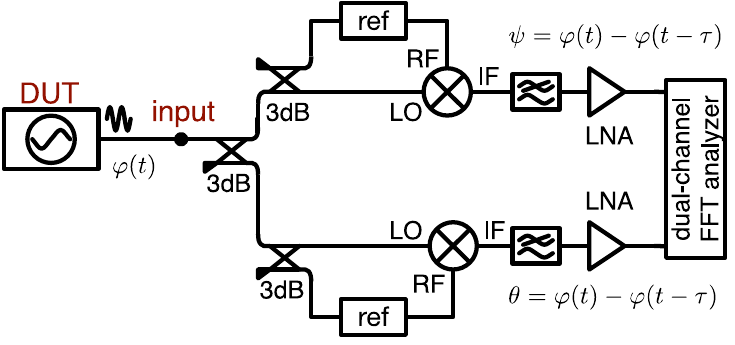}
  \end{subfigure}
  \begin{subfigure}[t]{\columnwidth}
    \vspace{0em}\centering
    \caption{\hspace*{0.8\columnwidth}}
    \includegraphics[bb = 0cm 0cm 2.7cm 11.3cm,angle=-90,scale=0.64]{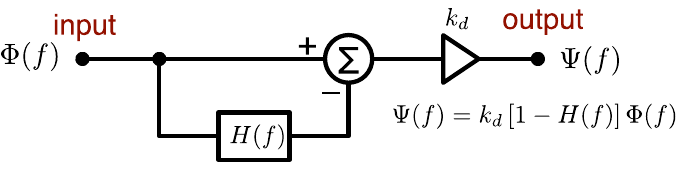}
    \vspace*{1ex}
  \end{subfigure}
  \caption{The discriminator method for the measurement of phase noise in oscillators. (a) Block diagram of the two-channel instrument.  (b) Phase space representation of one channel.}
  \label{fig:Discriminator}
\end{figure}

The use of a reference discriminator, either a resonator or a delay line, is another way to measure the phase noise of an oscillator by comparing the oscillator output to a delayed version of the same signal.

Fig.~\ref{fig:Discriminator} shows the principle and the equivalent scheme in phase space.  The latter follows the same approach used with the oscillator (Sec.~\ref{sec:Oscillators}).  Using the upper case for the Fourier transform of the lower-case function of time, the output is 
\begin{math}
  \Psi(f)=k_d\left[1-H(f)\right]\Phi(f)
\end{math},
where the discriminator's phase response is 
\begin{math}
  H(f)=(1/\tau)/(jf+1/\tau)
\end{math}
for a resonator characterized by the relaxation time $\tau$, and
\begin{math}
  H(f)=e^{-j2\pi\tau f}
\end{math}
for a delay line characterized by the delay $\tau$.  Accordingly, $\Phi(f)$ is evaluated from the analyzer readout as
\begin{align}
  \Phi(f) &=\frac{1}{k_d\left[1-H(f)\right]}\:\Psi(f)
\end{align}
Finally, using the two-channel configuration, the oscillator phase noise is evaluated as 
\begin{align}
  S_\varphi(f) &= \frac{1}{k_d^2\left|1-H(f)\right|^2}\:S_{\psi\theta}(f)\,.
\end{align}

This idea is for sure old, albeit we could not track the origin.  The scheme (Fig.~\ref{fig:Discriminator}) is suitable to implementations based on most or all the phase noise analyzers described in this Section.
The standard 1.55-$\mu$m optical fiber of appropriate length carrying a modulated beam is a good idea to delay a microwave signal.  We implemented a single-channel version \cite{Rubiola-2005-JOSAB}, and we further developed this concept adding the cross spectrum method \cite{Salik-2004,Volyanskiy-2008}.  To our knowledge, the OEwaves OE8000 is the one and only commercial phase noise analyzer based on the photonic delay.

\subsection{The Bridge (Interferometric) Method}\label{sec:Bridge}
\begin{figure}[t]\centering
  \includegraphics[angle=-90, width=0.9\columnwidth]{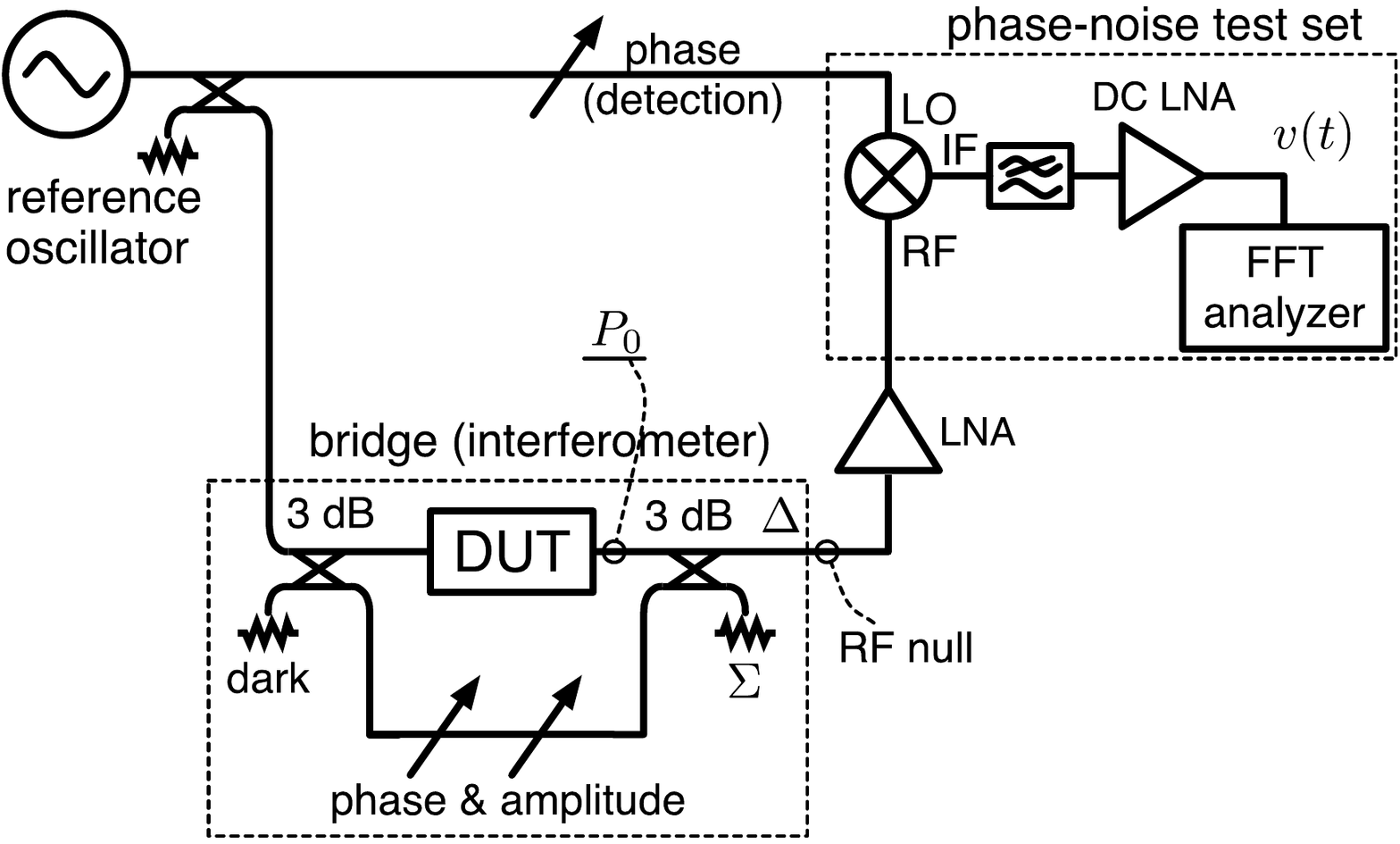}
  \caption{Scheme of the bridge (interferometric) phase noise measurement.}
  \label{fig:Bridge}
\end{figure}

The bridge (Fig.~\ref{fig:Bridge}) is an advanced method exhibiting extremely low background noise.  
The bridge must be at the equilibrium, achieved by setting the bridge's phase and amplitude.  At the equilibrium, the carrier is suppressed at the $\Delta$ port, and all the DUT power $P_0$ goes to the $\Sigma$ port.
The DUT noise sidebands, equally split between $\Sigma$ and $\Delta$, are amplified and synchronously detected by the mixer.  The variable phase at the mixer LO port sets the detection of PM, AM, or a combination of.  Unlike the previous schemes, the mixer works in linear regime, where the RF input is a small signal, and of course the LO input is saturated.

Low background flicker $\mathsf{b}_{-1}$ is the main benefit.  In fact, (i) the passive components used in the bridge (directional couplers, line stretchers, attenuators, etc.) feature extremely low flicker compared semiconductors and active devices, and (ii) the amplifier PM flicker is reduced proportionally to the carrier rejection because flicker comes from up-conversion of the near-dc noise.

The bridge exhibits low white noise as well.  Neglecting trivial losses, the background noise is $\mathsf{b}_0=2FkT_0/P_0$, where $F$ is the amplifier noise factor, $kT_0$ is the thermal energy at room temperature, and $P_0$ is the power at the DUT output.  The factor `2' means that only half the DUT noise goes to the amplifier, the other half goes to the $\Sigma$ port, but it can be reduced using an asymmetrical power combiner.  

Finally, low sensitivity to AC magnetic fields from the power grid is easily achieved because RF amplification (20-40 dB) takes place before down-converting to DC\@.

No commercial solutions exist, but the scheme of Fig.~\ref{fig:Bridge} can be built on top of a commercial phase noise analyzer, including the modern dual-channel instruments.

The main ideas derive from Sann \cite{Sann-1968}, but the RF amplification was introduced later by Labaar et al.~\cite{Labaar-1982}.
The method knew a sudden popularity in the late 1990s in Australia \cite{Ivanov-1998,Ivanov-1998a} and France \cite{Rubiola-1999-RSI}.  We used it for the measurement of frequency stability in piezoelectric quartz resonators \cite{Rubiola-2000-UFFC}, and of the phase noise of DACs and DDSs \cite{Calosso-2020}.  We demonstrated ultimate sensitivity adding the correlation for lowest white noise \cite{Rubiola-2000-RSI}, and a multi-stage bridge for lowest flicker \cite{Rubiola-2002-RSI}.

\subsection[Digital Phase Noise and AVAR Analyzer]{The Digital Phase-Noise and Allan-Variance Analyzer}\label{sec:Analyzer-digital}
\begin{figure}[t]\centering
  \includegraphics[bb = 0cm 15.2cm 13.9cm 29.7cm, angle=0, width=\columnwidth]{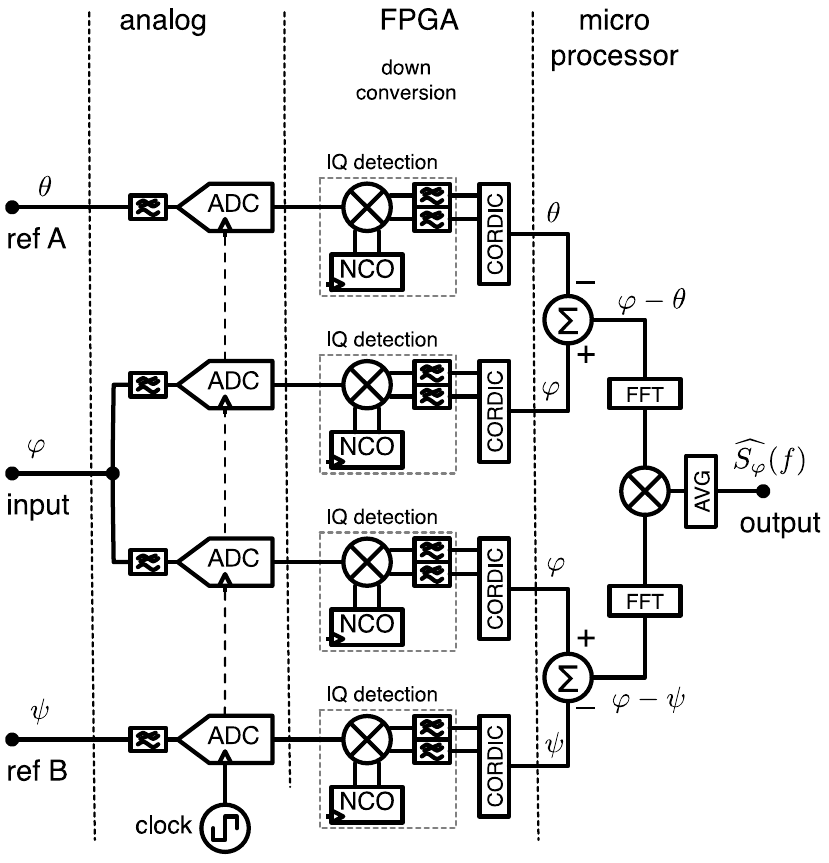}
  \caption{Digital phase noise and Allan variance analyzer.}
  \label{fig:Analyzer-Digital}
\end{figure}

Fig.~\ref{fig:Analyzer-Digital} shows the block diagram of a digital phase-noise analyzer.
The architecture differs from Fig.~\ref{fig:Analyzer-1} in the use of Software Defined Radio (SDR) techniques.  The input signal is digitized and down-converted to an I-Q stream at zero or near-zero frequency by multiplying the input data with sinusoidal signals from a NCO (Numerical Controlled Oscillator).  The CORDIC algorithm \cite{Volder-1959,Volder-2000,Meher-2009} is probably the best option to convert the IQ stream into polar coordinates, phase and amplitude.  The reference input signals A and B cannot be used as the sampling signal because the ADCs do not work well at arbitrary clock frequency.  Consequently, the measurement of $\varphi-\theta$ requires two ADCs clocked by a common-mode oscillator, whose fluctuation is rejected.  In turn, four ADCs are needed to reject the instrument noise. 

The digital architecture has following interesting features
\begin{itemize}
\item It is great at measuring large values of $\varphi(t)$, not limited to $\pm\pi$, and even millions of cycles are not a problem.
\item The use of a separate NCO at each input removes the requirement that the inputs are at the same frequency because all phases can be referred to the same $\nu_0$ after a trivial numerical conversion.
\item Full control on $f_H$ in the measurement of Allan variances.
\end{itemize}
These properties make the digital architecture great at measuring both phase noise and variances, and open new perspectives.

Reference \cite{Grove-2004} pioneered the measurement of phase noise and the Allan variance with direct digitization of the RF signal.  Later, \cite{Mochizuki-2007} provides a more detailed treatise, and \cite{Sherman-2016} focuses on SDR techniques.  
The experimental methods for the noise characterization of ADCs for AM/PM noise measurements are shown in \cite{Cardenas-2017}, and \cite{Calosso-2020} introduces a new method for the measurement of AM and PM noise in DACs and DDSs based on the amplification of the (random) modulation index, with optional AM/PM and PM/AM conversion.

Digital instruments are commercially available from Arcale (NoiseXT) and Microchip (Jackson Labs Technologies and Microsemi).  The maximum frequency is limited to 30--400 MHz, depending on the instrument.  Such instruments may have the reference oscillator(s) inside, or rely entirely on external oscillators. 
If the inputs A and B are accessed through a single connector instead of being available separately, the instrument cannot reject the noise of such oscillator.
A hobby project is available from Andrew Holme's home page\footnote{\url{http://www.aholme.co.uk/PhaseNoise/Main.htm}}, based on commercial boards, which even includes the source code and the binaries.

The Rohde \& Schwarz analyzers FSWP and FSPN are based on similar concepts, but they implement microwave-to-IF down-conversion to extend the input range to 8, 26 or 50 GHz, depending on the model.  They include two OCXOs and synthesizers as the references, and work with external references as well.
The internal architecture of the FSWP is explained in \cite{Feldhaus-2016}.

\subsection{Tracking DDS (TDDS)}\label{sec:TDDS}
\begin{figure}[t]\centering
  \hspace*{-4mm}%
  \includegraphics[bb = 0mm 192mm 101mm 296mm,width=0.8\columnwidth,angle=0]{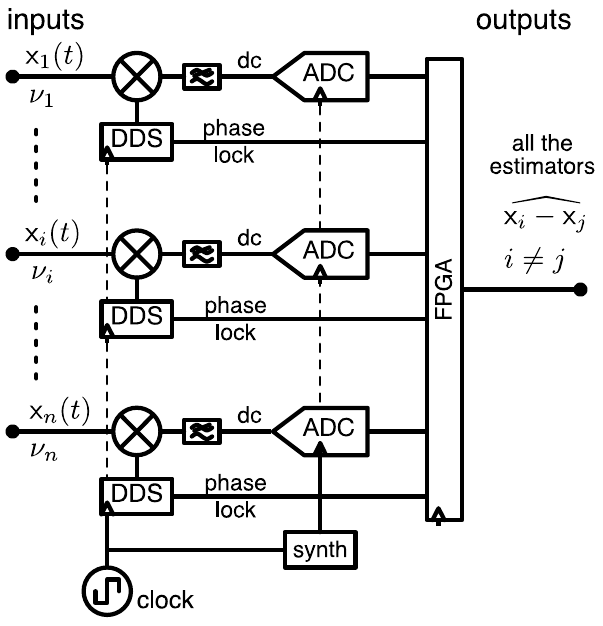}\\[1ex]
  \caption{Tracking DDS.}
  \label{fig:TDDS}
\end{figure}

The TDDS \cite{Calosso-2013a} is a phase locked loop where the voltage-controlled oscillator is replaced with a Direct Digital Synthesizer (DDS).  
The lock differs from a regular PLL in that the FPGA acts on the numerical phase of the DDS, instead of on the frequency of a voltage-controlled oscillator (VCO).  Thus, an additional integrator is needed in the control loop.
Interestingly, the TDDS locks in a wider frequency range than most PLLs, being limited by the frequency range of the mixer and of the DDS\@.  The background noise is limited by the DDS (see \cite{Calosso-2012-IFCS} for the DDS noise), noisier than double-balanced mixers.

Fig.~\ref{fig:TDDS} shows a multichannel TDDS\@.  At the start of operation, the FPGA sets the numerical frequency of all the DDSs to the frequency of the respective input.  Then, the FPGA phase-locks each DDS to the input.  At the same time, the FPGA provides the estimation of all the phase-time differences $\widehat{\mathsf{x}_i-\mathsf{x}_j}$ by combining the numerical phases and dc errors. 

A multichannel TDDS was succesfully used for the direct measurement of the frequency stability at the 100 MHz output of cryogenic sapphire oscillators \cite{Calosso-2019}.
The Italian institute of metrology INRiM has developed the \emph{Time Processor}
a multichannel TDDS for the continuous monitoring and stability measurement of high-end oscillators and atomic frequency standards.  Albeit some prototypes have already been transferred under contract to qualified users (European Space Agency and the FEMTO-ST Institute), until now there is only a conference publication \cite{Calosso-2021}.   The Principal Investigator informed us\footnote{Claudio E. Calosso, private communication, October 2022.} that a startup Company is being created, under the provisional name Kairos TeX\@.  
The PicoPak\footnote{Hamilton Technical Services, \url{http://www.wriley.com/7510A.pdf}.}, a single channel TDDS, was produced by Precision Time and Frequency, LLC, but it is unsure whether it is still available.

\subsection[Multichannel Analog AVAR Analyzer]{Multichannel Analog Allan Variance Analyzer}
\begin{figure}[t]\centering
  \includegraphics[bb = 0cm 0cm 11.3cm 13cm, angle=-90, width=0.9\columnwidth]{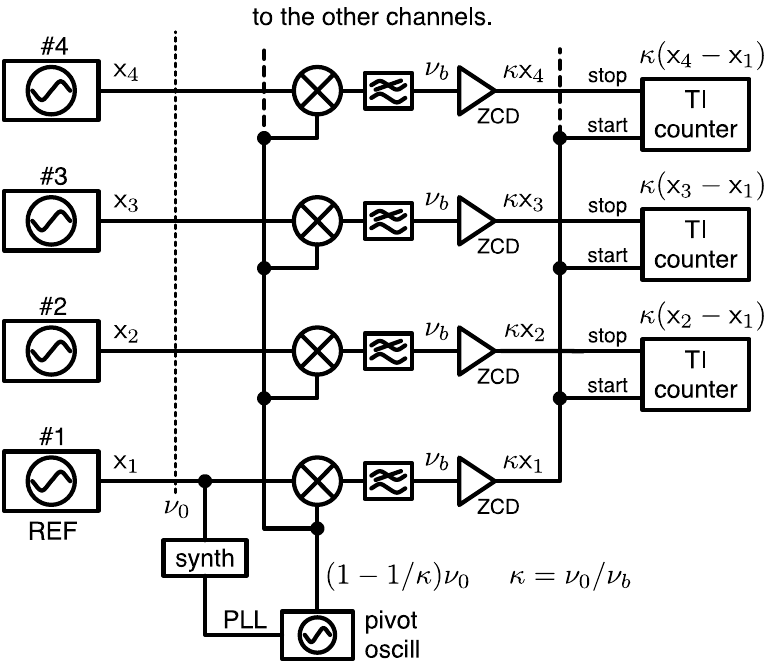}
  \caption{Multichannel analog Allan variance analyzer.}
  \label{fig:Multi-mixer}
\end{figure}

This type of analyzer is a highly specialized instrument intended to simultaneously monitor multiple oscillators in a time scale (Cs beams, fountains, and H masers), usually comparing the 10 MHz or the 100 MHz outputs.  Fig.~\ref{fig:Multi-mixer} shows an example, easily extended to more than 4 inputs.

The machine exploits the leverage effect seen in Section \ref{sec:Beat}, but in this case 
the pivot frequency $\nu_p$ is just below $\nu_0$, so that $\nu_b$ is in the sub-audio or low audio range.  
Accordingly, $\mathsf{x}(t)=\varphi(t)/2\pi\nu$ is stretched by a factor $\kappa=\nu_0/\nu_b$.  With $\kappa=10^6$, a 100 MHz signal can be measured with 10 fs resolution using a simple counter that has a resolution of 10 ns.  However, the actual resolution is limited by other factors.

The dual-mixer method, later extended to the multichannel system shown in Fig.~\ref{fig:Multi-mixer}, is introduced in \cite{Allan-1975}.  Reference \cite{Brida-2002} discusses the design of the dual-mixer system, providing interesting experimental data.
The best choice for the zero-crossing detector (ZCD) is a multistage amplifier where the first stage has narrow bandwidth for low noise, and the bandwidth increases progressively towards the output to allow high slew rate in the saturated signal \cite{Dick-1990,Collins-1996}.

Major metrology labs have been using instruments based on this principle for many years.  However, those instruments are house-built prototypes.  Commercial units are available from Quartzlock (UK), Time Tech (Germany), and VREMYA-CH (Russia).

\ifSubfilesClassLoaded{\bibliography{Ref-short,References,Ref-local}}{}
\end{document}

\section{Software Tools\label{sec:Software}}

\ifSubfilesClassLoaded{\setcounter{section}{10}\section{Software Tools}}{}

\begin{table}[t]
  \centering
  \caption{{\hspace*{-8em}}\\[1ex]\centering\textsc{Software Tools for the Two-Sample Variances}}
  \label{tab:Software-tools}
  \includegraphics[trim={30mm 90mm 62mm 22mm},clip,width=0.98\columnwidth]{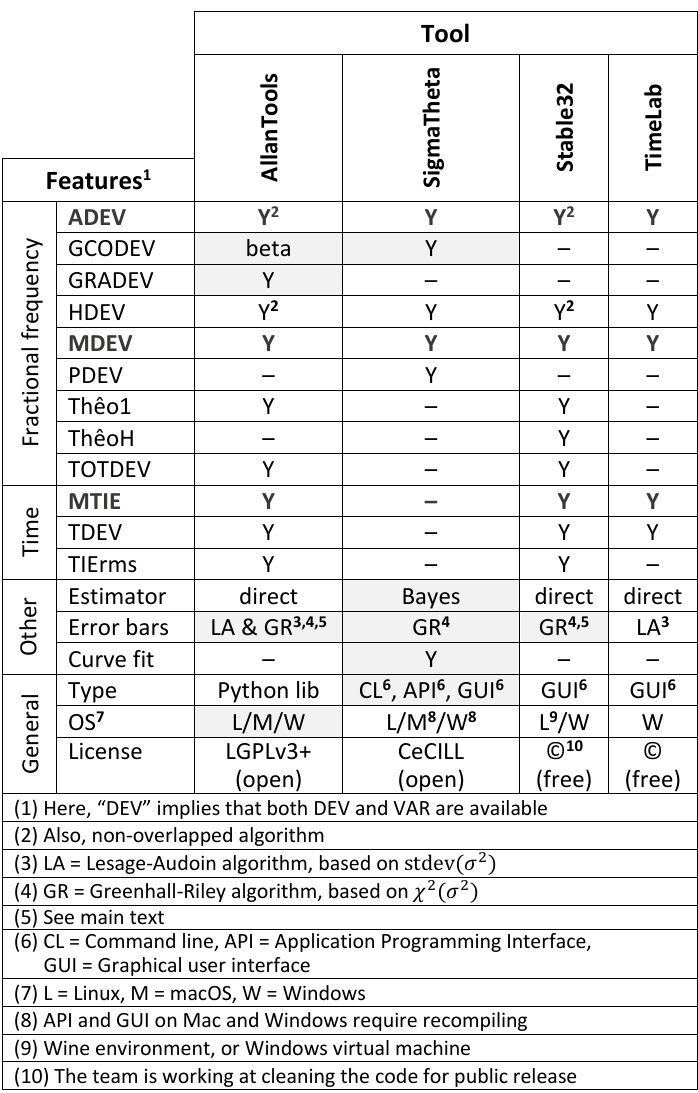}
\end{table}

A quick Internet search reveals that there is a rather broad choice of software packages for the Allan variance, but almost none for phase noise.  This relates to the fact that the variances are used to investigate slow phenomena, thus they require rather low sampling rate and small digital storage space.  A value of 1 kS/s is generally sufficient for all practical cases, but 1 S/s or 10 S/s is most often used in atomic time scales.  Conversely, it is quite common to plot the phase noise up to 1 MHz Fourier frequency.  This requires a practical minimum of 2.5 MS/s, allowing a mere 250 kHz for the anti aliasing filter to roll off.  In turn, a transfer rate of 10 MB/s is necessary for a cross-spectrum system under the hypothesis that the phase is encoded on 16 bits.  Thus, a measurement lasting 100 s takes 1 GB disk space.  To the best of our knowledge, noise analyzers do not save or transfer raw data of this size and at this rate.  
TimeLab\footnote{Information provided by John Miles, September 2022.} is no exception, to the extent that continuous gap-free IQ data are transferred from the 3120A/5330A/53100A instruments to the computer at reduced bandwidth (100 kHz).  Faster IQ data required to plot spectra up to the maximum frequency of 1 MHz are transferred in bursts with 10\% duty cycle.  Only the spectra bins are saved in the .TIM files, together gap-free data at significantly lower rate (1 kS/s) for time measurements and Allan variances.

We present a selection of software packages (Table~\ref{tab:Software-tools}), chosen for their scientific value or for their wide use.  The grayed area highlight some interesting features.

\subsection{Features}

\subsubsection{Main purposes} AllanTools, SigmaTheta and Stable32 are intended for data analysis.  By contrast, TimeLab is a tool for data acquisition, with limited analysis capabilities.  It supports the phase noise analyzers from Jackson Lab and Microchip (formerly Microsemi), and a few frequency counters.

\subsubsection{Graphical interface versus scripting}  Albeit elderly, Stable32 has an efficient graphical interface, which makes it a great choice for occasional users.  For this reason, it is by far the most widely used.  AllanTools and SigmaTheta require programming skills.  On the other hand, scripting is great in that it enables the analysis of a bulk of data sets at once.

\subsubsection{Mathematical functions}
All the packages provide ADEV, MDEV and the other mostly-used functions.  
For simplicity, DEV in Tab.~\ref{tab:Software-tools} stands for DEV or VAR\@.
When it comes to more exotic functions like Thêo and the Groslambert covariance, the choice of packages is smaller.

\subsubsection{Estimator}
The unique feature of SigmaTheta is that it uses the Bayesian statistics (inverse problem) to estimate the variances.  The other packages evaluate the \emph{average} value using the classical formulae like \eqref{eqn:AVAR-evalx} for AVAR\@.

\subsubsection{Missing data}
The Gap Resistant ADEV (GRADEV), included only in AllanTools, enables the evaluation of ADEV in the case of missing data during the measurement.

\subsubsection{Error bars} The basic choice is between the classical Lesage-Audoin algorithm \cite{Lesage-1973}, and the more sophisticated Greenhall-Riley algorithm \cite{Greenhall-2003}.  The LA algorithm uses the square root of the 4th moment, assuming that the error distribution is Gaussian centered around the estimate. 
By contrast, the GR algorithm estimates the degrees of freedom using the $\chi^2$ distribution.  For each value of $\tau$, the number of samples and the dominant noise type are taken into account.

However, Wallin points out that the identification of noise processes is also needed \cite{Wallin-2018b}.  In the same blog, it is said that AllanTools has no `intelligent' top-level algorithm that takes a time-series as input, and would choose between lag-1-ACF, $B1$ and $R(n)$, and automatically determine the power-law noise type.  Wallin suggests that Stable32 may use a combination of lag-1-autocorrelation \cite{Riley-2004}, $B1$-ratio \cite{Howe-2000a,Barnes-1969}, and $R(n)$.

Finally, the principles used inside SigmaTheta are discussed in Sec.~\ref{sec:Bayes}.

\subsection{Software Availability}
The availability is summarized in Tab.~\ref{tab:Availability}.  AllanTools and SigmaTheta are open code released to public domain under very similar licenses.  Stable32 is \copyright\ IEEE\@.  The developers informed us\footnote{Magnus Danielson, private email.} that the code itself is covered by the MIT license, but compiling needs code from other sources.  They are trying to clean the code for it to be released into the public domain, provided the IEEE agrees, with the ultimate goal of having at least Linux, macOS and Windows distributions available.

\begin{table}[t]\centering\sffamily
\caption{{\hspace*{-3em}}\\[1ex]\centering\textsc{Software Availability and Developpers}\label{tab:Availability}}
\vspace{-1em}
\newcommand{\vrA}{\rule[-1ex]{0pt}{3.5ex}}
\newcommand{\vrB}{\rule[-1.2ex]{0pt}{3ex}}
\newcommand{\vrC}{\rule[0ex]{0pt}{1.8ex}}
\newcommand{\vrD}{\rule[-1.2ex]{0pt}{1.2ex}}
\begin{tabular}{ll}\\\hline
\textbf{AllanTools} &\vrA\\
\multicolumn{2}{l}{Anders E. E. Wallin, Danny Price, Cantwell G. Carson,}\\
\multicolumn{2}{l}{Frédéric Meynadier, Yan Xie, and Eric Benkler}\\
Anaconda& \url{https://anaconda.org/conda-forge/allantools}\vrA\\
GitHub  & \url{https://github.com/aewallin/allantools}\vrB\\
PyPI    & \url{https://pypi.org/project/AllanTools/}\vrB\\\hline
\textbf{SigmaTheta} &\vrA\\
\multicolumn{2}{l}{François Meyer, François Vernotte, and Attila Kinali, Benoit Dubois}\\
Cmd line  & \url{https://gitlab.com/fm-ltfb/SigmaTheta}\vrA\\
API       & \url{https://gitlab.com/bendub/libsigmatheta}\vrC\\
          & Not tested on Mac/Win, some funct.\ still missing.\vrB\\
GUI       & \url{https://gitlab.com/bendub/pysigmatheta}\vrC\\
          & $\beta$-test on Linux, lives on API via Python wrapper\vrD\\\hline
\textbf{Stable32} &\vrA\\   
\multicolumn{2}{l}{Magnus Danielson, Vivek D. Dwivedi \& al.}\\
\multicolumn{2}{l}{Originally, William J. Riley (retired)}\\
GitHub  & \url{https://github.com/IEEE-UFFC/stable32}\vrA\\
IEEE    & \url{https://ieee-uffc.org/software}\vrB\\ 
Stable32& \url{http://www.stable32.com/162Stable32.exe}\vrB\\\hline
\textbf{TimeLab} &\vrA\\
\multicolumn{2}{l}{John Miles, KE5FX}\\
Miles Design  & \url{http://www.miles.io/timelab/beta.htm}\vrA\\\hline
\end{tabular}
\end{table}

\ifSubfilesClassLoaded{\bibliography{References,Ref-local}}{}
\end{document}

\section*{Acknowledgments}
\addcontentsline{toc}{section}{Acknowledgments}
We express gratitude to the Nobel laureate John Lewis ``Jan'' Hall for his authoritative and friendly encouragement\footnote{Private email, February 5, 2020.}
\begin{quotation}
  \ldots\emph{But the best thing for me is the link to your new Chart, collecting the multiple information about how to analyze measured phase noise spectra to begin to grasp the underlying causes.  I have been using a 10th-generation photocopy of an early precursor to your chart, which came many years ago from either Don Halford or Helmut Hellwig at the NBS/NIST}\ldots
\end{quotation}

We thank the users of the earlier versions of the Chart for feedback and corrections.

\bibliography{References}

\appendix
\section*{Notation and Symbols}
\addcontentsline{toc}{section}{Notation, Symbols and Acronyms}

\ifSubfilesClassLoaded{\appendix\setcounter{section}{0}\section{Notation and Symbols}}{}

\begin{description}
  \item[dot ($\dot{a}$)]\quad derivative over time
  \item[bar ($\bar{a}$)]\quad mean, or weighted average
  \item[hat ($\hat{a}$)]\quad estimation
  \item[$\leftrightarrow$] Fourier or Laplace transform inverse-transform pair
  \item[$\left<~\right>$] mathematical expectation, same as $\mathbb{E}\{\:\}$
  \item[$\left<~\right>_\tau$] average over a specified time interval $\tau$
  \item[$\left<~\right>_n$] average over an integer number $n$ of samples
  \item[$B$] bandwidth of the baseband signal, and of AM/PM noise (the bandwidth of the RF signal is $2B$)
  \item[$\mathsf{b}_n$] coefficients of the polynomial law of $S_\varphi(f)$
  \item[$\mathsf{D_y}$] fractional-frequency drift
  \item[$\mathfrak{D}$] degrees of freedom (often denoted with $\nu$, dof or DOF in the literature about statistics)
  \item[$\mathsf{d}_n$] coefficients of the polynomial law of $S_{\nu}(f)$
  \item[$\mathbb{E}\{\:\}$] mathematical expectation
  \item[$F$] noise factor (see NF below)
  \item[$f$] Fourier frequency, in spectral analysis
  \item[$f_c$] corner frequency, where flicker noise equals white noise
  \item[$f_L$] resonator's Leeson frequency (baseband bandwidth), half of the RF bandwidth
  \item[$\mathsf{h}_n$] coefficients of the polynomial law of $S_\mathsf{y}(f)$
  \item[$H(f)$] transfer function, typically in $|H(f)|^2$
  \item[$\mathsf{k}_n$] coefficients of the polynomial law of $S_\mathsf{x}(f)$
  \item[$\mathscr{L}(f)$] phase noise, usually $10\log_{10}[\mathscr{L}(f)]$ [dBc/Hz]
  \item[$N$] white noise PSD [W/Hz]
  \item[NF] Noise Figure, $\unit{NF}=10\log_{10}(F)$
  \item[$P$] carrier power
  \item[$Q$] quality factor, in resonators
  \item[$S_x(f)$] one-sided power spectral density of the random variable $x(t)$.  For us, chiefly $\alpha$, $\varphi$, $\mathsf{x}$, $\mathsf{y}$ and $\nu$
  \item[$T$] time interval, or period
  \item[$T$] acquisition time (data record for one FFT)
  \item[$T$] (subscript) signal truncated over a duration $T$, as in $x_T(t)$ and $X_T(f)$
  \item[$\mathcal{T}$] total acquisition time (full data record used in the measurement)
  \item[$t$] time
  \item[$V_0$] peak amplitude (of the clock signal)
  \item[$w(t;\tau)$]\quad weight function (counters), or wavelet-like function (variances)
  \item[$\mathsf{x}(t)$] time fluctuation
  \item[$\mathsf{y}(t)$] fractional frequency fluctuation
  \item[$\alpha(t)$] fractional amplitude fluctuation
  \item[$(\Delta\nu)(t)$]\quad frequency fluctuation
  \item[$\epsilon(t)$] amplitude fluctuation
  \item[$\theta(t)$] random phase, replacement for $\varphi(t)$, when needed
  \item[$\Lambda$] triangular average, or a frequency counter implementing triangular average
  \item[$\nu$] carrier frequency
  \item[$\Pi$] uniform average, or a frequency counter implementing uniform average
  \item[$\sigma^2_\mathsf{x}(\tau)$]\quad same as TVAR, used in formulas
  \item[$^{\scriptscriptstyle A\!}\sigma^2_\mathsf{y}(\tau)$]\quad same as AVAR, used in formulas
  \item[$^{\scriptscriptstyle M\!\!}\sigma^2_\mathsf{y}(\tau)$]\quad same as MVAR, used in formulas
  \item[$^{\scriptscriptstyle P\!\!}\sigma^2_\mathsf{y}(\tau)$]\quad same as PVAR, used in formulas
  \item[$^{\scriptscriptstyle H\!\!}\sigma^2_\mathsf{y}(\tau)$]\quad same as HVAR, used in formulas
  \item[$\tau$] delay (delay line)
  \item[$\tau$] measurement (integration) time
  \item[$\tau$] relaxation time, in resonators
  \item[$\tau$] time shift (correlation, convolution, etc.)
  \item[$\varphi(t)$] random phase
  \item[$\psi(t)$] random phase, replacement for $\varphi(t)$, when needed
  \item[$\omega$] shorthand for $2\pi\nu$
  \item[$\Omega$] shorthand for $2\pi f$
  \item[$\Omega$] parabolic-weight average, or a frequency counter implementing such average
\end{description}

\section*{Acronyms}
\begin{description}
  \item[ADC] Analog to Digital Converter
  \item[AM] Amplitude Modulation (also AM noise)
  \item[ADEV]\quad Square root of AVAR
  \item[AVAR]\quad Allan VARiance
  \item[BER] Bit Error Rate (digital electronics and telecom)
  \item[BPF] Band Pass Filter
  \item[CORDIC]\qquad COordinate Rotation DIgital Computer, or Volder's algorithm 
  \item[DAC] Digital to Analog Converter
  \item[DBM] Double Balanced Mixer
  \item[DDS] Direct Digital Synthesizer
  \item[DUT] Device Under Test (oscillator or two-port component)
  \item[FFT] Fast Fourier Transform
  \item[FM] Frequency Modulation (also FM noise)
  \item[FPGA] Field Programmable Gate Array
  \item[FS] FemtoSecond (laser)
  \item[GRADEV]\qquad Gap Resistant ADEV
  \item[HDEV]\quad Square root of HVAR
  \item[HVAR]\quad Hadamard VARiance
  \item[IF] Intermediate Frequency (superheterodyne receiver).  Also IF output of a mixer
  \item[IQ] In-phase and Quadrature (detection, modulation)
  \item[LO] Local Oscillator (superheterodyne). Also LO input of a mixer
  \item[LPF] Low Pass Filter
  \item[LSB] Lower Side Band, in modulated signals
  \item[MDEV]\quad Square root of MVAR
  \item[NCO]\quad Numerically Controlled Oscillator
  \item[MVAR]\quad Modified [Allan] VARiance
  \item[PLL] Phase Locked Loop
  \item[PM] Phase Modulation (also PM noise)
  \item[PDF] Probability Density Function
  \item[PSD] Power Spectral Density
  \item[PDEV]\quad Square root of PVAR
  \item[PVAR]\quad Parabolic VARiance
  \item[RF] Radio Frequency.  Also RF input of a mixer
  \item[SDR] Software Defined Radio
  \item[TDDS] Tracking Direct Digital Synthesizer
  \item[TIE] Time Interval Error \cite{ITU-G.8260}
  \item[TVAR] Time VARiance
  \item[USB] Upper Side Band, in modulated signals
  \item[VCO] Voltage Comtrolled Oscillator.  Also voltage-control input of an oscillator 
  \item[ZCD] Zero Crossing Detector
\end{description}

\ifSubfilesClassLoaded{\bibliography{Ref-short,References,Ref-local}}{}
\end{document}


\clearpage
\section*{\color{BrickRed}Corrections to Previous Versions\label{sec:Corrections}}
\addcontentsline{toc}{section}{\color{BrickRed}Corrections to previous versions}

\ifSubfilesClassLoaded{\section*{Corrections to Previous Version}}{}

\subsection*{Equation \eqref{eqn:HA}}
In the original version, the square was missing in the denominator \((\pi\tau f)^2\) (Thanks to Robert Jördens).  The correct equation is
\begin{equation*}
  |H_A(f;\tau)|^2=2\,\frac{\sin^4(\pi\tau f)}{(\pi\tau f)^2}
  \tag{\ref{eqn:HA}}
\end{equation*}

\subsection*{Equation \eqref{eqn:All-variances}}
In the original version, the square brackets were erroneously placed inside the integral.  
The correct equation is
\begin{equation}
  \sigma^2_\mathsf{y}(\tau) = 
  \mathbb{E}\biggl\{\biggl[\int_0^\infty\mathsf{y}(t)\,w(t;\tau)\:dt\biggr]^2 \biggr\}\,.
  \tag{\ref{eqn:All-variances}}
\end{equation}
The same error is present in some earlier versions of the Enrico's Chart, region 2.2.

\subsection*{Equation \eqref{eqn:TVAR}}
In the original version, the square was missing in \({}^{M\!\!}\sigma^2_\mathsf{y}(\tau)\) on the right-hand side (Thanks to Robert Jördens).  The correct equation is
\begin{align*}
  \sigma^2_\mathsf{x}(\tau) = \dfrac{1}{3}\: \tau^2 \; {}^{M\!\!}\sigma^2_\mathsf{y}(\tau)\,.
  \tag{\ref{eqn:TVAR}}
\end{align*}

\subsection*{A More Subtle Error}
There exist two definitions of the Hadamard variance HVAR (see next Section for a discussion), which differ for a normalization coefficient.  So, we have introduced a factor of 1.5 (+1.76 dB) in order to comply with the most widely used software (Stable 32 and AllanTools), and the package SigmaTheta has been updated accordingly.  Thanks to Nils Nimitz for reporting the problem. 

Accordingly, the 3rd unnumbered equation in Sec.~\ref{sec:Appropriate-var} (page~3010 of the IEEE MTT article) is replaced with
\begin{align*}
  ^{H\!\!}\sigma^2_\mathsf{y}(\tau) 
  &= \frac{1}{2}\:\frac{\mathsf{h}_0}{\tau}
    \simeq 0.50\:\frac{\mathsf{h}_0}{\tau} 
  &&\text{(Hadamard)}
\end{align*}
Here, the coefficients are changed as \(1/3\rightarrow1/2\), thus \(0.33\rightarrow0.50\).  
Sadly, this correction propagates in numerous places of the Enrico's chart, namely: 
\begin{itemize}
  \item The weight functions \(w_\Delta\) plotted in region 2.1, and \(w_H\) plotted in region 2.2.
  \item All the coefficients in region 2.7, column ``HVAR.''
  \item All the coefficients of asymptotic plot of HVAR in region 2.5.  In contrast, value of \(\tau\) of the corners are unchanged.
  \item The transfer function \(\left|H(f)\right|^2\) of HVAR plotted in region 1.5.  Of course, only the red curve is affected, which relates to HVAR.
\end{itemize}
The changes are highlighted in Fig.~\ref{fig:Corrections} with a yellow background.  
The corrected version of the Chart is now available from Zenodo, \doi{10.5281/zenodo.4399218}.
This DOI will always resolve to the latest version.

\begin{figure}[t]
  \centering
  \includegraphics[angle=0,width=0.7\columnwidth]{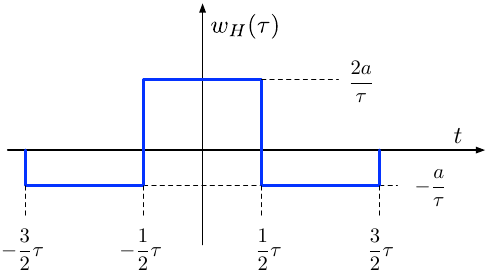}
  \caption{Measurement pattern for the Hadamard variance.}
  \label{fig:Hadamard}
\end{figure}

\subsection*{The Origin of the Subtle Error}
The Hadamard variance — as the term is currently used in the community of time and frequency — can be defined as
\begin{align}
  \sigma_\mathsf{y}^2(\tau)=E\left\{a^2 \left[-\bar{\mathsf{y}}_1+2\bar{\mathsf{y}}_2-\bar{\mathsf{y}}_3 \right]^2 \right\}
  \label{eqn:Hadamard}
\end{align}
where \(E\{\;\}\) is the mathematical expectation, \(\bar{\mathsf{y}}\) is the fractional frequency fluctuation averaged on a time \(\tau\), and \(a^2\) is a normalization factor.  The corresponding measurement pattern is shown in Fig.\ref{fig:Hadamard}.  In practice, \eqref{eqn:Hadamard} is estimated from a data stream of \(M\) samples denoted with \(\mathsf{y}_i\), \(i=1\ldots M\)
\begin{align}
  \widehat{\sigma_\mathsf{y}^2} 
  &= a^2 \frac{1}{(M-2)}\sum_{i=1}^{M-2}
  \left[-\mathsf{y}_{i+2}+2\mathsf{y}_{i+1}-\mathsf{y}_i\right]^2
  \label{eqn:Hadamard-estimate}
\end{align}

The normalization used in Stable 32 and AllanTools, and in Riley's 2008 booklet \cite{Riley-2008} is \(a^2=1/6\). However, the same variance is found in Rutman 1978 \cite{Rutman-1978} under the name modified three-sample variance, with the normalization factor \(a^2=1/9\), and later called Picinbono variance.  It turns out that Gagnepain was the PhD advisor of one of us (FV) in the early 1990s.  So, we have been using the French normalization \(a^2=9\) since, and at some point, we replaced the name Picinbono with Hadamard without realizing that rest of the world started using \(a^2=1/6\).  It is worth mentioning that the factor of 1.5 (+1.76 dB) between the two normalizations is hard to spot on the plots found at conferences and in journals.

The frequency response associated to \eqref{eqn:Hadamard} is
\begin{align}
  \left|H(f;\tau)\right|^2 
  &= 16 a^2\, \frac{\sin^6(\pi \tau f)}{(\pi \tau f)^2}
\end{align}
thus,
\begin{align}
  \int_0^\infty |H(f;\tau)|^2 \, df
  &= \int_0^\infty 16 a^2 \, \frac{\sin^6(\pi \tau f)}{(\pi \tau f)^2}\,df
  = \frac{3a^2}{\tau}
\end{align}
Accordingly, the response to white FM noise \(S_\mathsf{y}(f)=\mathsf{h}_0\) is
\begin{align}
  \sigma_\mathsf{y}^2(\tau)
    &=\frac{1}{3}\frac{h_0}{\tau} \simeq 0.33\frac{\mathsf{h}_0}{\tau}
    \qquad
    \text{with \(a^2=1/9\)}
\intertext{in the original Companion, and}
  \sigma_\mathsf{y}^2(\tau)
    &=\frac{1}{2}\frac{h_0}{\tau}=0.5\frac{\mathsf{h}_0}{\tau}
    \qquad
    \text{with \(a^2=1/6\) (current version)}
\end{align}  
for the corrected version.
Obviously, the response to the other noise types differs by the same factor of \(3/2\).

\begin{figure*}[t]
  \centering
  \includegraphics[angle=0,width=0.94\textwidth]{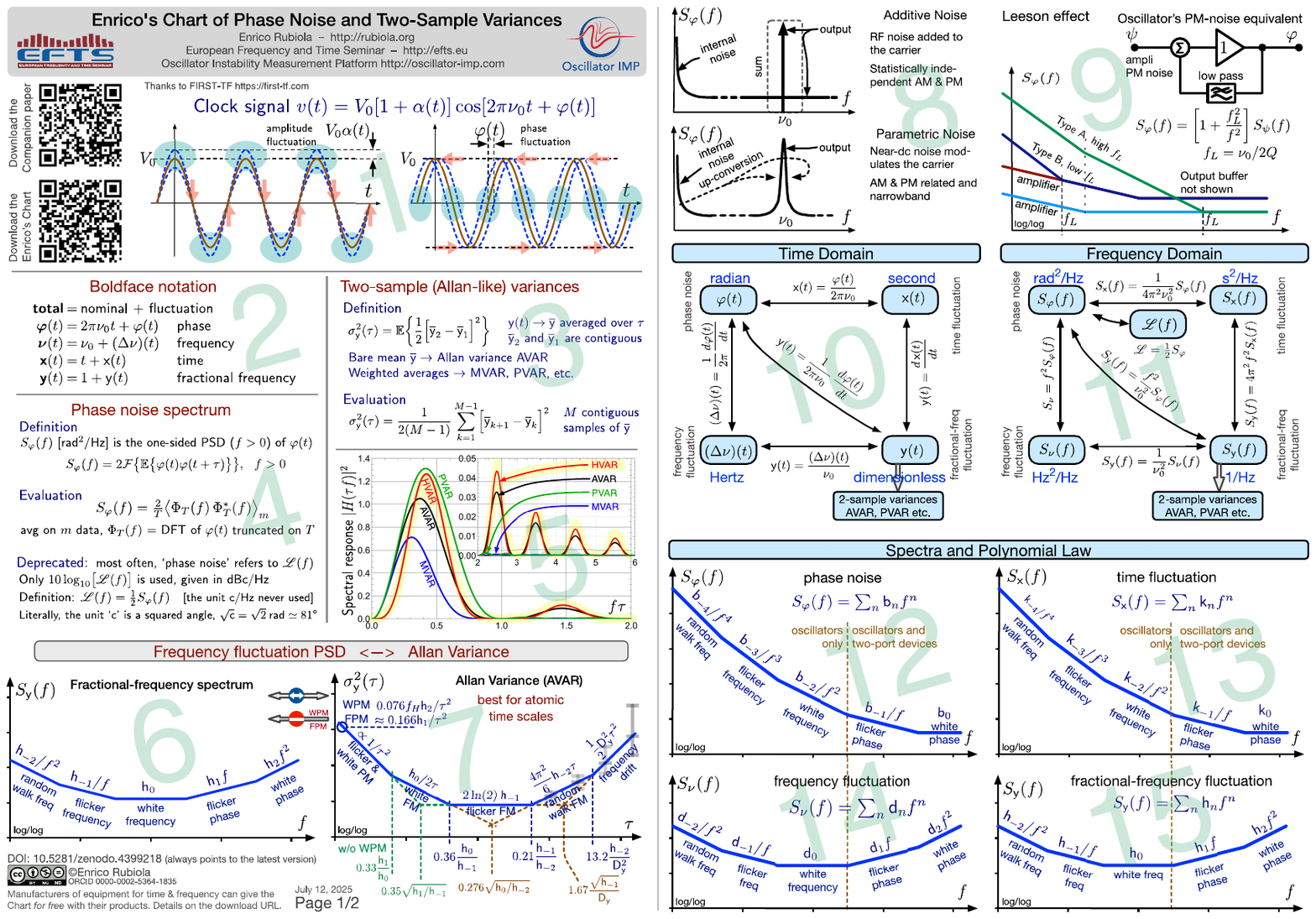}
  \includegraphics[angle=0,width=0.94\textwidth]{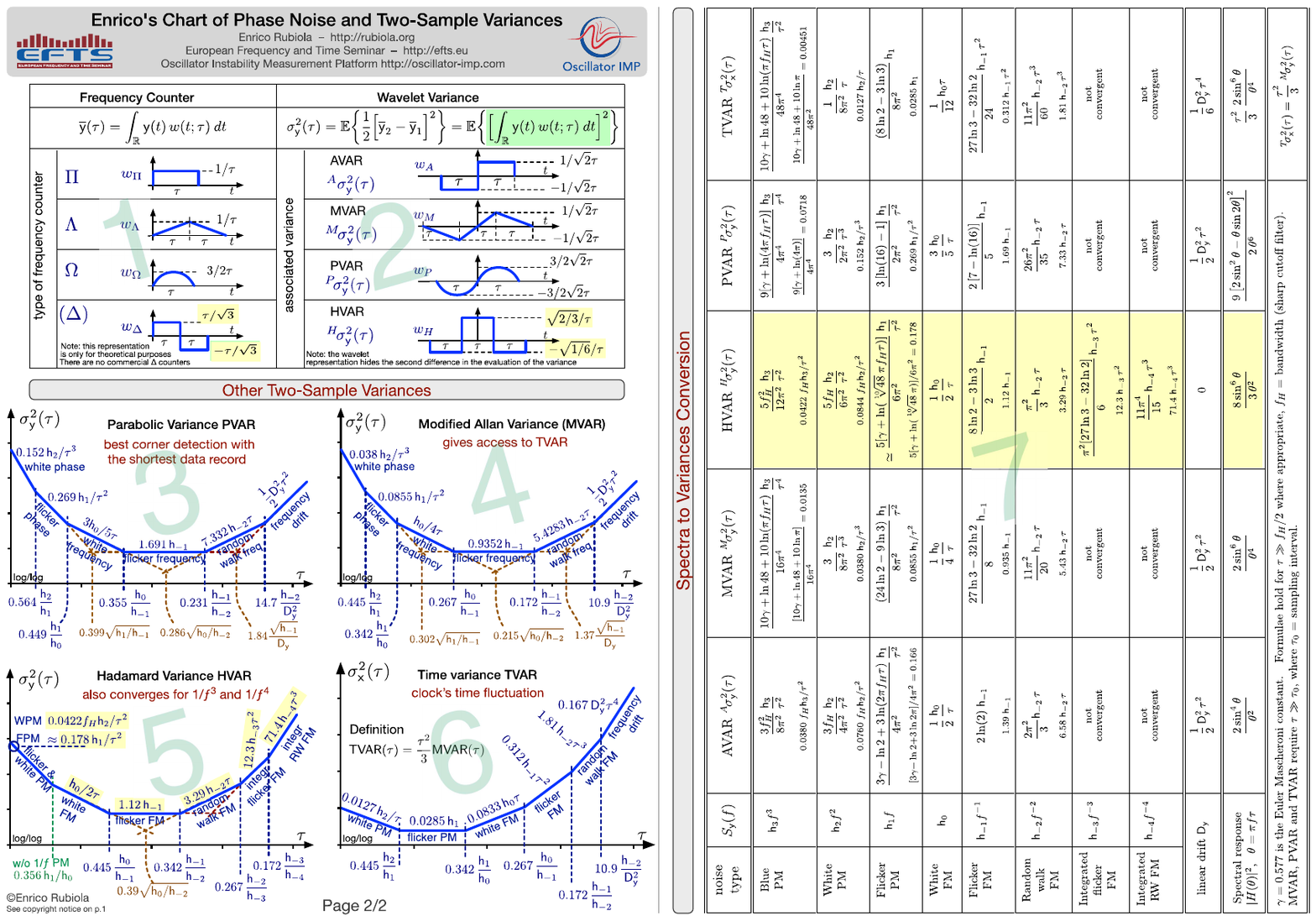}
  \caption{Corrections to the Enrico's Chart, highlighted in yellow and green.}
  \label{fig:Corrections}
\end{figure*}

\ifSubfilesClassLoaded{\bibliography{Ref-short,References,Ref-local}}{}
\end{document}


\end{document}